\documentclass[aps,prl,twocolumn, groupedaddress,showpacs,showkeys, floatfix]{revtex4-1}
\usepackage{graphicx}
\usepackage[dvipsnames]{xcolor}
\usepackage{amsmath,amssymb,latexsym}
\usepackage[english]{babel}
\usepackage{dcolumn}
\usepackage{bm}
\usepackage{float}
\usepackage[normalem]{ulem}
\usepackage{epstopdf}
\usepackage{mathrsfs}
\usepackage{amsfonts}
\usepackage{cancel}

\usepackage{subfigure}
\usepackage{wasysym}
\usepackage{soul} 
\usepackage{comment}
\usepackage{enumerate}
\usepackage{helvet}

\newcommand{\va}[1]{{\color{black}{{#1}}}}
\newcommand{\pa}[1]{{\color{black}{{#1}}}}
\newcommand{\el}[1]{{\color{black}{{#1}}}}

\begin{document}

\title{Active polymer rings: activity-induced collapse and dynamical arrest}

\author{Emanuele Locatelli$^{1,2}$, Valentino Bianco$^3$, Paolo Malgaretti$^{4,5,6}$}

\affiliation{$^1$Faculty of Physics, University of Vienna, Vienna, Austria}
\affiliation{$^2$Institut f{\"u}r Theoretische Physik, TU Wien, Vienna, Austria}
\affiliation{$^3$ Faculty of Chemistry, Chemical Physics Department,
   Complutense University of Madrid, Plaza de las Ciencias, Ciudad Universitaria, Madrid 28040, Spain}
\affiliation{$^4$Max Planck Institute for Intelligent Systems,   Heisenbergstr.\ 3, 70569 Stuttgart,   Germany}
\affiliation{$^5$IV\!\! Institute for Theoretical Physics, University of Stuttgart, Pfaffenwaldring 57,\! 70569 Stuttgart, Germany}
\affiliation{$^6$Helmholtz Institut Erlangen-N\"urnberg  for Renewable Energy (IEK-11), Forschungszentrum J\"ulich, F\"urther Str. 248, 90429, N\"urnberg, Germany}

\begin{abstract}
We investigate, using numerical simulations, the conformations of isolated active ring polymers. We find that the their behaviour depends crucially on their size: short rings ($N \lesssim$ 100) are swelled whereas longer rings ($N \gtrsim$ 200) collapse, at sufficiently high activity. By investigating the non-equilibrium process leading to the steady state, we find a universal route driving both outcomes; we highlight the central role of steric interactions, at variance with linear chains, and of topology conservation. We further show that the collapsed rings are arrested by looking at different observables, all underlining the presence of an extremely long time scales at the steady state, associated with the internal dynamics of the collapsed section. 
\pa{Finally, we found that is some circumstances the collapsed state spins about its axis.}
\end{abstract} 

\begin{titlepage}
\maketitle
\end{titlepage}

Active matter systems, such as synthetic and biological swimmers, show remarkable single particle \va{ and} collective dynamics that are completely different from their equilibrium counterparts~\cite{BechingerRMP}. 
For example, their active motion leads single active particles to accumulate at walls~\cite{Lauga2008,Elgeti_2013} or at fluid interfaces~\cite{DiLeonardo2011,Malgaretti2018} and to the onset of Motility Induced Phase Separation (MIPS) for dense suspensions~\cite{Cates2015}.
Up to now, the majority of the studies have focuses on ``simple'' active systems that lack ``internal'' degrees of freedom, such as colloids. 
However, recent works \va{on more complex active systems, like active polymers,  have shown rich and counter-intuitive dynamics}~\cite{Kaiser2015, Eisenstecken2016, Isele-Holder2015,Yan2016,Vutukuri2017,Gonzalez2018, bianco2018,Sunil2018, foglino2019, das2019deviations}. For example, tangentially-active polymers (i.e. polymers for which the active force acts tangentially to their backbone) 
undergo a coil-to-globule transition upon increasing the activity~\cite{bianco2018} and show a size-independent diffusion~\cite{bianco2018,Sunil2018}.\\
Such systems are far from being a purely theoretical speculation. Chains of active colloids can be assembled using state-of-the art synthesis techniques\cite{Yan2016}; further, experiments with living worms (regarded as tangentially-active polymers) have shown the onset of phase separation~\cite{deblais2020phase} akin to active colloids. Moreover, biological filaments such as DNA, RNA, actin and microtubules experience the force of molecular motors~\cite{Albers}. 
Notably, diverse biological scenarios feature closed structures, i.e.  rings or loops, as it happens for DNA and RNA\cite{schleif1992, allemand2006, semsey2005}, extruded loops in chromatin\cite{goloborodko2016, schwarzer2017}, bacterial DNA\cite{worcel1972, postow2004, wang2013}, kinetoplast networks\cite{borst1979, diao2015, klotz2020} and actomyosin rings\cite{sehring2015,Pearce2018}. Finally, topological constraints facilitate packing of long linear macromolecules, a process of capital importance in eukaryotic chromosomes\cite{grosberg1993,rosa2008,Rosa2013,halverson2014}. Since the dynamics of ring molecules differs dramatically from that of linear chains~\cite{frank1975, cates1986, rubinstein1986,grosberg1988,grosberg1993,micheletti2011} the question about the dynamics of active rings \va{arises} naturally.

In this Letter we characterize, by means of numerical simulations, the conformation and the dynamics of active self-avoiding polymer rings, whose monomers are self-propelled in the direction tangent to the polymer backbone; the rings are unknotted and their topology is preserved at all times.\\
Our results \va{ on active self-avoiding rings show a non-monotonous dependence of the  gyration radius on the ring size, in contrast with the monotonous behavior found in both  passive  rings  and  active  self-avoiding  linear  chains, highlighting a dramatic change in their dynamics.}
\pa{Moreover} we identify the general pathway leading to either inflation (small rings) or collapse (large rings), along with the critical size that  separates the basins of attraction of these two steady states.\\
Since these features are absent for active self-avoiding linear chains clearly they are induced by the topological constraints. \va{We } prove this 
\va{by} \va{comparing} the dynamics of active self-avoiding rings against that of active ghost-rings (that do not conserve topology). Interestingly, our results show that active ghost-rings swell for all ring sizes\va{, implying} that the collapse of active rings is due to activity and collisions among non near-neighboring monomers. Such a feature reminds that of \va{MIPS} for Active Brownian Particles (ABP). Finally, focusing on collapsed rings, we find that their internal dynamics \pa{shows the hallmarks of dynamical arrest and the onset of a spinning state.}

We consider fully flexible bead-spring polymer rings, suspended in an homogeneous fluid in three dimensions. We perform standard Langevin Dynamics simulations neglecting 
hydrodynamic interactions\footnote{ 
\pa{Such an approximation is made in the spirit of "dry active matter" that has identified the onset of activity-induced phase separation (known as MIPS), later observed also in experiments (hence where hydrodynamic coupling is present). Similarly, we expect that our results will qualitatively persist even in the presence of hydrodynamic coupling }}.
The bead diameter $\sigma$ sets the unit of length, and $m =$ 1 sets the unit of mass. 
The active force $\mathbf{f}^{\rm act}$ acts with constant magnitude $f^{\rm act}$ along  
the vector tangent to the polymer backbone~\cite{bianco2018}; such construction \va{applies} to all monomers. 
We quantify the strength of the activity  
via the P{\'e}clet number ${\rm Pe} \equiv f^{\rm act} \sigma / (k_B T)$, where $k_B T$ is the thermal energy of the heath bath (being $k_B$ the Boltzmann constant and $T$ the absolute temperature), in which the ring polymer is suspended. 
\va{Following} Ref.~\cite{Stenhammar2014}, we choose to fix  $f^{\rm act} \sigma = $ 1 and increase the P{\'e}clet number by decreasing the thermal energy of the heath bath. 

\va{We} employ a modified Kremer-Grest model \va{to avoid crossing events and knots} (\va{Sup.} Section 1). 
Hence, we simulate ring polymers of length 70 $< N <$ 800, at 1 $< {\rm Pe} < $ 100; the data reported are averaged over 250 $< M <$ 2850 independent configurations.

For tangentially-active linear polymers~\cite{bianco2018,Sunil2018} the average gyration radius $R_g \equiv \sqrt{ \langle \sum_{i=1}^N \left( \mathbf{r}_i - \mathbf{r}_{com} \right)^2/N \rangle}$ -- where $\mathbf{r}_{i}$ and  $\mathbf{r}_{com}$ are the positions of the monomer $i$ and center of mass of the polymer, respectively --  grows with $N$  with a smaller scaling exponent, compared to the passive case, whose value depends on $\text{Pe}$.

\begin{figure}[t!]
\includegraphics[width=0.48\textwidth]{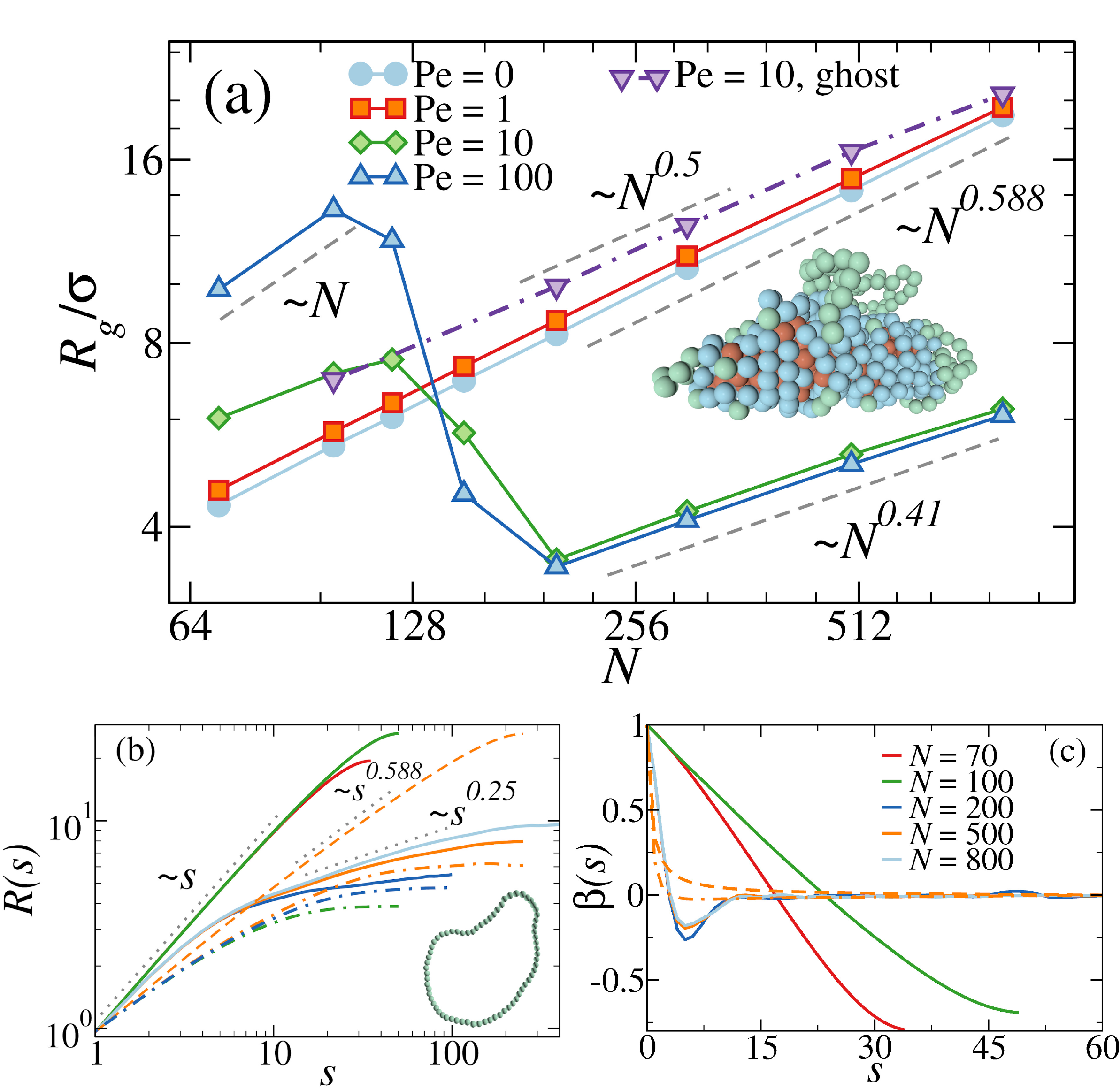}
\caption{a) Gyration radius as function of $N$, for active self-avoiding rings (different values of ${\rm Pe}$) and active ghost rings (${\rm Pe} =$ 10) b) Mean internal distance as function of the distance $s$ along the contour. c) Bond-bond correlation function as function of $s$. {In panel b) and c), full lines refer to active rings at ${\rm Pe}$ = 100; orange dashed lines refer to the passive case in good solvent for $N =$ 500\va{;} dash-dotted lines refer to the passive case in bad solvent for $N =$ 100, 200, 500. \va{Panels} b) and c) \va{share the legend}.} In panel a) and b) snapshots of short $N=$ 70 (inflated) and long $N=$ 500 (collapsed) rings, with colors referring to beads: i) in the dangling sections or in the inflated state (green), ii) on the surface of the collapsed structure (blue), iii) in the interior of the collapsed core (red). }
\label{fig:Rg1}
\end{figure}
\noindent In contrast, for active polymer rings $R_g$ shows a more complex dependence on $N$ (Fig.~\ref{fig:Rg1}a).
In particular, while for ${\rm Pe}\lesssim 1$ the scaling exponent matches the equilibrium one $\nu=\nu_{\rm eq}=0.588$ for all values of $N$, 
for $\text{Pe} > 1$ 
two distinct regimes emerge\va{,} each of which is characterized by a specific scaling of $R_g$ with $N$.
For short rings $N \lesssim $100, \va{$R_g$} of active rings becomes larger than $R_g$ of \va{passive rings}.
A power-law fit in this region leads to an exponent, $\nu_{short}$, that depends on ${\rm Pe}$: at the \va{largest} activity considered (${\rm Pe}$ = 100), \va{we find}  $\nu_{short} \approx$ 1, similar to the behaviour of fully rigid rings.
Hence, for short rings the activity induces an effective bending rigidity\va{, with} a persistence length comparable to the \va{ring's} size.\\
Upon increasing the length of the polymers (Fig.~\ref{fig:Rg1}a), activity induces a structural collapse. In this regime, the scaling exponent is 
$\nu=\nu_{long} \approx$ 0.41 and, for $\text{Pe} \geq 10$,  
it is independent on $\text{Pe}$. The small value of $\nu_{long}$ indicates that the rings assume a very compact conformation. 
It is worth noting that the value of $\nu_{long}$ is close to, but not exactly the one expected in bad solvent 
conditions $\nu_{BS} = $0.33~\cite{Doi1996}.
Indeed, as shown in the snapshot in Fig.~\ref{fig:Rg1}a, the collapsed structure is quite complex \va{, being} composed of a compact self-wrapped core and few \textit{dangling sections} \va{fluttering} on \pa{its} surface. These dangling sections, absent in the case of ring polymers in bad solvents~\cite{rubinstein2003book}, are responsible for the larger value of $\nu_{long}$ as compared to $\nu_{BS}$ {(\va{Sup.} Section 6)}.
\va{Moreover}, upon increasing ${\rm Pe}$, the transition between inflated and collapsed rings becomes progressively sharper. We \va{elucidate} the role of self-avoidance in this phenomenon by simulating active ghost rings  \va{ that, by contrast,} maintain their passive scaling $N^{0.5}$ for all values of $N$ investigated (see the violet curve in Fig.~\ref{fig:Rg1}a); further, activity swells the rings, without further altering their configurational properties (\va{Sup.Fig.~7}).\\
\va{To} understand the physical origin of the scaling regimes observed, we analyze the conformations attained by long and short active rings.
Accordingly, we compute, {in} the steady state, {the root} mean square distance among monomers that are $s$-th neighbors along the backbone $R(s) \equiv \sqrt{\langle \left( \mathbf{r}_{s+s_0} - \mathbf{r}_{s_0} \right)^2 \rangle}$ 
where $s_0$ is the starting bead. 

 Fig.~\ref{fig:Rg1}b shows that for short rings $R(s)$ displays a single power law trend  $R(s)\sim s$ whose exponent is compatible with the one estimated from the scaling of the gyration radius in Fig.~\ref{fig:Rg1}a.
The \textit{single} power law \va{fitting $R(s)$ up to $s\simeq N/2$} implies \va{the self-similarity of} active rings. For reference, we report $R_{eq}(s)$ for fully-flexible  passive rings in good solvents, for which $R_{eq}(s)\simeq s^{0.588}$~\cite{rubinstein2003book} (orange dashed lines in Fig.~\ref{fig:Rg1}b).
In contrast, longer active rings show a richer behaviour for $R(s)$. Indeed, for $s \lesssim 10$, $R(s)$ shows a universal power law $R(s)\sim s$ whose prefactor does not depend on \va{$N$} (all curves collapse on a master curve in Fig.~\ref{fig:Rg1}b). Then, for $s \gtrsim 10$, the scaling of $R(s)$ is size-dependent and, for intermediate values of $s$, is fitted by $R(s)\simeq s^{0.25}$. This change is the signature of the collapsed structure: monomers very far away along the backbone end up being very close in real space. This behaviour is qualitatively similar to that observed for passive rings in bad solvents \va{(dashed-dotted curves} in Fig.~\ref{fig:Rg1}b\va{)}.

In order to characterize the local arrangement of monomers in the \va{inflated/collapsed states} we measure the bond-bond spatial correlation function $
\beta(s) \equiv \langle \mathbf{b}_{s+s_0} \cdot \mathbf{b}_{s_0} \rangle
$, where $\mathbf{b}_i \equiv \mathbf{r}_{i+1} - \mathbf{r}_i$.
 
As shown in Fig.~\ref{fig:Rg1}c, short active rings $N=$ 70,100 develop a strong anti-correlation over the scale of the whole polymer, akin of rigid passive rings. In contrast, for long active rings, the bond-bond correlation function shows a very fast decay at small contour separations, followed by an anti-correlation region that eventually fades to a complete decorrelation. 
In the collapsed state, part of the chain wraps on itself (\va{Sup.Fig.}~11) and such wrappings are characterized by a ``pitch'' of $\sim 5$ beads which is, roughly, at the same contour distance $s\simeq 5$ for all $N$ {and ${\rm Pe}$} investigated. 
These behaviors are in contrast to those observed for passive rings in both good \va{(dashed orange curve)} and bad solvents \va{(dotted orange curve)}\va{ that} display no minimum at short contour separations.

Next we investigate the pathway from \va{a} passive, equilibrated \va{ring} configuration to the inflated/collapsed steady state, \va{ by considering} \el{the time evolution  of the bond correlation function $\beta(t)$ and the square average contour distance between pairs of beads that are close in space $T(t) = \langle \left((j-j_1)^2+ \ldots +(j-j_m)^2 \right) / m \rangle$ that measures the  ``tangleness'' of a polymer chain (see Sup. Eq.(7)).}
In particular,  Fig.~\ref{fig:Rg1}c shows that $\beta(s)$ is characterized by a minimum, either at small $s$ or $s = N/2$. We follow the evolution in time of \el{the contour length at which} such minimum \el{appears}, \el{${\rm argmin}(\beta(s))$}, as it marks the characteristic size of the local structures that form along the polymer backbone 
\va{(Fig.~\ref{fig:bondcorr_time}a)}. For passive rings, the minimum \va{is} at $\sim N/2$ for all times. This coincides with the value obtained for active rings 
at early times {($t\leq 10\tau_0$)}; during such time frame, comparable to the diffusion time $\tau_0=\sigma^2/D$ of a monomer over its size, activity has not yet affected the conformation of the ring.
\begin{figure}[t!]
\centering
\includegraphics[width=0.49\textwidth]{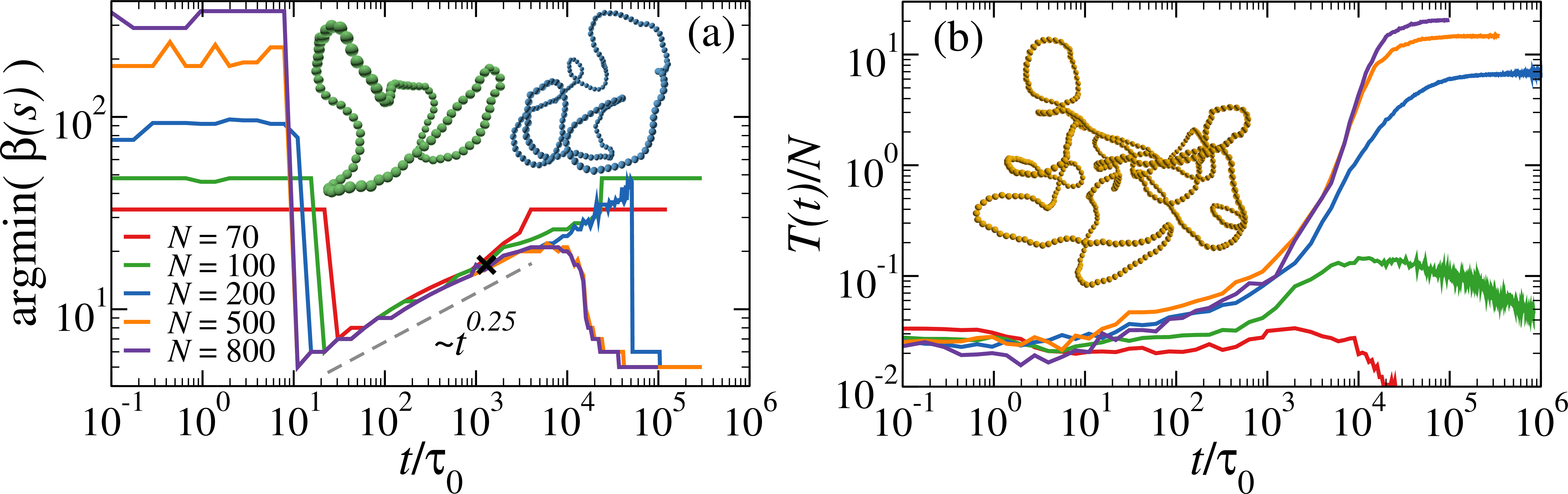} 
\caption{Route to the steady state: \el{(a) position of the minimum of the bond correlation (b) Tangleness $T(t)/N$ as a function of time, normalized by $\tau_0$, for rings of different length at $\rm Pe = $100. The black symbol in panel (a) marks the time at which the snapshots shown are taken. Snapshots: $N =$ 100 green, $N =$ 200 blue, $N =$ 500 orange.\\
} }
\label{fig:bondcorr_time}
\end{figure}
At intermediate times, $t\simeq 10\tau_0 - 1000\tau_0$ \va{(Fig.~\ref{fig:bondcorr_time})}, small ``loops'' appear, highlighted by a drop of the minimum {of} the bond correlation to $s \simeq$ 5, roughly constant for all $N$.  Their \va{sharp} onset takes place at earlier times upon increasing the polymer size ($\sim 20\tau_0$ for $N=70$, $\sim 10\tau_0$ for $N=500$) \el{and it is weakly dependent on ${\rm Pe}$, for ${\rm Pe} \gtrsim$ 10}.
The size of the loops grows up to a characteristic size  $s$, reached in $t\in[\sim10^3\tau_0:10^4\tau_0]$, with a growth rate {$\propto (t/\tau_0 )^{1/4}$} essentially independent of $N$ and ${\rm Pe}$ (\va{Sup.Fig.}~9b), hence setting an universal route towards the steady state.
{This universal growth $\propto t^{1/4}$ reminds of the coarsening of $2D$ ABPs undergoing MIPS~\cite{Stenhammar2013,Patch2017}}.
Snapshots of rings  
taken during this stage are reported in Fig.~\ref{fig:bondcorr_time}; loops are clearly visible in all cases irrespective of $N$.
At later times, the dynamics is {no more} universal and the size of the ring matters. 

\va{For} $N \lesssim 100$, loops of size $s\simeq 20$ are relatively close to their equilibrium value $N/2$. \va{When} two loops meet they merge giving raise to a larger loop (see the jumps in $\beta(s)$ in Fig.~\ref{fig:bondcorr_time}{.a} for $N=70,100$). At variance, for $N > 200$ (Sup. Video 1) when two loops of size $s\simeq 20$ get closer they can thread one into the other, 
triggering a cascade of collisions that drives sections of the backbone to tangle, \va{inducing} the collapse of the entire chain. \va{After} such \va{a} catastrophic event, the rest of the ring is progressively recruited in the {main} tangle (see Sup. Video 1).
\pa{Such a scenario is also supported by the tangleness $T(t)$.}
\el{Indeed, we observe the tangleness per size $T(t)/N$, at very short times, i.e. when activity has not yet affected the ring, has a characteristic value, dependent on the chosen cut-off radius $r_c$ (defining spatial neighbours) and independent on $N$. Afterwards, the behaviour of $T(t)/N$ depends on the final steady state. For short, inflating rings, the tangleness shows a shallow maximum and then decreases towards a small constant value. In contrast, for long collapsing rings, $T(t)/N$ monotonically increases \va{toward} a large constant value. This increment develops on roughly the same timescales as the growth of ${\rm argmin}(\beta(s))$ but, further, shows two regimes, characterised by a mild increase first and a steeper slope later on. The increase of $T(t)$ can be thus directly connected with an increase of the steric interactions and with collisions. 
$T(t)$ and $\text{argmin}(\beta)$ appear complementary to each other: the tangleness better captures the \va{``two-steps'' collapse}, while ${\rm argmin}(\beta(s))$ \va{better highlights the universality of this route}. The \va{polymers} conformations  along the pathway can be further characterised \va{with} the torsional order parameter (\va{Sup.Fig.}~11).}
We remark that the collapsed state is, in its origin, akin to \va{MIPS}~\cite{cates2015motility}, as both are initiated by collisions and maintained by self-avoidance. This confirms the result reported for ghost rings: without self-avoidance, tangles can't form and the rings are effectively composed by non-interacting loops (\va{Sup.Fig.}~5). \va{The described pathway is common for sufficiently high ${\rm Pe}$, while for ${\rm Pe}\lesssim 5$  sufficiently long rings may end up in a collapsed state following a much smoother route (\va{Sup.Fig.}~10).}
\begin{figure}
\centering
\includegraphics[width=0.4\textwidth]{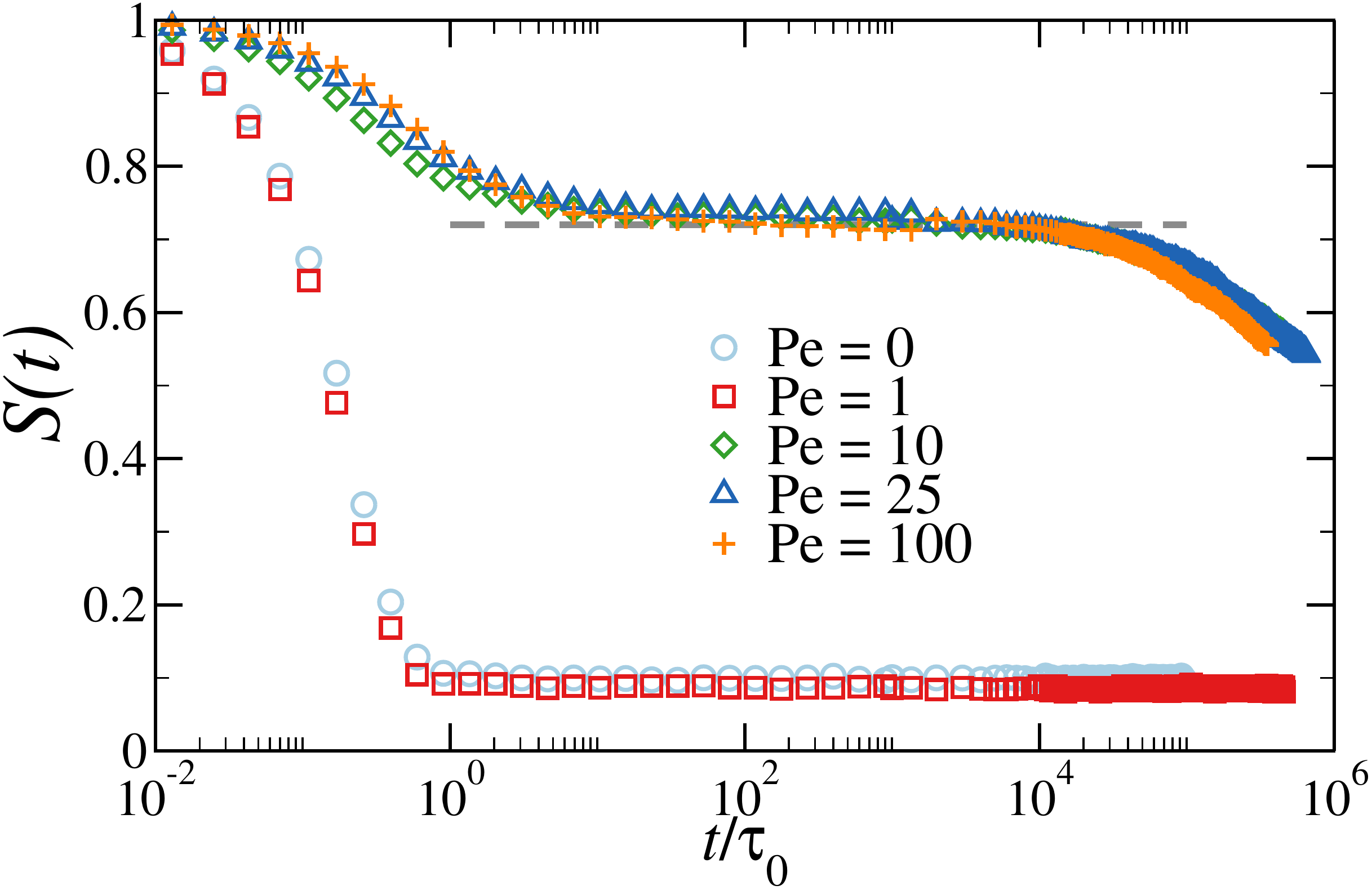}\\
\caption{Fraction of survived neighbours as function of time; data refer to rings of size $N =$ 500 and several ${\rm Pe}$ at the steady state. The gray dashed line is a guide for the eye, highlighting the intermediate plateau.}
\label{fig:arrest0}
\end{figure}

After the collapse, monomers in the tangle find themselves in a complex and tight structure whose dynamics, in the steady state, is arrested. \va{This can be verified by computing} the surviving fraction of neighbours $S(t)$, defined as the fraction of \va{monomer's} neighbours { within a radius $r_c$ } (excluding the first-neighbors along the backbone), chosen at any arbitrary time $t_0$ during the steady state, that are still neighbours of the same monomer \va{at} $t>t_0$. \va{We fix the} neighbouring cut-off $r_c =$ 1.2$\sigma$. Since $S(t)$ provides a measure of the permanence of the collapsed configurations we expect $S(t)\sim 1$ for a completely frozen system\va{, otherwise}
$S(t)$ \va{decays} to a small non-vanishing value after a characteristic time.
Fig.~\ref{fig:arrest0} shows $S(t)$ for  $N=$ 500 and increasing activity (\va{Pe$\in[0:100]$}). For ${\rm Pe}<= 1$ rings are not collapsed and $S(t)$ displays a fast decay and plateaus to a small value akin of passive rings~
\footnote{We remark that such value is not strictly zero, as relatively close monomers can randomly get in and out of the threshold range chosen ($r_c = $1.2 $\sigma$.)}.
As soon as the rings collapse ($\rm Pe\ge10$) $S(t)$ shows a strikingly different behaviour. First, at short times ($t<=\tau_0$), $S(t)$ decays mildly \va{due to the highly mobile dangling sections}. \va{This} decay occurs on time scale comparable to that of passive rings, \va{but with reduced magnitude}.
The initial decay is followed by a plateau that lasts several decades and whose time-span is  slightly dependent on $N$ (\va{Sup. Fig.} 16).
Afterwards, a second decay is observed, which  possibly plateaus at later times (outside the  time frame of the simulations). 
\va{This} double decay can be found also in  the intermediate scattering function and in the time correlation of the characteristic vectors of the ring (\va{Sup.Fig.}~18-21). Overall, for every observable considered, the first decay is due to the the contribution of the dangling sections, whereas the second, much slower, is related to the complete rearrangement {of the beads neighbouring environment within a length scale of the order of $\sigma$.} 
\begin{figure}[t]
\includegraphics[width=0.49\textwidth]{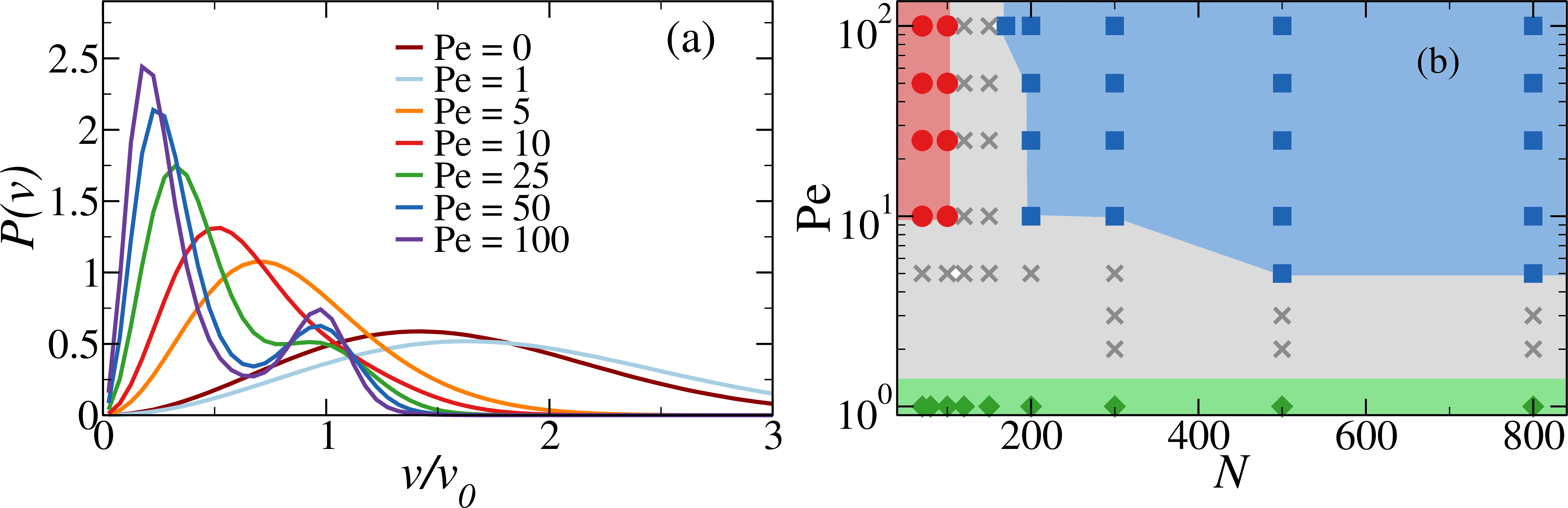}
\caption{a) Monomer velocity distributions at the steady state for fixed $N =$ 500, several ${\rm Pe}$. b) Dynamic phase diagram for active rings. Symbols refer to state points sampled by means of numerical simulations: collapsed rings are reported as blue squares, inflated rings as red circles, passive-like rings as green diamonds, system showing a more complex behaviour as grey crosses.}
\label{fig:ring_vel}
\end{figure}

A hallmark of an arrested dynamics appears also in the 
distribution of the instantaneous velocities of the monomers at steady state (Fig.~\ref{fig:ring_vel}a).
As expected, at equilibrium (${\rm Pe} =0$) the distribution is Maxwell-Boltzmann. Interestingly, for ${\rm Pe} \lesssim 1$ the distribution is again Maxwell-Boltzmann but with an effective temperature $T_{eff} \simeq 1.3$ $k_B T^{\star}$ \va{, ``heating up''} the rings.
\va{At} sufficiently high ${\rm Pe}$, \va{the distributions} exhibit two peaks, \va{at $v<<v_0$ and $v\sim v_0$ rispectively}. 
{In Fig.~\ref{fig:ring_vel}a , the peak at small velocities is given by the monomers trapped in the collapsed section (\va{Sup.Fig.}~22), whereas the peak at $\nu/\nu_0\sim1$ is due to monomers in the dangling sections.}
{Such velocity distributions} remind those observed in MIPS, where active particles inside the dense phase-separated region experience a reduced mobility with respect to their counterparts in the gas phase~\cite{SANCHEZ2018,Loewen2019,MarconiArxiv}. {In particular, comparing Fig.~\ref{fig:ring_vel}  with Figs.~\ref{fig:Rg1},\ref{fig:arrest0} we note that the velocity distribution varies continuously upon increasing ${\rm Pe}$, whereas nor the $R_g$ neither $S(t)$ are sensitive to such a change. This implies that the configuration of the polymers, and hence the onset of MIPS-like transition, is robust to changes in the velocity distribution provided that both ${\rm Pe}$ and $N$ are large enough.}

{We \va{collect} our data into a phase diagram shown in Fig.~\ref{fig:ring_vel}b, \va{where} four regions can be identified according to the scaling of $R_g$ with $N$. At small ${\rm Pe}$, active rings retain their equilibrium scaling for all values of $N$. For $\text{Pe} \gtrsim$ 10, the scaling of $R_g$ with $N$ depends on the active ring size: smaller active rings ($N \lesssim 100$) swell ($R_g\propto N$), whereas larger active rings ($N\gtrsim 200$) display arrested, collapsed configurations with dangling sections. The transition between these two ``phases'' occurs via a transition region, of finite extent, in which the dependence of $R_g$ on $N$ is more complex (see Fig.~\ref{fig:Rg1})} \el{and can be either jumping in between fairly compact and fairly open conformation (typical for 100 $< N <$ 200) or quasi-collapsed but not arrested conformations (typical for ${\rm Pe} <$ 5 and $N >$ 300)\va{(Sup. Videos 3 and 4).}} \el{Finally, the supplemental video 1 \va{and 2} \va{show a rotating motion and self-propulsion of} the collapsed state, due to an active torque and a non-zero net active force on the centre of mass, respectively } \pa{\va{(Sup.Fig.~23 and~20a)}. }

In summary, we have presented the effects of tangential activity on the conformation of fully flexible self-avoiding active ring polymers. We have shown that, upon increasing the activity, there is a conformational transition between short rings ($N \lesssim$ 100) that swell and assume a disk-like shape, and long rings ($N \gtrsim$ 200) that exhibit a structural collapse. The non-equilibrium evolution towards the steady state follows a general 
route featuring loop formation  growing in size up \va{to} a characteristic size $\sim 20$. Then, for sufficiently long rings, the collapse is triggered by clashes between monomers belonging to non-neighbouring loops.
Finally, the extremely slow structural relaxation \va{of} different observables, indicates that the collapsed rings represent a unique example of arrested  macromolecule.  
These features are typical of tangentially-active ring polymers and may disappear \va{in case of isotropic~\cite{Mousavi2019} or scalar \cite{smrek2020active} activity}.
Neglecting hydrodynamics has allowed us to robustly investigate rings of large size \va{for} long time scales. 
Nevertheless, hydrodynamic interactions will be crucial to investigate the dynamics and the stability of the open conformations at small values of $N$\cite{liebetreu2020hydrodynamic}. Granted the phenomenology observed in this paper is robust, active rings may be \va{exploited to wrap, protect and deliver drugs.} Further, topology-based  materials have been already proposed\cite{krajina2018}, for which activity may change the macroscopic properties, as happens for biopolymer networks\cite{koenderink2009active}. Possibly, the most exciting application concerns the modeling of {self-propelled filaments in gliding assays\cite{liu2011loop} and of biophysical systems, exploring the effect of activity in chromatin\cite{terakawa2017condensin, SaintillanMotor, haushalter2003chromatin}, in bacterial DNA and in the cytoskeleton\cite{danuser2013mathematical, murrell2015forcing, fletcher2010cell}, where actomyosing ring may play a key role in cell division\cite{litschel2020reconstitution}, or in purified protein networks\cite{koenderink2009active, Stam2017, burla2019mechanical}.}     

\begin{acknowledgments}
The authors thank L. Tubiana for helpful discussions and would like to acknowledge the contribution of the COST Action CA17139. V. Bianco acknowledges the support the European Commission through the Marie Sk\l{}odowska–Curie Fellowship No. 748170 ProFrost. The computational results presented have been achieved using the Vienna Scientific Cluster (VSC).
\end{acknowledgments}

\newpage \newpage \newpage

\title{Supplemental Material}

\begin{titlepage}
\maketitle
\end{titlepage}

\renewcommand{\thesection}{\arabic{section}}

\onecolumngrid

\section{Model}
We model ring polymers as bead-spring chains in three dimensions; the bead diameter, $\sigma$, sets the unit of length, and $m$  sets the unit of mass. We perform standard Langevin Dynamics simulations using a custom modified version of LAMMPS\cite{LAMMPS}, neglecting at this stage hydrodynamic interactions; we set the friction coefficient $\gamma =$ 1 and the elementary timestep $\Delta t = 10^{-3}$. The requirement of the topology conservation, i.e. maintaining the rings unknotted, requires special care in modeling the activity and in choosing the interaction potentials. First, each monomer is self-propelled by a force ${\bf f}^{\rm act}$, with constant magnitude $f^{\rm act}$. The direction of ${\bf f}^{\rm act}_{i}$ is always parallel to ${\bf r}_{i+1,i-1} \equiv {\bf r}_{i+1} - {\bf r}_{i-1}$, i.e the vector connecting the first neighbours of monomer $i$ along the polymer backbone; such construction applies to all monomers. As standard in active systems, we quantify the activity of each monomer by the P\'eclet number ${\rm Pe}$
\begin{equation}
  {\rm Pe} \equiv \frac{f^{\rm act} \sigma}{k_B T}
  \label{eq:def-Pe}
\end{equation} 
where $k_B T$ is the thermal energy of the heath bath, in which the polymer is suspended. In the spirit of \cite{Stenhammar2014}, we choose to fix  $f^{\rm act} = $ 1 and we change the P\'eclet number by decreasing the thermal energy. This is equivalent to set a reference temperature $T^{\star}$, such that $k_B T^{\star} =$ 1.

\begin{figure}[!h]
\centering
\includegraphics[width=0.45\textwidth]{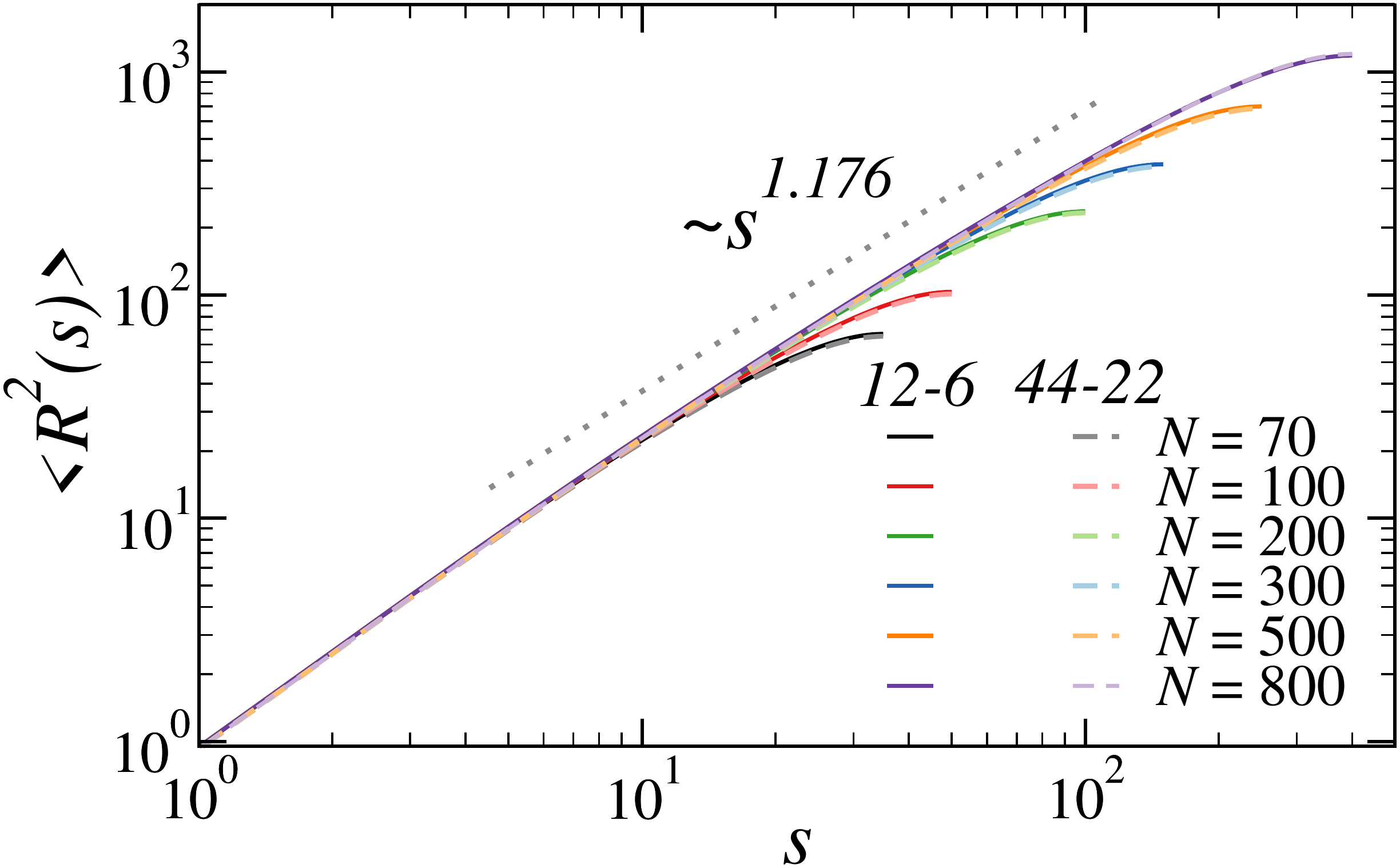}
\caption{Mean square internal distance as function of the contour separation $s$ for the standard Kremer-Grest model ($12-6$ in the legend) and the model employed in this study ($44-22$ in the legend) for passive rings of different length in equilibrium.}
\label{fig:models_comparison}
\end{figure}

This simulation strategy allows us to avoid the introduction of huge forces in the system and to use a reasonably small timestep. Further, in order to avoid crossing events we choose to employ a modified Kremer-Grest model.\\ 
The original Kremer-Grest model\cite{Kremer1990} preserves the topology in absence of activity and is widely used as generic force field for self-avoiding polymers. However, as already mentioned, in presence of activity long rings tend to collapse; we find that this state is particularly prone to knot formation as the backbone is rather stressed and bonds tend to stretch. The Kremer-Grest model, employing a standard FENE potential, allows for considerable stretching, as the maximum distance is $R_{max} =$ 1.5$\sigma$ whereas the typical distance (typical bond length) is $b \simeq$ 0.97 $\sigma$. In principle, crossings are possible in absence of activity but, in practice, in equilibrium the energy barrier is so high ($\sim$ 70 $k_B T$) that crossings are negligible. This is, clearly, not true in the collapsed state considered, because of the self-avoidance as well as of the active forces. In principle, the use of a much smaller time-step (e.g. $\Delta t \leq 10^{-6}$) would solve the problem at the cost of making the long time dynamics inaccessible. In order to avoid crossing events at the chosen time step $\Delta t = 10^{-3}$, we choose to employ a WCA-like potential between any pair of monomers
\begin{equation}
 V_{mm}(r) = 
\begin{cases}
4 \epsilon 
\left[ \left(\frac{\sigma}{r} \right)^{44} -
  \left(\frac{\sigma}{r} \right)^{22} \right] 
+ \epsilon; &{\text {for}}\,\, r<2^{1/22}\,\sigma, \cr
0; &{\text {for}}\,\, r \geq 2^{1/22}\,\sigma,
\end{cases}
\label{eq:LJ}
\end{equation}
with $\epsilon = $1; further, neighbouring beads along the backbone are bonded by a FENE potential
\begin{equation}
V_{\rm {FENE}}(r) = 
\begin{cases}
-150 \epsilon \left( \frac{R_{\rm max}}{\sigma} \right)^2  \ln \left[ 1 - \left( \frac{r}{R_{\rm max}} \right)^2 \right] ; &{\text {for}}\,\, r \leq R_{\rm max}, \cr
\infty; &{\text {for}}\,\, r > R_{\rm max},
\end{cases}
\label{eq:FENE}
\end{equation}
with $R_{\rm max} = $1.05. Such potential is much steeper and imposes a much smaller range of fluctuations than the standard Kremer-Grest model; the bonds never stretch more than 1-2 \% of their typical value $b \simeq$ 0.96 $\sigma$ (see Fig.~\ref{fig:bonddistr}). Having a typical bond length as similar to the original as possible was the rationale behind the choice of the parameters for this force field. We point out that such hard and tight bonds are necessary only for very long rings; short rings do not collapse and standard Kremer-Grest potentials with $\epsilon =$ 10 is sufficient to preserve the topology for $N \leq$ 400 at any activity considered.\\
We show now, in Fig.~\ref{fig:models_comparison}, that indeed the new model does not change the properties of passive rings in good solvent conditions.

As visible, the mean square internal distance (defined in the following Section) shows negligible differences, indicating that the conformational properties are the same for both models, as expected. Caution shall be used, though, when employing this model in dense suspensions as the monomers are effectively slightly smaller.\\
As final remark we notice that, given the choice of employing tangent vectors of fixed length, the total active force, acting on the centre of mass, is not zero: the sum of the tangent vectors is zero by definition, but the corresponding unit vectors do not add up to a perfect zero. This does not affect neither the inflation, the collapse or the arrest, but affects the dynamics of the centre of mass, which may be relevant when considering suspensions of active rings at finite density.

\section{Observables}
\label{sec:obs}
We define here the observables considered in this work: the gyration radius of a polymer of length $N$ is defined as
\begin{equation}
R_g^2 = \frac{1}{N} \sum_{i=1}^N{ \left( \mathbf{r}_i - \mathbf{r}_{com} \right)^2}
\end{equation}
where $\mathbf{r}_{com}$ is the position of the center of mass of the polymer; the average value reported in this work has been average over time (at steady state) as well as over a number 250 $< M <$ 2850 independent realisations.\\The root mean square internal distance is defined as 
\begin{equation}
\langle R(s) \rangle = \sqrt{\langle \left( \mathbf{r}_{s+s_0} - \mathbf{r}_{s_0} \right)^2 \rangle}
\end{equation}
where $s$ is the distance (in beads) along the backbone. The average is performed as above and in addition over the starting bead $s_0$.\\
The bond-bond correlation function is defined as
\begin{equation}
\beta(s) = \langle \mathbf{b}_{s+s_0} \cdot \mathbf{b}_{s_0} \rangle
\end{equation}
where $\mathbf{b}_i = \mathbf{r}_{i+1} - \mathbf{r}_i$ (averages are performed as above).\\{The tangleness $T(t)$ will be defined as  
\begin{equation}
    T(t) = \left\langle \frac{(j-j_1)^2 + \ldots + (j-j_m)^2}{m} \right\rangle
    \label{eq:tangl}
\end{equation}
where $j$ is the index of the beads along the contour, $j_i$ stand for the indices of the spatial neighbours of bead $j$ (first neighbours along the contour are excluded) and the average is performed over all the beads as well as over several realizations.The distance along the contour accounts for the ring topology (i.e. we consider the minimal distance along the contour).}\\
\el{The torsional order parameter is defined following \cite{chelakkot2012a} as $U_T = \sum_{i=0}^{N} \cos(\gamma_i)$ where
\begin{equation}\label{eq:tors-ord-param}
    \cos(\gamma_i) = \frac{(\mathbf{b}_{i-1} \times \mathbf{b}_i) \cdot ( \mathbf{b}_i \times \mathbf{b}_{i+1}) }{|\mathbf{b}_{i-1} \times \mathbf{b}_i| |\mathbf{b}_i \times \mathbf{b}_{i+1}| }
\end{equation}
where $\mathbf{b}_{i-1}$, $\mathbf{b}_i$ and $\mathbf{b}_{i+1}$ are three subsequent bond vectors; the periodicity of the ring is taken into account.
}\\The self-intermediate scattering function is defined as 
\begin{equation}
F_s(k^*,t) = \langle \, \left( \sum_{i}^{N}{e^{i \mathbf{k^*} \cdot (\mathbf{r}_{i}(t) - \mathbf{r}_i(t_0) )} }\right) \, \rangle
\end{equation}
where $k^*$ is a fixed value and $\mathbf{k^*}$ is any vector in reciprocal space of magnitude $k^*$.\\Finally, a standard way to measure the "characteristic" time of a ring in equilibrium is to compute the time correlation function of the vectors normal to the surface of the ring
\begin{equation}
\mathbf{c} = \mathbf{a} \times \mathbf{b}
\label{eq:v_ring}
\end{equation} 
where $\mathbf{a} \equiv \mathbf{r}_{i+N/2} - \mathbf{r}_{i}$ and $\mathbf{b} \equiv \mathbf{r}_{i+3N/4} - \mathbf{r}_{i+N/4}$; alternatively the time correlation of the vector $\mathbf{a}$ (the "half-ring`` vector) may also be considered.

\section{Clustering algorithm and neighbour search}

We describe here how we identify neighbours and how we define the collapsed section of the ring by means of a clustering algorithm. We always consider configurations at steady state. Concerning the permanence of the neighbours, we perform a standard neighbour search, based on their euclidean distance; we consider as neighbours particles within a certain cut-off radius $r_c$. As mentioned in the main text, we exclude first neighbours along the backbone.\\Further, we employ a clustering algorithm to identify the collapsed section in Fig.~(4) of the main text. We perform first a neighbour search; we consider all the monomers who are interacting with at least 4 monomers, i.e. whose distance from at least 4 monomers is less than $r_c = $ 1.2$\sigma$, as belonging to a cluster. Joining the clusters defines the collapsed section. The rest of the monomers belong to the dangling sections.      



\section{Unknottedness}

\begin{figure}[!h]
\centering
\includegraphics[width=0.45\textwidth]{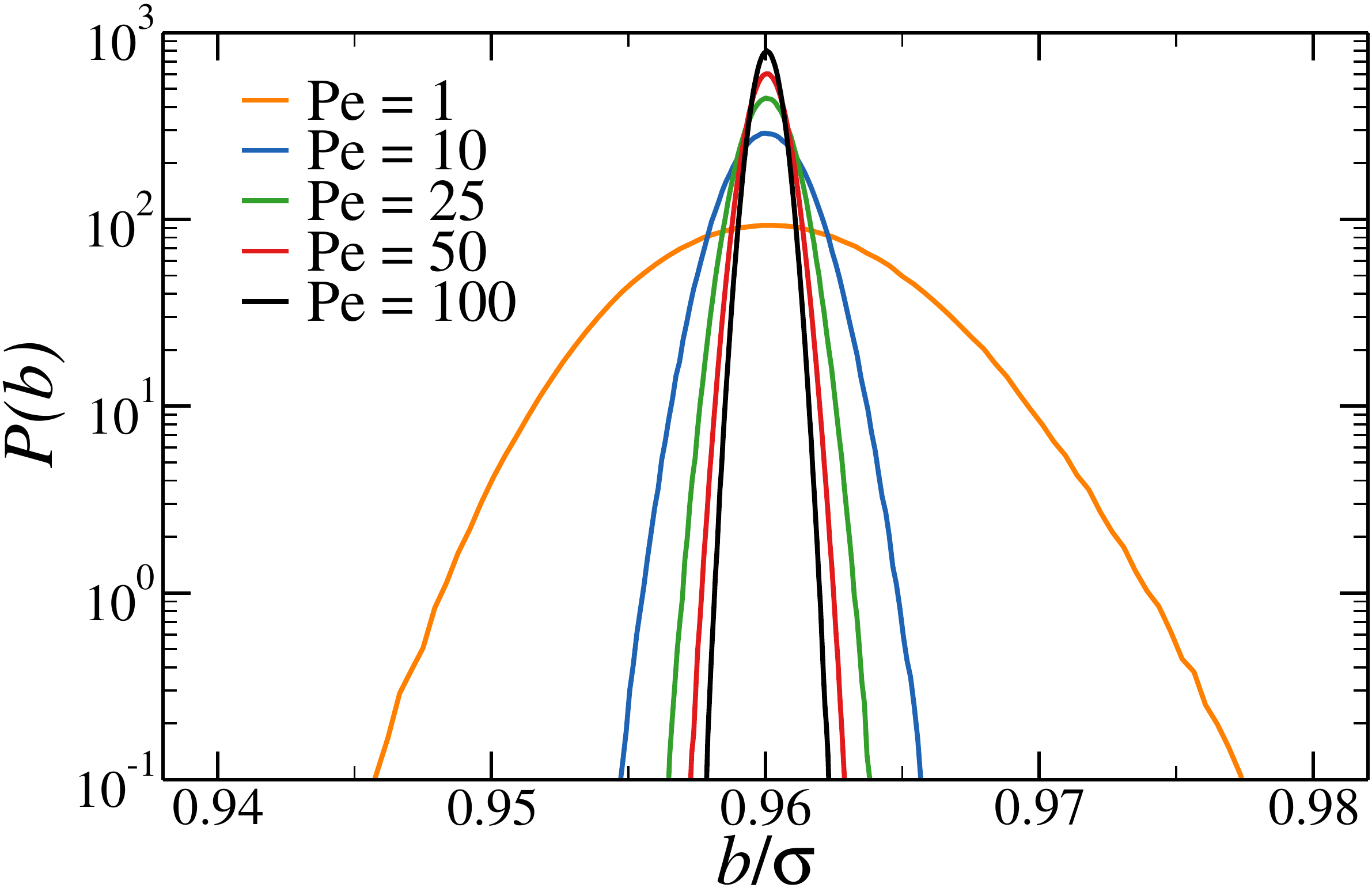}
\includegraphics[width=0.45\textwidth]{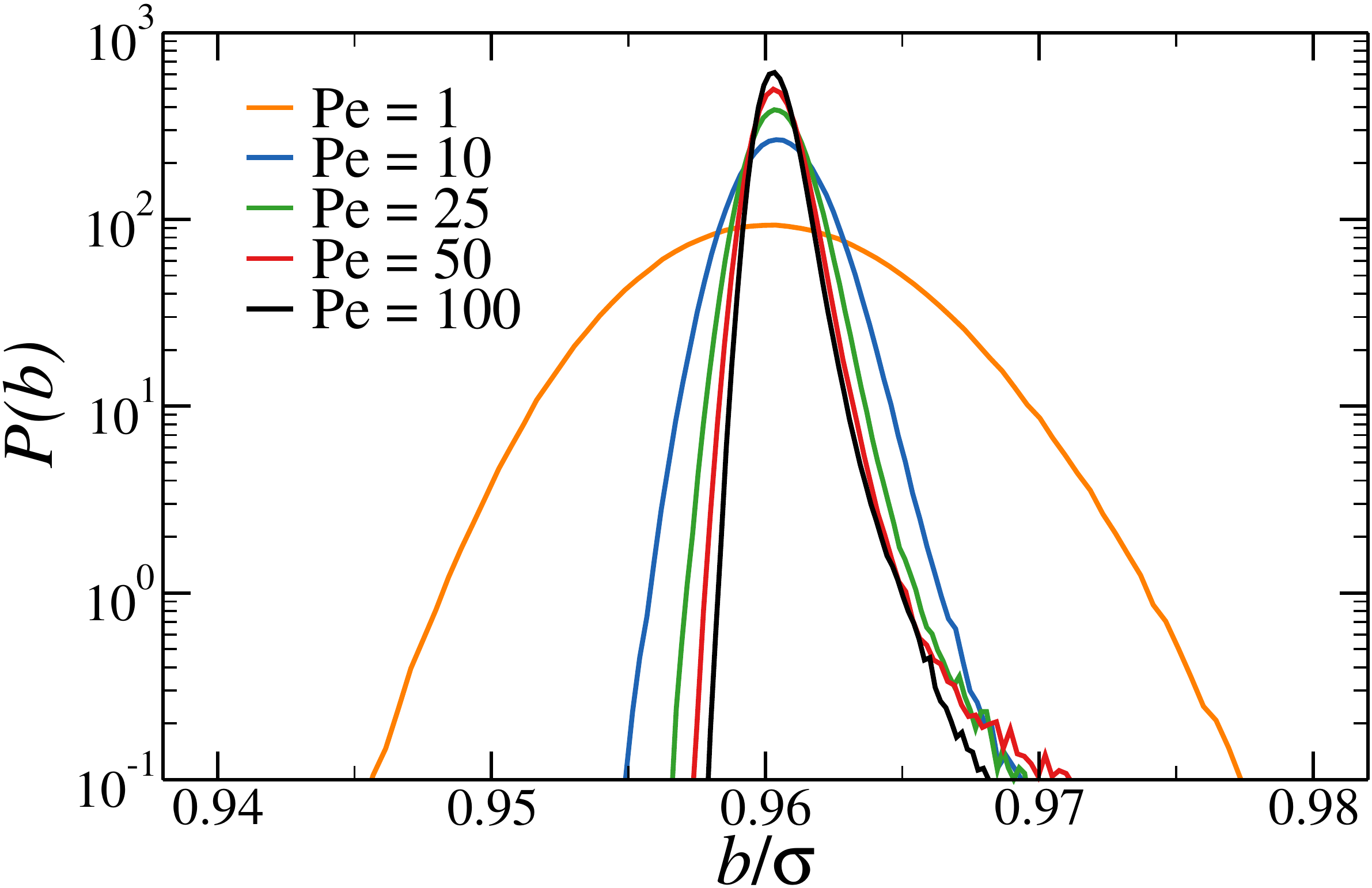}
\caption{Distribution of the bond length of active rings for A) $N =$ 70, B) $N =$ 500, and several values of ${\rm Pe}$.}
\label{fig:bonddistr}
\end{figure}

The study reported in this paper concerns unknotted ring polymers. In order to assess the role of the topology on the conformational and (especially) on the dynamical properties, one has to ensure that the topology is preserved, i.e. the rings must remain unknotted at all times. We tested the unknottedness of the rings using Kymoknot\cite{Tubiana2018}; no knot have been found by means of this test. Further, we check the distributions of the bond length, visible in Fig.~\ref{fig:bonddistr}

We check rings of length $N=$ 70 and $N =$ 500; we notice that the distributions are very similar at fixed ${\rm Pe}$ for both cases (the former inflated, the latter collapsed). Further, we notice that the maximum extension of the bonds is 1-2 \% of the characteristic bond length $b =$ 0.96 $\sigma$ and it is sufficient to prevent crossings.  

\section{Ghost rings}

\begin{figure}[!h]
\centering
\includegraphics[width=0.45\textwidth]{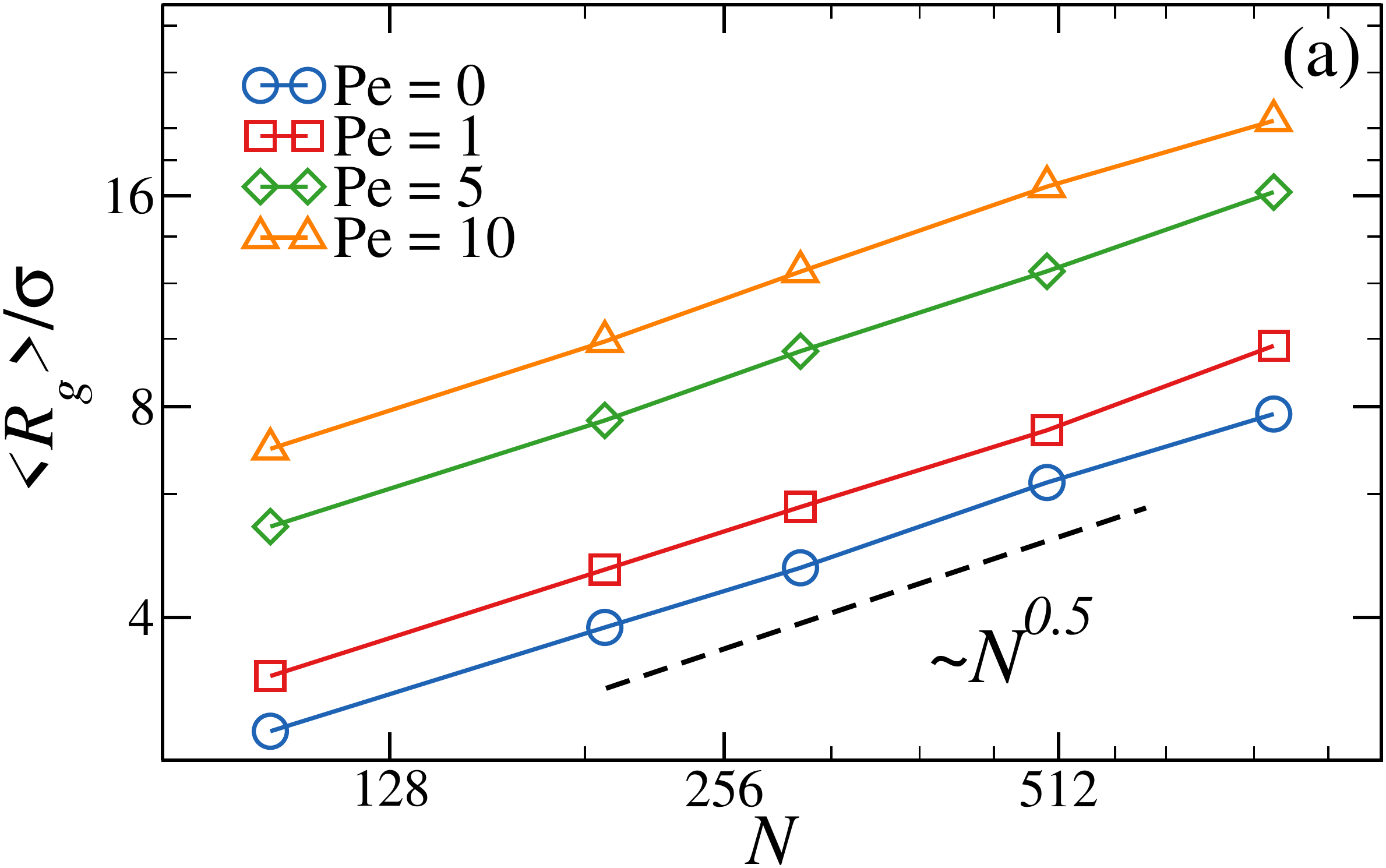}
\includegraphics[width=0.45\textwidth]{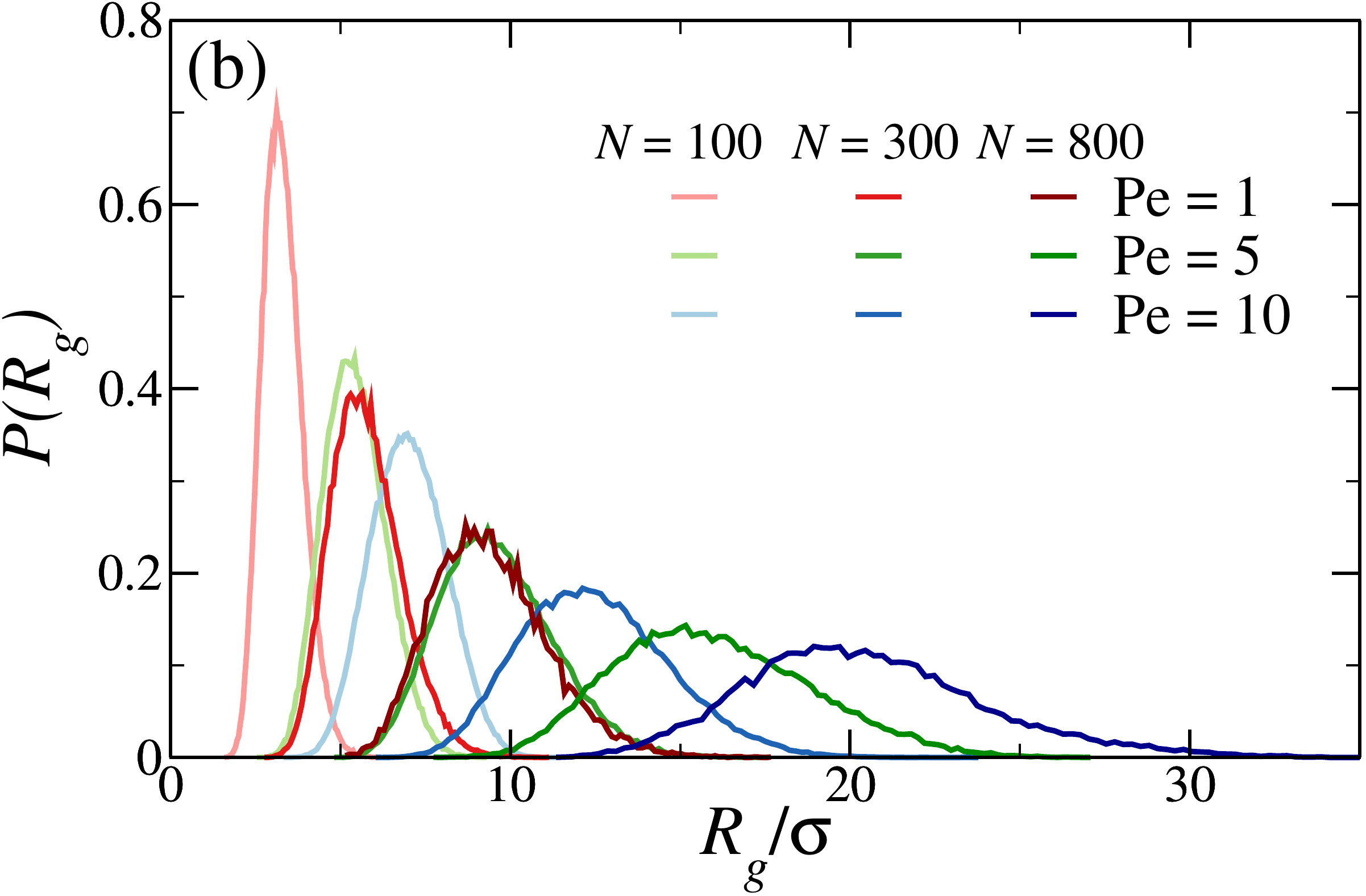}\\
\includegraphics[width=0.45\textwidth]{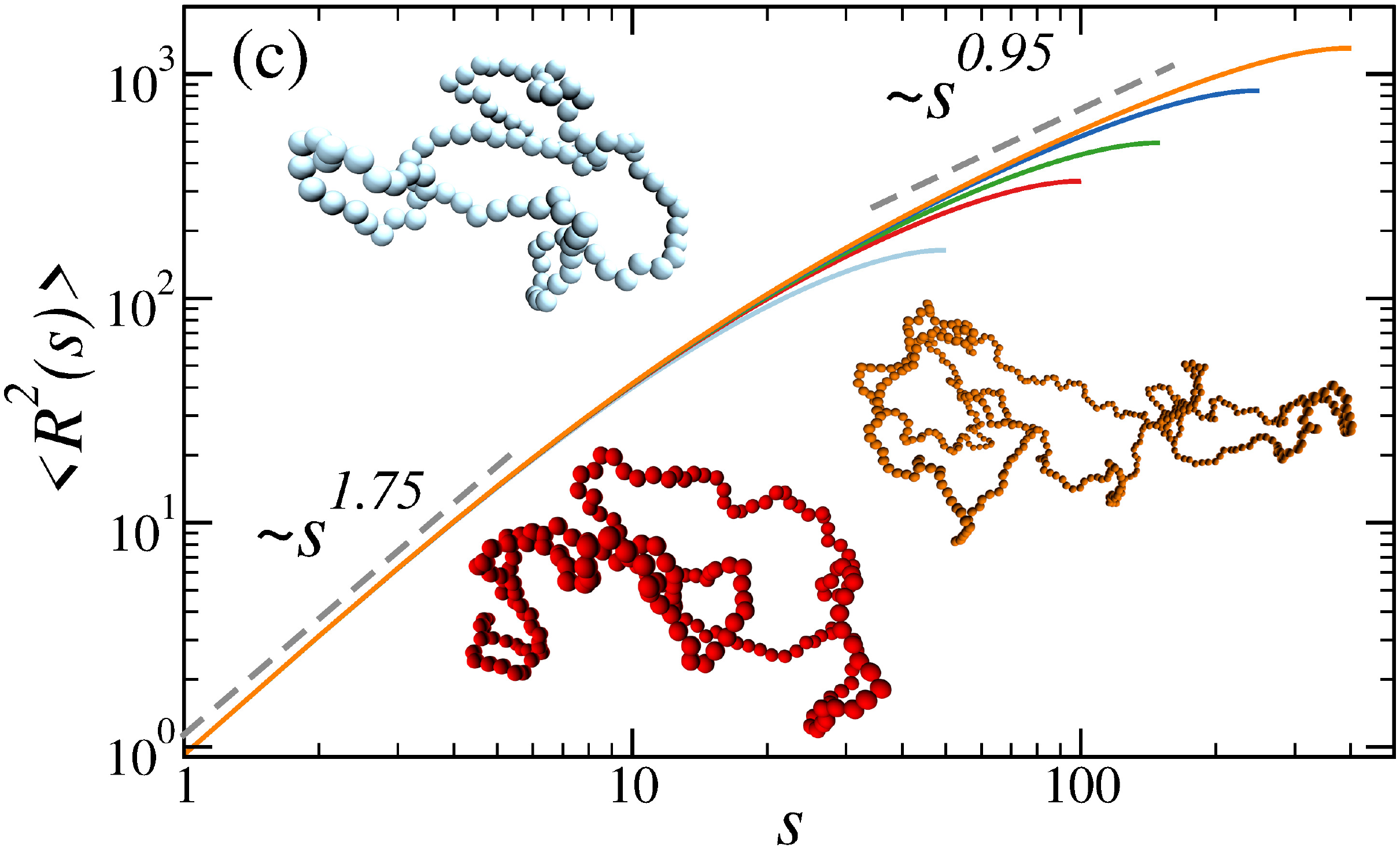}
\includegraphics[width=0.45\textwidth]{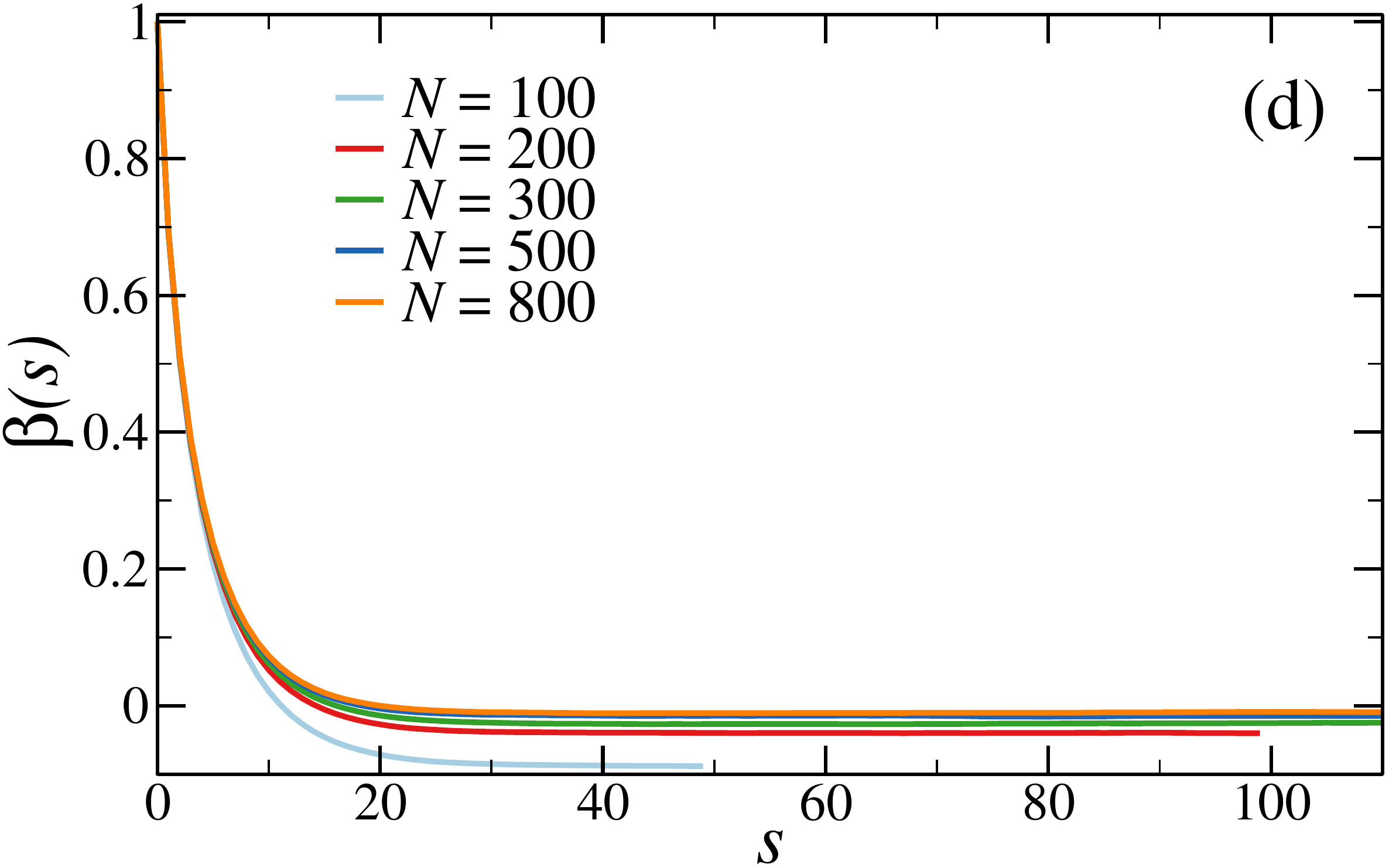}\\
\includegraphics[width=0.3\textwidth]{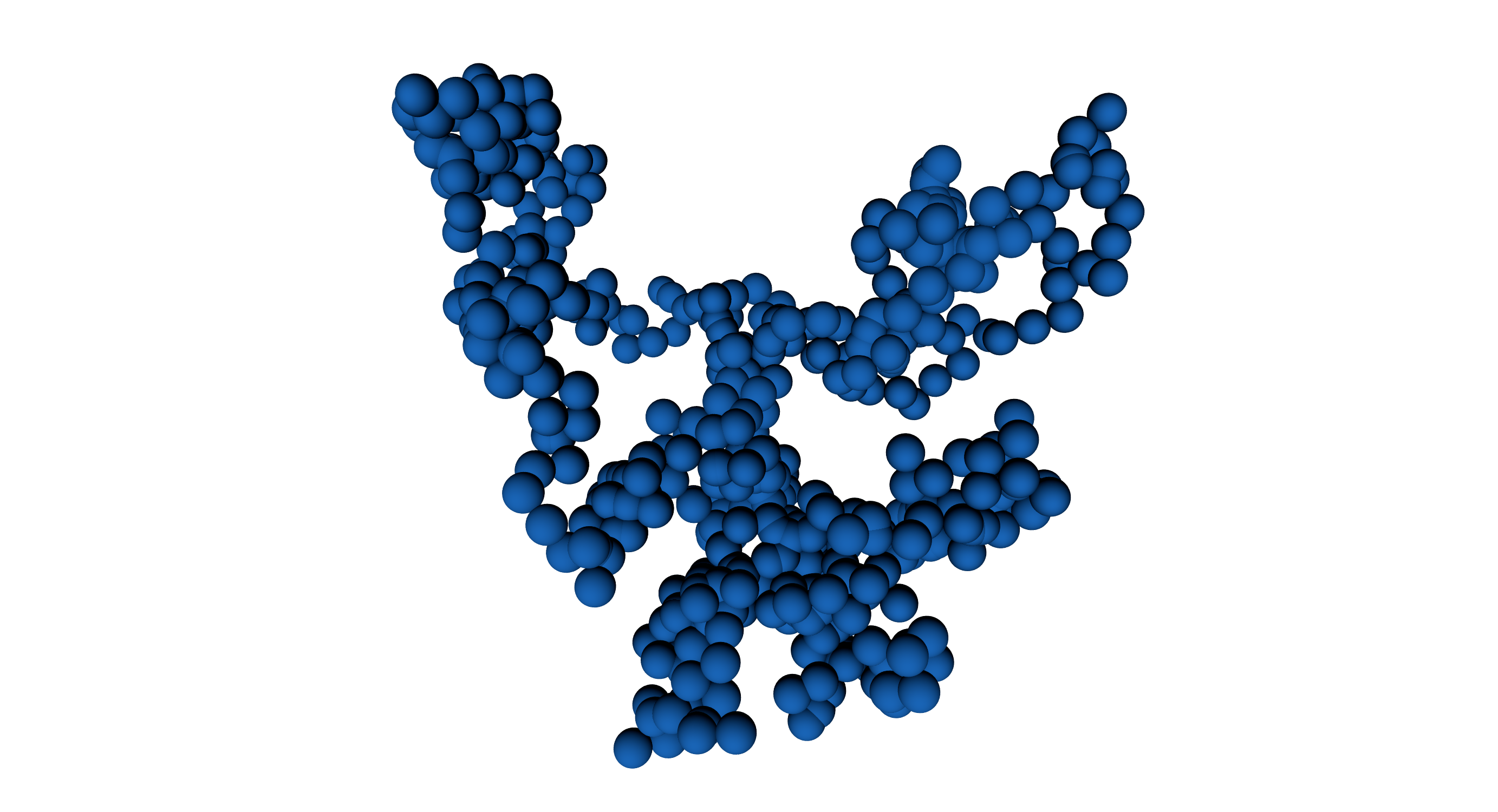}
\includegraphics[width=0.3\textwidth]{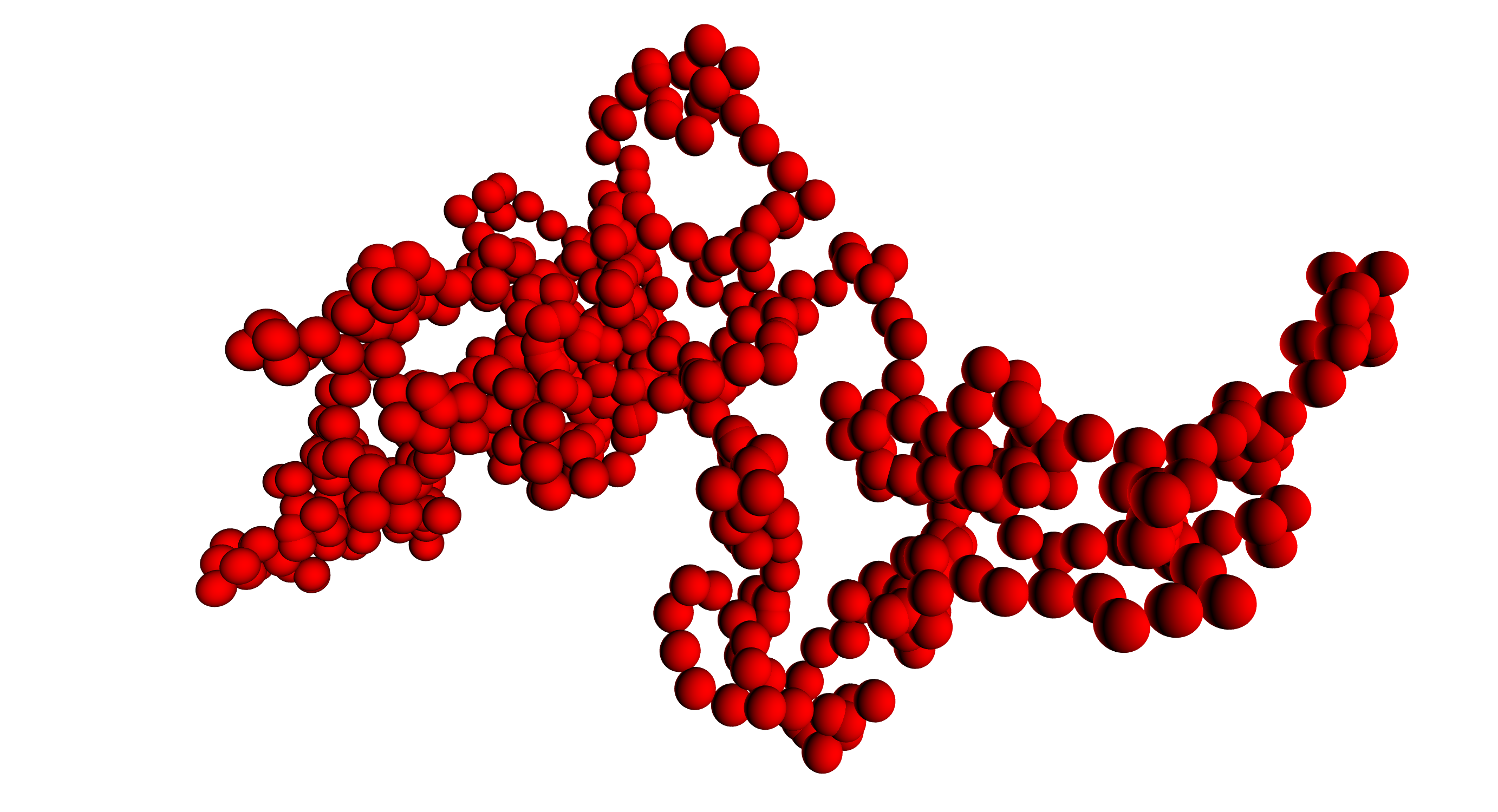}
\includegraphics[width=0.3\textwidth]{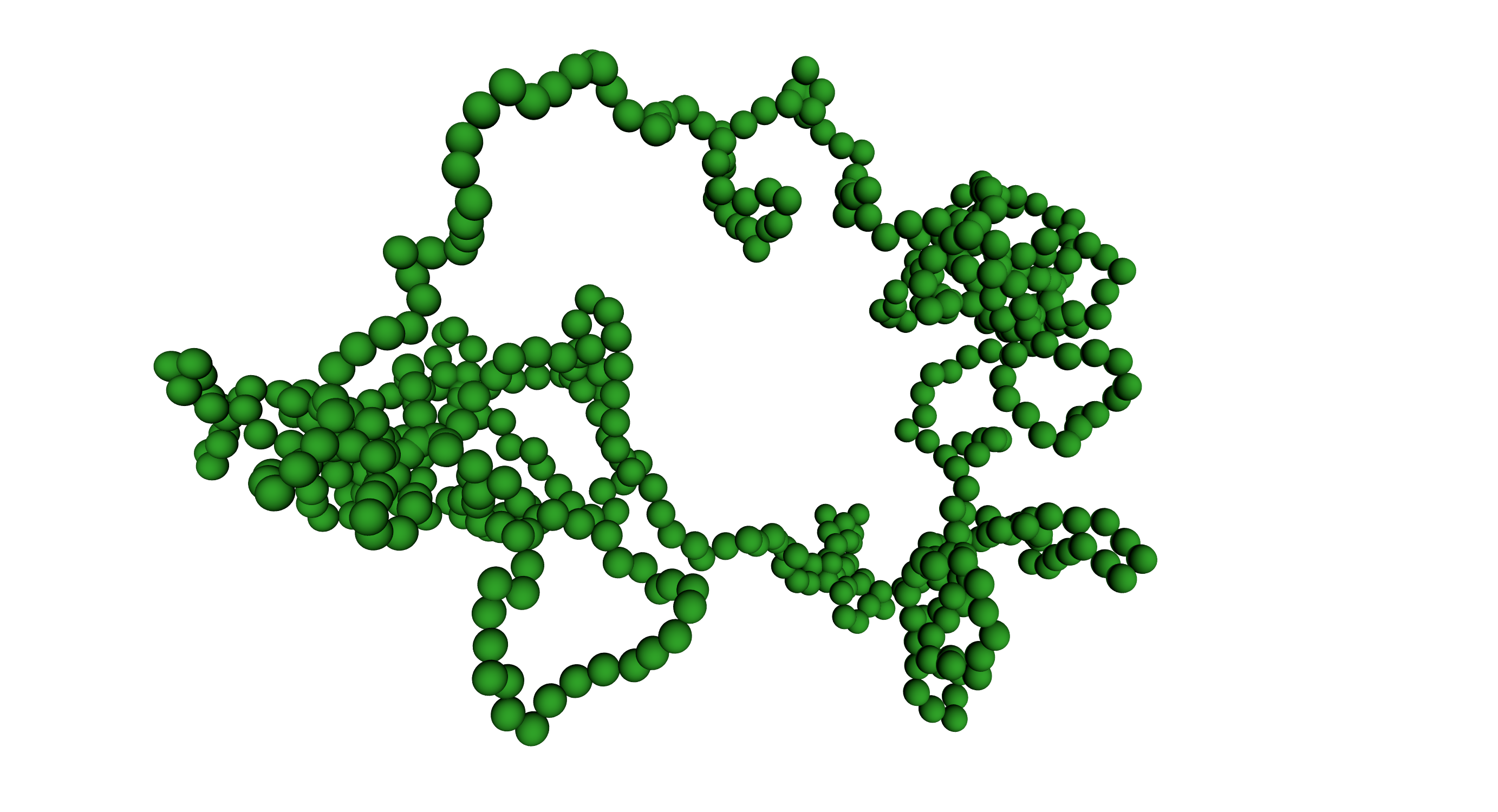}
\caption{Configurational properties of ghost rings: A) Gyration radius as function of the ring length $N$; B) Distributions of the gyration radius; C) Mean square internal distance as function of the contour distance $s$; D) Bond-bond correlation as function of the contour distance. Snapshots: panel c) light blue $N =$ 100, red $N =$ 200, orange $N =$ 500, all at ${\rm Pe} =$ 10; bottom row blue ${\rm Pe} =$ 0, red ${\rm Pe} =$ 1, green ${\rm Pe} =$ 5, all of length $N =$ 500. }
\label{fig:ghostrings}
\end{figure}

We consider, in this section, the conformational properties of ghost (or phantom) rings, i.e. rings where non-nearest neighbours along the backbone do not interact with each other. For such rings, topology is not preserved; we check the effect of the activity in such case, in order to highlight the role of topology preservation and steric interactions over the phenomenology observed in the main text (see Fig.~\ref{fig:ghostrings}).  

The gyration radius (panel (a)) scales now as $N^{0.5}$ for all the systems considered; activity only swells the rings. This behaviour is different from what observed for gaussian linear chains, that also showed a progressive decrease of the exponent of the gyration radius, at the values of ${\rm Pe}$ considered here\cite{bianco2018}. Notice that the exponent is the right one in the passive case ($\nu =$ 0.5), because topology is not conserved. Indeed this is confirmed in panel (b) and (c); notice that the fluctuations of $R_g$ become stronger as we increase ${\rm Pe}$. Further, in panel (d) the bond-bond correlation does not show the presence of any minimum, indicating that the bonds do not have a preferred organization, with a characteristic size (see snapshots). All these quantities support the observation that, without self-avoidance and topology preservation, formation of tangles is not possible. With respect to gaussian linear chains, buckling happens also there, but cyclisation reduces the space of possible configurations to a subset where buckling does not lead to a size reduction. 

\section{Theoretical analysis of the scaling of $R_g$ with $N$ shown in Fig.\ref{fig:Rg1}}

{We derive a simple argument to understand the scaling exponent of the gyration radius $R_g$ with $N$ shown in Fig.~\ref{fig:Rg1}.
In order to get analytical insight we split the monomers into two sets: those that belong to the compact collapsed main structure and those that belong to the dangling sections. 
We further assume that the $\mathcal{N}$ dangling sections are all composed by $M$ monomers and that their size $M$ is independent on $N$. The idea behind the latter assumptions is that in the scaling regime i.e., for very long polymers, the size of the dangling sections is determined by some ``microscopic'' parameters rather then by the polymer length $N$. 
Moreover we approximate both the collapsed structure and the dangling regions with spheres of radii $R_C$ and $R_M$. 
The gyration radius is defined as 
\begin{align}
 R_G^2=\sum_{i=1}^{N} \left( x_i-x_\text{CM}\right)^2 
\end{align}
where $x_\text{CM}$ is the position of the center of mass that, in practice, coincides with the center of mass of the collapsed structure.
The last expression can be rewritten as 
\begin{align}
 R_G^2 &=\sum_{i=1}^{N-\mathcal{N}M} \left(x_i-x_\text{CM}\right)^2+\sum_{i=N-\mathcal{N}M+1}^{N} \left( x_i-x_\text{CM}\right)^2\\
 &=b^2\left(N-\mathcal{N}M\right)^\frac{2}{3}+\mathcal{N}\sum_{i=1}^{M}\left(\sqrt{x_M^2+R_i^2-2\cos\alpha_i x_M R_i}-x_\text{CM}\right)^2
 \label{eq:RG-2}
\end{align}
where the first term on the r.h.s. accounts for the contribution from the monomer is the collapsed structure and the latter from all monomers belonging to the $\mathcal{N}$ identical dangling sections. In particular, in the latter term we have decomposed the distance from the center of mass into the distance to the center of mass of the dangling region plus the distance from it (see Fig.~\ref{fig:dist-scaling}).
\begin{figure}
 \includegraphics{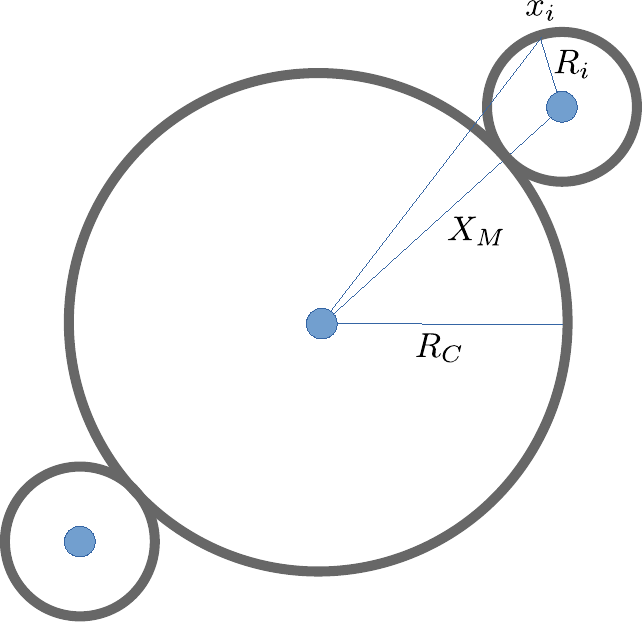}
 \caption{cartoon of a spherical collapsed structure with two equal dangling regions.}
 \label{fig:dist-scaling}
\end{figure}

Without loss of generality we shift the reference frame such that $x_\text{CM}=0$. Moreover we approximate the second term in Eq.~\eqref{eq:RG-2} as
\begin{align}
\sum_{i=1}^{M} x_M^2+R_i^2-2\cos\alpha_i x_M R_i\simeq M(x_M^2+R_M^2)
\label{eq:eq12}
\end{align}
where we have assumed that all the monomers belonging to any single dangling sections are a at distance $\sim R_M$ from its center of mass and that the term proportional to the cosine becomes negligibly small, once we sum over all the monomers.
Next, as shown in Fig.~\ref{fig:dist-scaling}, we have that 
\begin{align}
x_M=R_C+R_M=    b \left(N-\mathcal{N}M\right)^\frac{1}{3}+R_M
\end{align}
Accordingly, when the number of monomers involved in the dangling sections is small ($\mathcal{N}M\ll N$) Eq.~\eqref{eq:eq12} can be approximated by
\begin{align}
    M(x_M^2+R_M^2)=M(R_C^2+2R_cR_M+R_C^2)\simeq MR_C^2\simeq M b^2 \left(N-\mathcal{N}M\right)^\frac{2}{3}
\end{align}
Accordingly, the expression for the gyration radius can be be approximated by
\begin{align}
 R_G^2\simeq b^2 \left(N-\mathcal{N}M\right)^\frac{2}{3}(\mathcal{N}M+1)\simeq  b^2 N^\frac{2}{3}\mathcal{N}M\,.
 \label{eq:RG-3}
\end{align}
where have used that $1\ll \mathcal{N}M \ll N$. We note that
Eq.~\eqref{eq:RG-3} is indeed function of  $\mathcal{M}=\mathcal{N}M$ that is the total number of monomers in the dangling regions. 
Predicting the scaling of $\mathcal{M}$ with $N$ is a formidable task. Indeed, due to the active nature of the system, we cannot define a free energy and hence we cannot compare any free energy cost with thermal energy as usual in scaling models of passive polymers at equilibrium~\cite{DeGennes_book}. Accordingly, we compare Eq.~\eqref{eq:RG-3} with the outcome of the numerical simulations
\begin{align}
 b^2 N^\frac{2}{3}\mathcal{M}\simeq b^2_\infty N^{2\nu_\text{long}}
 \label{eq:RG-3-1}
\end{align}
where $b_\infty \approx 0.39-0.40$ is a prefactor with dimension of a length that can be extracted from Fig.~\ref{fig:Rg1}. 
From the last expression we obtain the scaling of $\mathcal{M}$
\begin{align}
 \mathcal{M}\simeq \frac{b^2_\infty}{b^2} N^{2\nu_\text{long}-\frac{2}{3}}\,.
 \label{eq:RG-4}
\end{align}
Upon substituting 
$\nu_\text{long}\simeq 0.41$ in Eq.~\eqref{eq:RG-4} we obtain
\begin{align}
 \mathcal{M}\simeq \frac{b^2_\infty}{b^2} N^{0.15}
 \label{eq:RG-5}
\end{align}
therefore we expect $\mathcal{M}$ to grow very slowly with $N$. }

\section{Non-equilibrium route to steady state}

{We report here more information about the non-equilibrium route to the steady state.}
\begin{figure}[!h]
\centering
\includegraphics[width=0.4\textwidth]{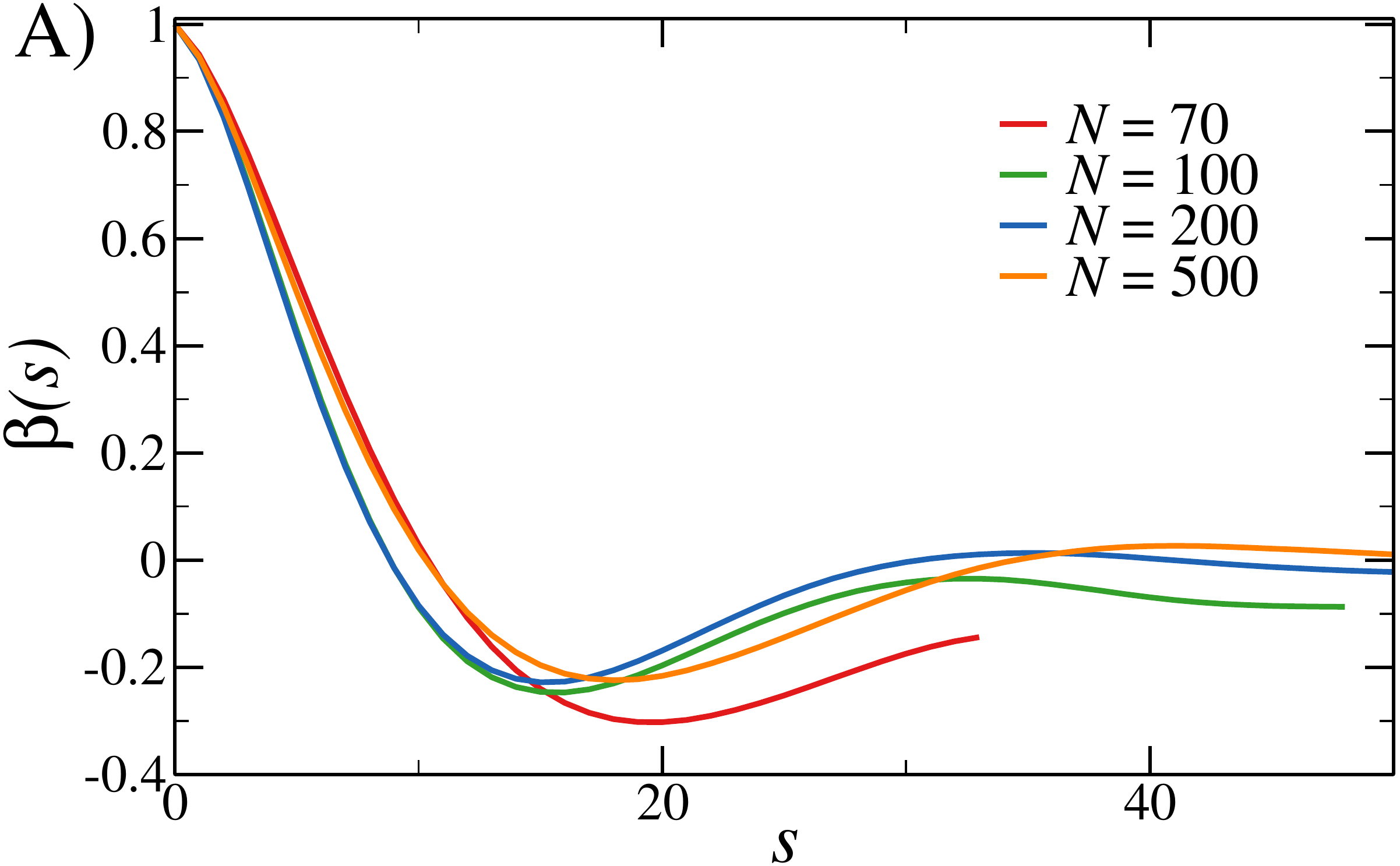}
\includegraphics[width=0.38\textwidth]{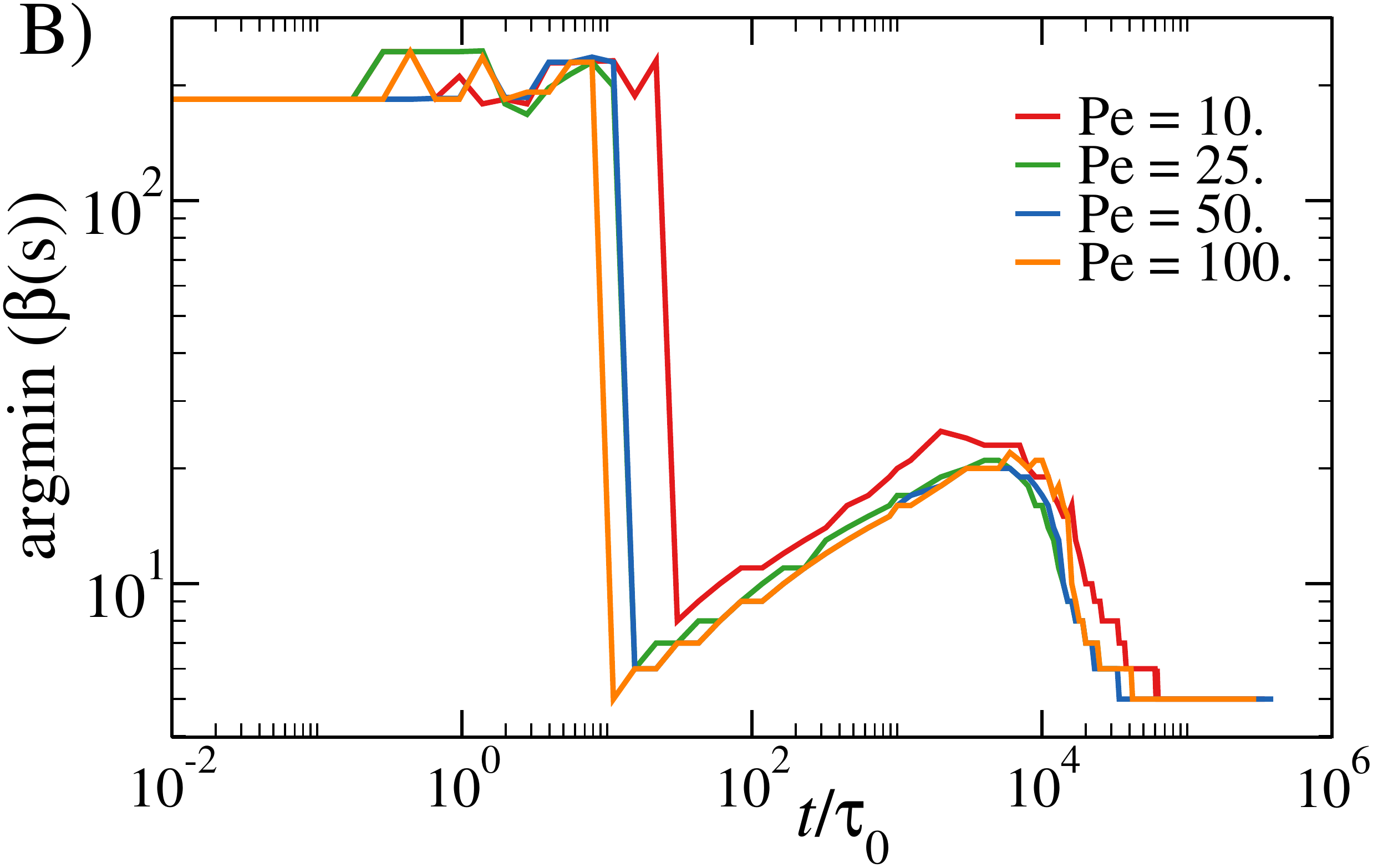}
\put(0,0){A)}
\caption{A) Bond-bond correlation as function of the contour distance $s$ for rings of different length, fixed activity ${\rm Pe} =$ 100, measured at the (arbitrary) fixed time $t/\tau_{0} =$ 1300. B) Position of the minimum of the bond correlation as a function of time, normalized by $\tau_0$, for rings of length $N =$ 500 and different values of $\rm Pe$.} 
\label{fig:reecorr70}
\end{figure}
We report, {in Fig.~\ref{fig:reecorr70}a,} examples of bond-bond correlation functions during the non-equilibrium route to the steady state, taken at the (common) time $t/\tau_{0} =$ 1300, where $\tau_0$ is the typical diffusion time of a single monomer. An average over 400 $< M < $ 2850 independent configurations has been carried out. We notice that, as suggested by Fig.~2 of the main text, the position of the minimum is roughly the same for all the values of $N$ considered. {Moreover, we consider, in Fig.~\ref{fig:reecorr70}b, the minimum of $\beta(s)$ for fixed length $N =$ 500 and various ${\rm Pe} >$ 10. We observe that the curves display the same features as the curves reported in the main text. In particular, the intermediate growth of $\beta(s)$ is almost independent on ${\rm Pe}$ (one can observe a slight difference between the curve for ${\rm Pe} =$ 10 and the other curves), while the growth exponent remains the same. We also note that, upon reaching the steady state,  the minimum of $\beta(s)$ is independent on ${\rm Pe}$, for sufficiently high values of ${\rm Pe}$.  }

\subsection{Route to steady state at low Pe}

\begin{figure}[!h]
\centering
\includegraphics[width=0.32\textwidth]{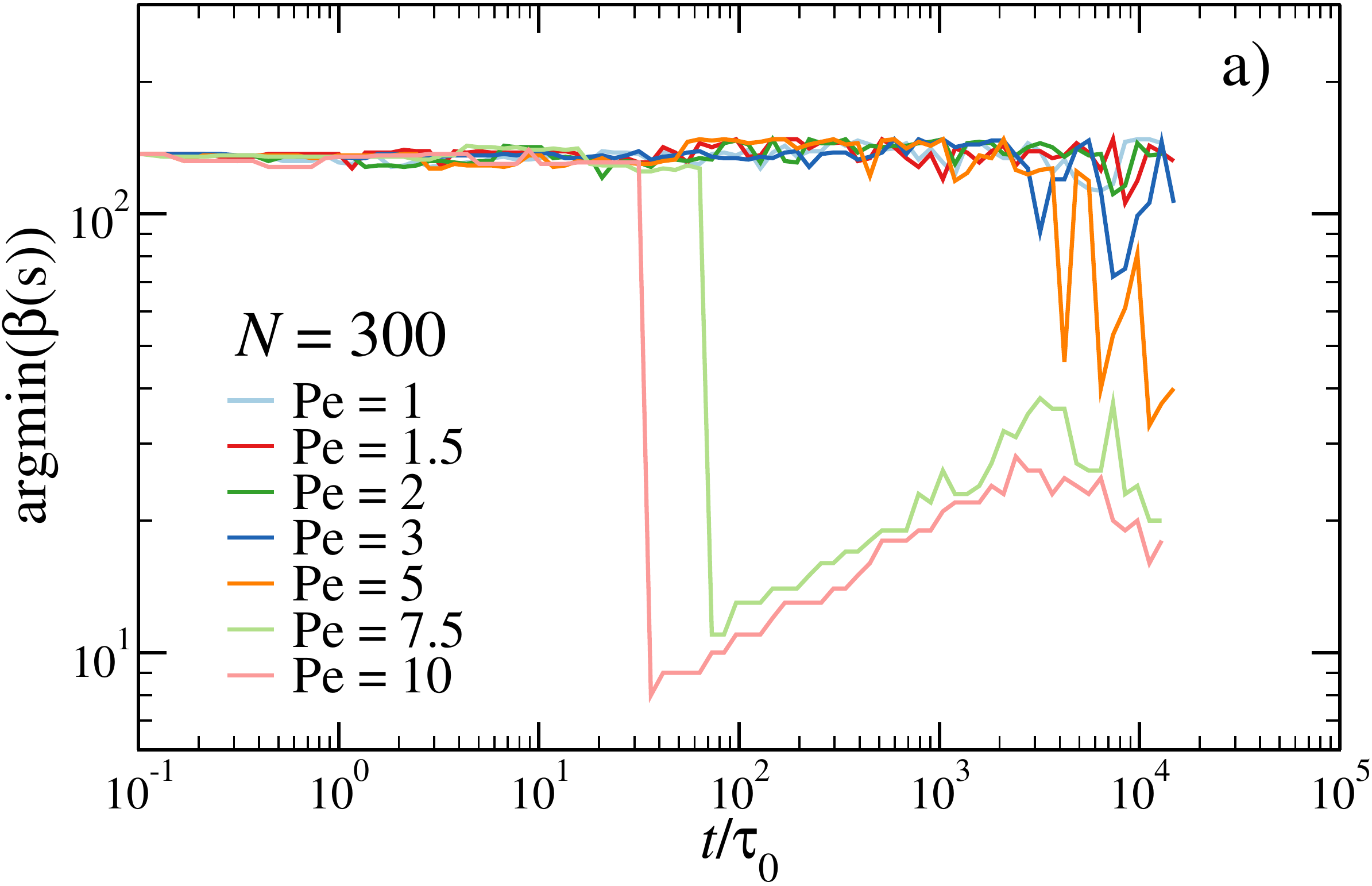}
\includegraphics[width=0.32\textwidth]{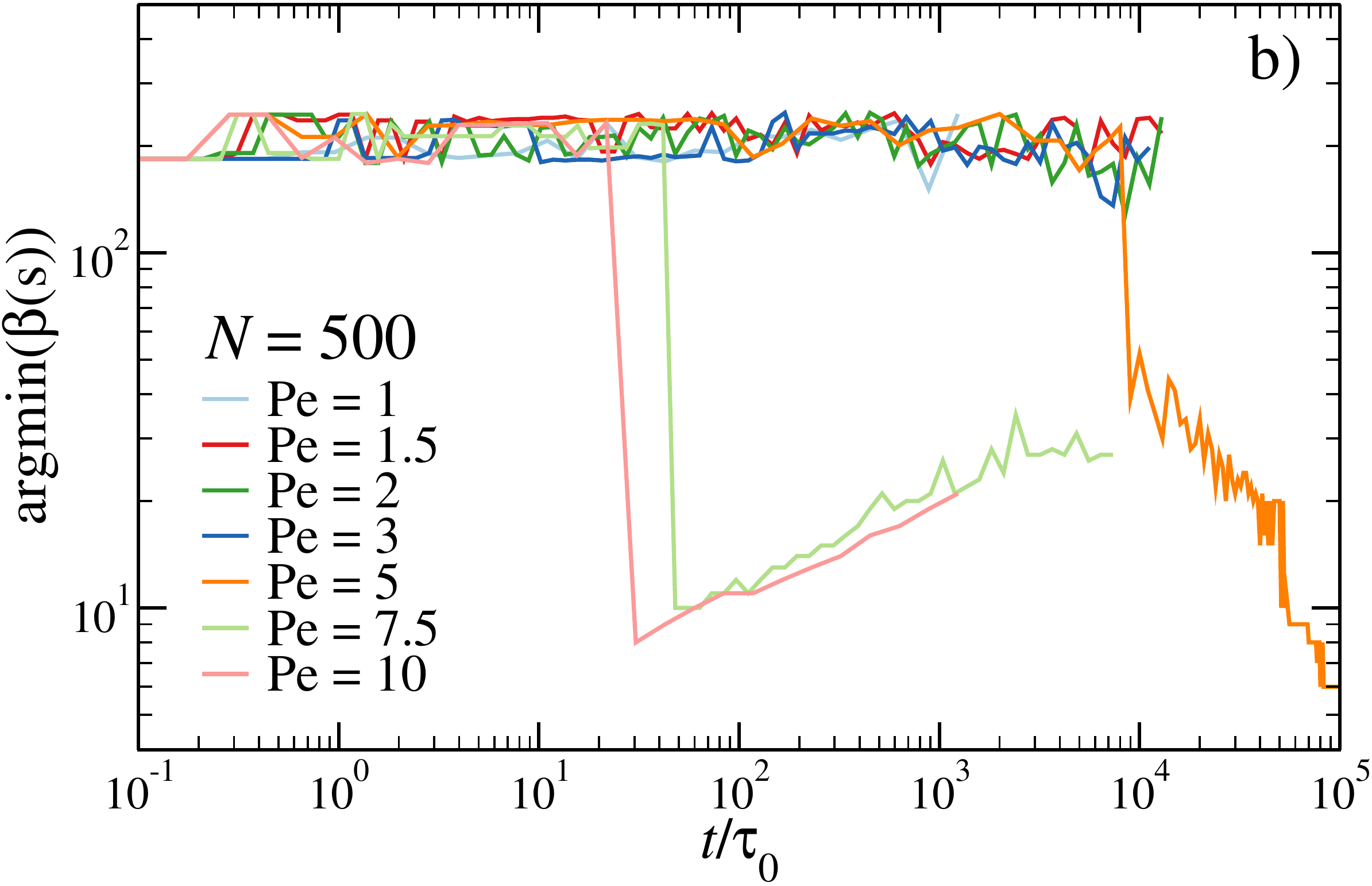}
\includegraphics[width=0.32\textwidth]{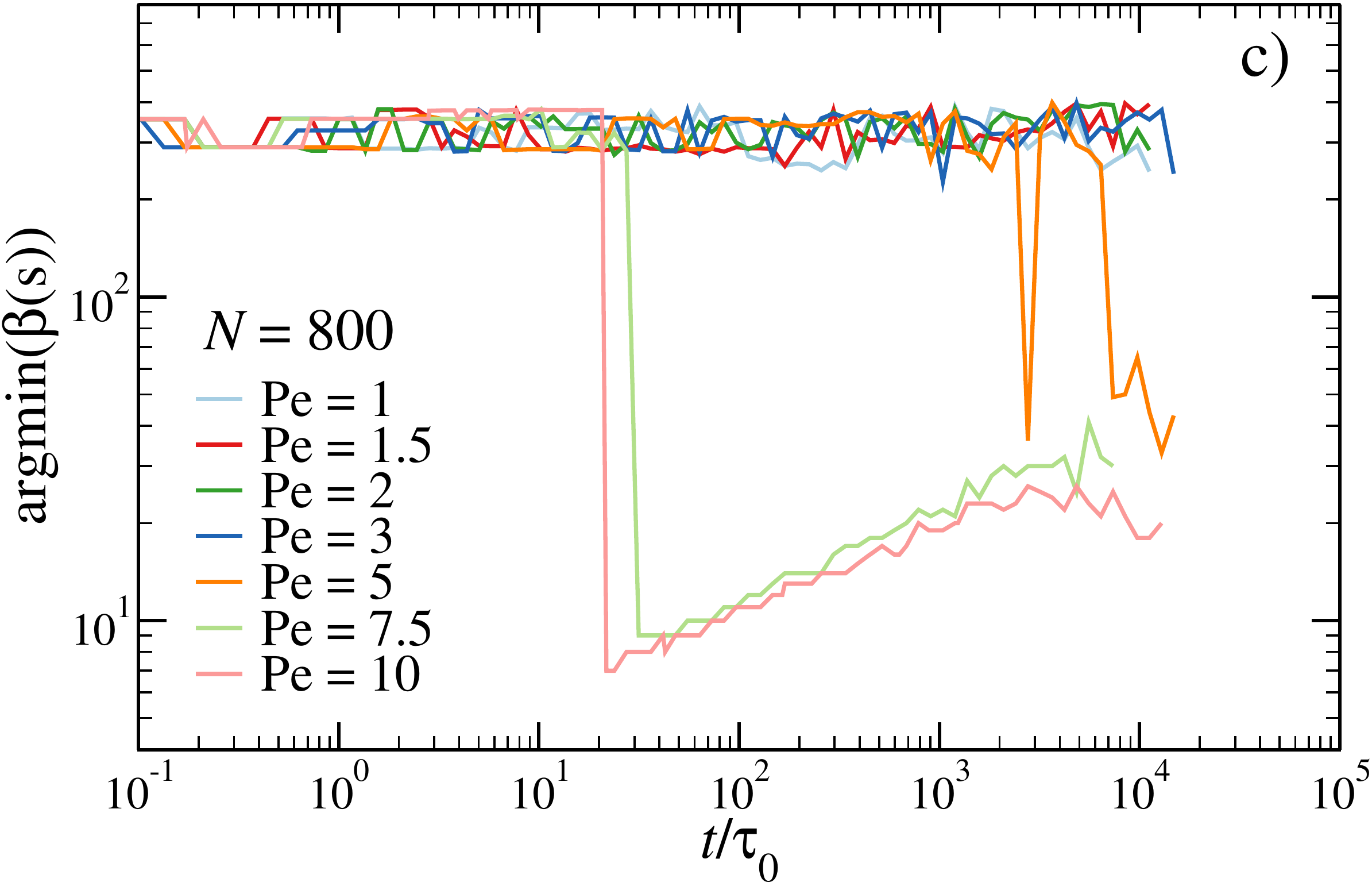}
\caption{Loci of the minima of the bond-bond correlation ${\rm argmin} \beta(s)$ as function of time for several ${\rm Pe} \leq$ 10 and a) $N =$ 300, b) $N =$ 500, c) $N =$ 800.}
\label{fig:bcorrminima_lowpe}
\end{figure}

\el{In order to better characterize the low-Pe regime, we performed short simulations specifically at 1$< {\rm Pe} <$ 10 and $N \geq$ 300. As reported in Fig.~\ref{fig:bcorrminima_lowpe}, we observe that the sudden drop observed at larger ${\rm Pe}$ disappears. Instead, we observe a less abrupt decrease or no decrease at all (within the investigated time window). Thus, it appears there is an additional pathway at low ${\rm Pe}$, where rings still end up in a collapsed state but that does follow the pathway described in the main text. Overall, the fact that the characteristic bends do not develop at low ${\rm Pe}$ is consistent with a previous result on active linear chains\cite{bianco2018},
where similar "bends" were really marked only at sufficiently high ${\rm Pe}$.}

\subsection{Torsional order parameter}

\begin{figure}[!h]
\centering
\includegraphics[width=0.45\textwidth]{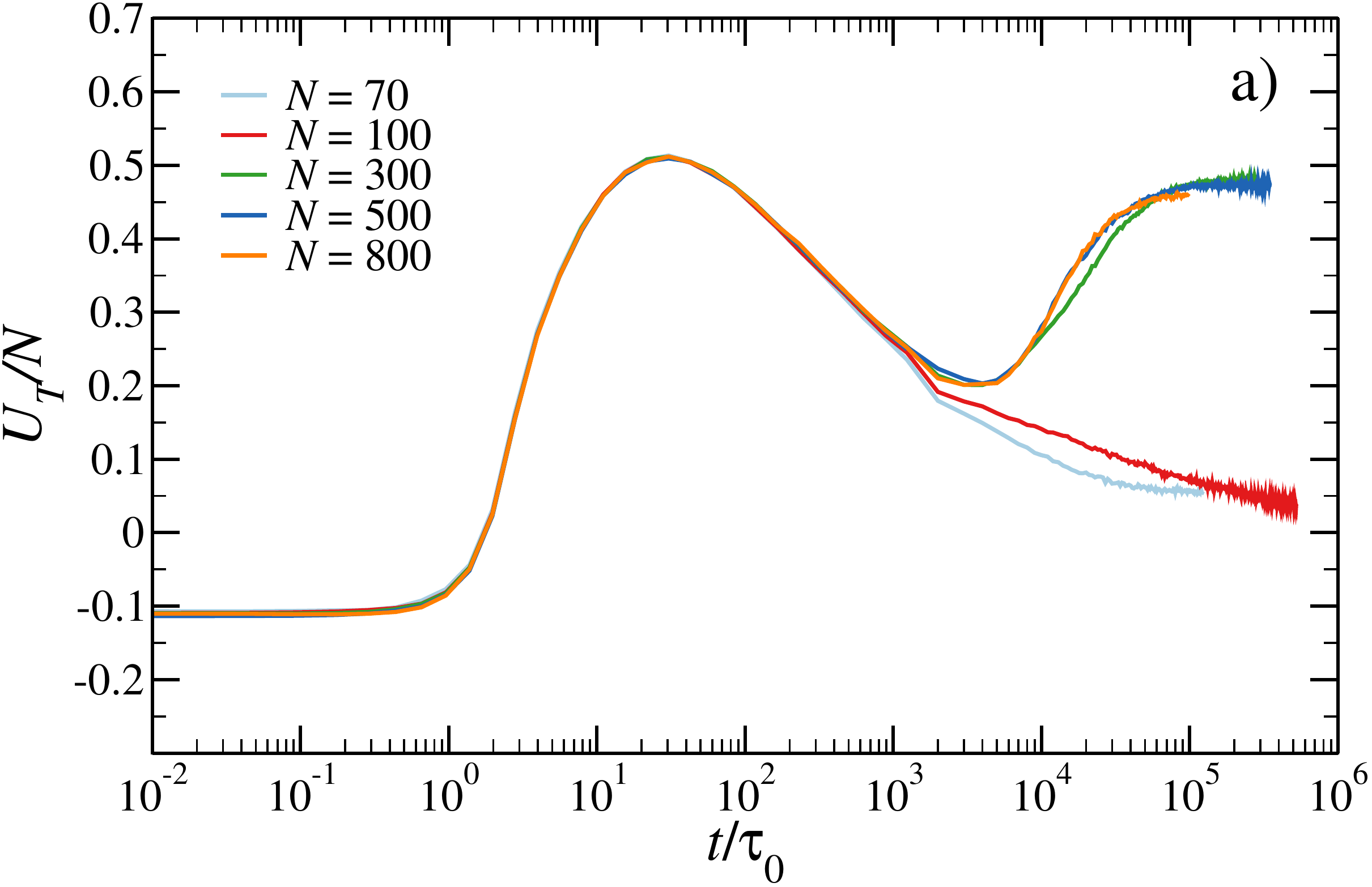}
\includegraphics[width=0.45\textwidth]{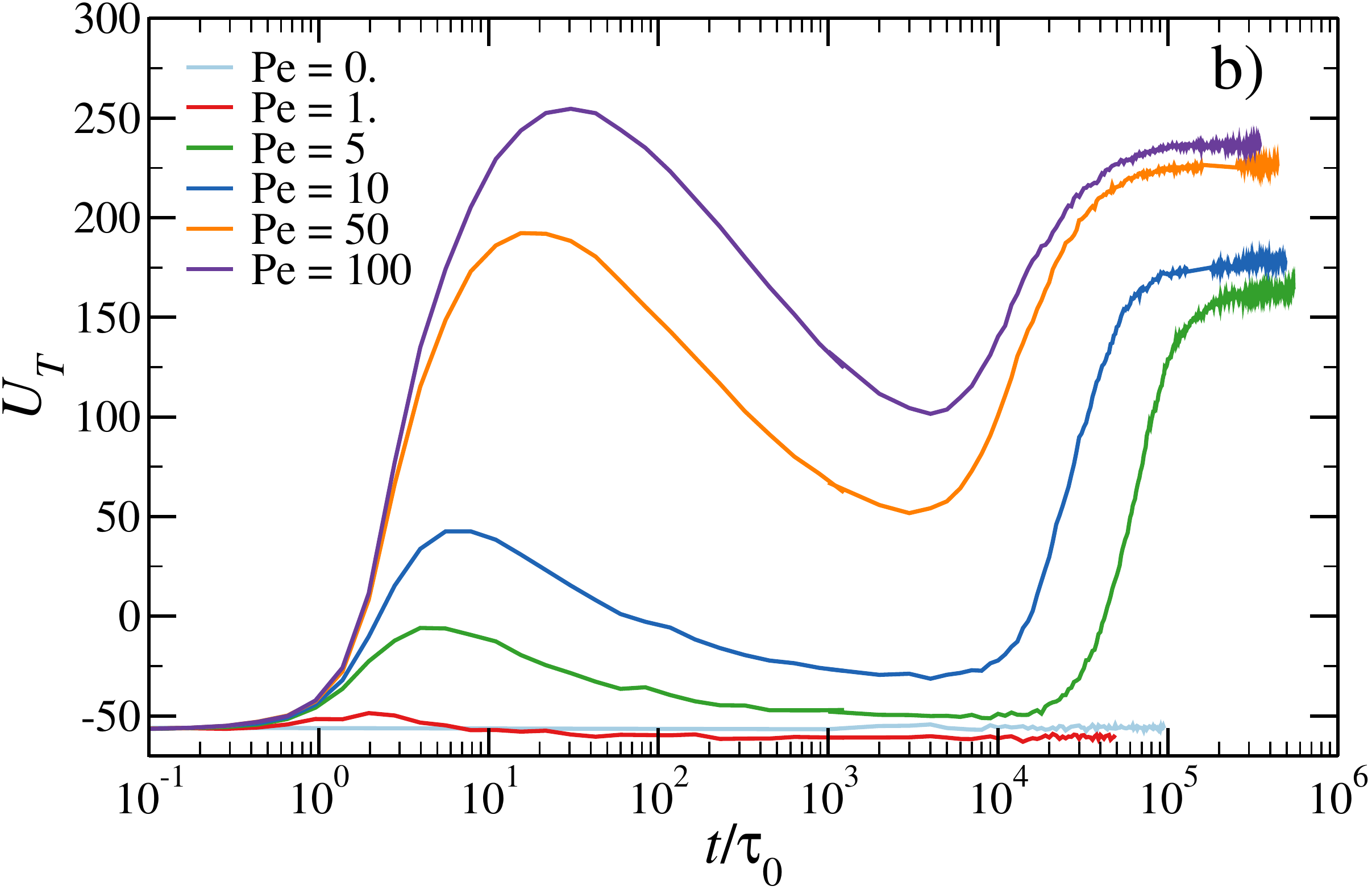}
\caption{Torsional order parameter $U_T$ as function of time for A) fixed ${\rm Pe} =$ 100, several $N$; B) fixed $N =$ 500, several values of ${\rm Pe}$.}
\label{fig:torsionalparameter}
\end{figure}

\el{In order to characterize further the organization in the collapsed state and, possibly, the pathway to the steady state, we compute the torsional order parameter, defined in Eq.~\eqref{eq:tors-ord-param} in Section "Observables" and we report it in Fig.~\ref{fig:torsionalparameter}. The quantity $U_T$ increases as the polymer (or parts of it) assumes an helical conformation and can be used to characterise the helical buckling of polymers in strong flow, where the backbone reacts to a compression/extension in the direction of the channel's main axis\cite{chelakkot2012a}. This compression/extension happens, to some extent, also in tangentially active polymers. Indeed, in Fig.~\ref{fig:torsionalparameter}, we report $U_T$ as function of time, from a passive, equilibrated conformation to the steady state. We observe that, for all rings, $U_T$ is proportional to $N$; it also assumes a negative value for passive rings, which is constant in equilibrium (see Fig.~\ref{fig:torsionalparameter}b), compatibly with the results reported in \cite{chelakkot2012a} in the same conditions. For active rings, $U_T$ shows a strong increase within the time window $\tau_0 < t \lesssim 100 \tau_0$; the magnitude of the peak as well as its position in time grow larger upon increasing ${\rm Pe}$. After the peak, at $t/t_0 \approx 10^4$, $U_T$ continues to decrease for short rings, while it increases again for long rings until it finally reaches a plateau; as visible in  Fig.~\ref{fig:torsionalparameter}b, the value of this plateau depends non-trivially on ${\rm Pe}$. Notice that the position in time of this minimum for $N >$ 300 is strikingly compatible with the timescale at which we observe the onset of the final collapse in Fig.~2 of the main text. Indeed, the emergence of helical-like structures is perfectly compatible with the anti-correlation observed in the bond correlation function. Further, notice that the fact that $U_T/N$ assumes a large value at the steady state for long rings brings forward a more qualitative description of the organization in the collapsed state, as a structure formed by an internal "core", wrapped all around by other chain segments.}

\section{Ring configurational properties at steady state}

\subsection{Gyration radius}

\begin{figure}[!h]
\centering
\includegraphics[width=0.45\textwidth]{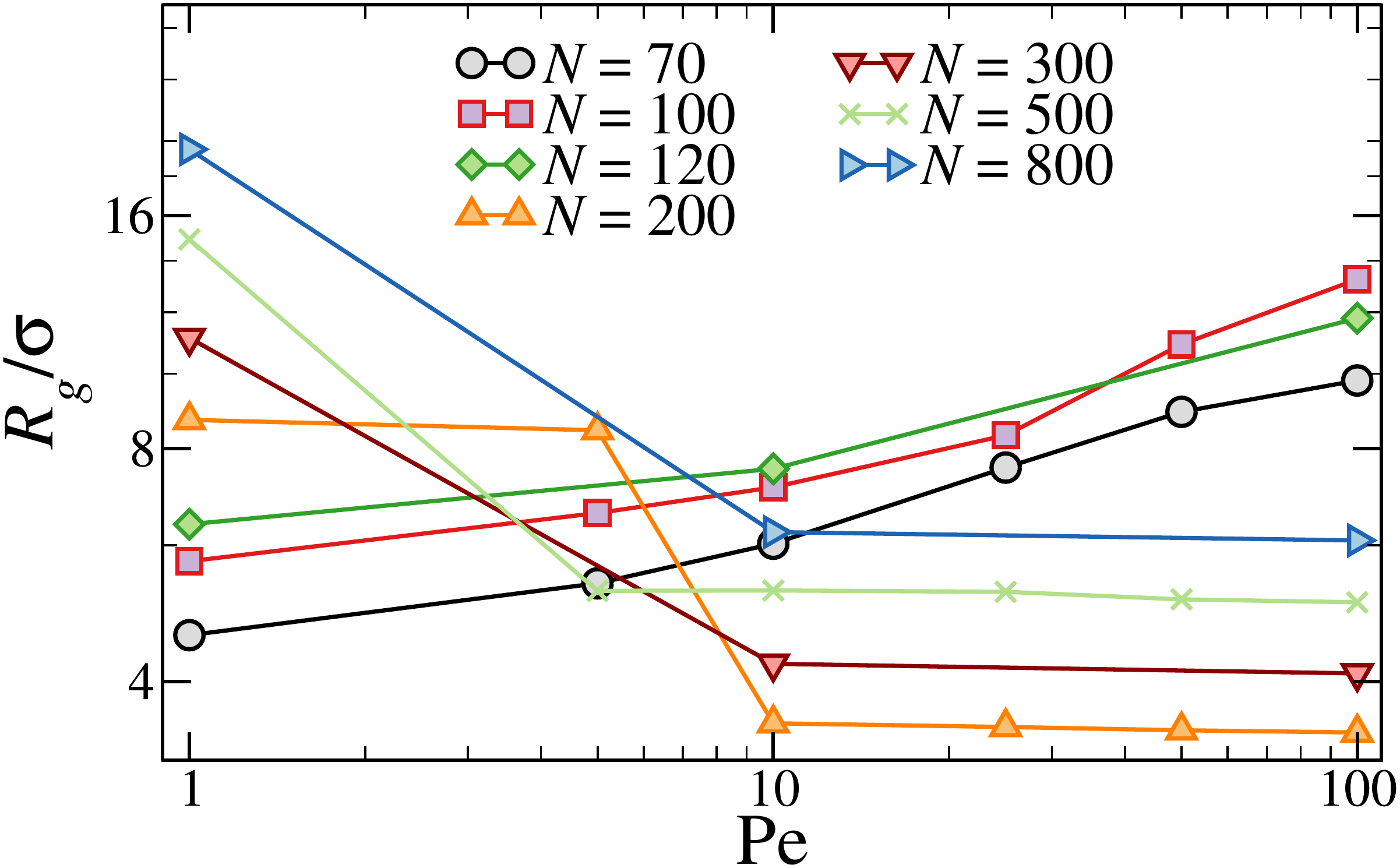}
\caption{Gyration radius of as self-propelled ring as function of ${\rm Pe}$.}
\label{fig:Rgscaling_Pe}
\end{figure}

We report, in Fig.~\ref{fig:Rgscaling_Pe}, the average gyration radius of self-propelled ring polymer as function of ${\rm Pe}$. We notice that for short chains, up to $N =$ 120, $R_g$ increases upon increasing ${\rm Pe}$. Above $N \approx$ 300, $\langle R_g \rangle$ become insensitive of ${\rm Pe}$, for all ${\rm Pe} >$ 1 investigated.      

\begin{figure}[!h]
\centering
\includegraphics[width=0.40\textwidth]{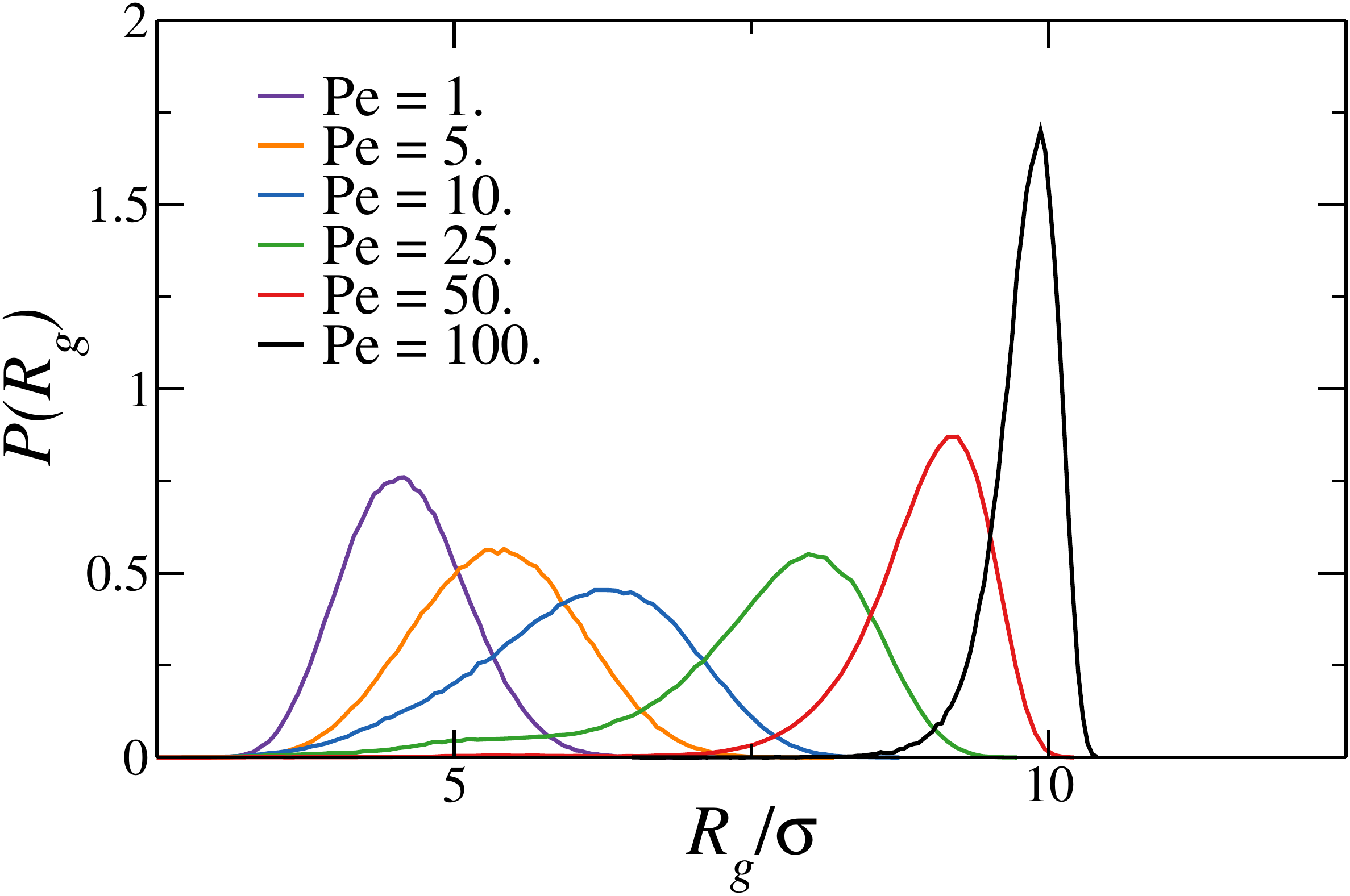}
\includegraphics[width=0.40\textwidth]{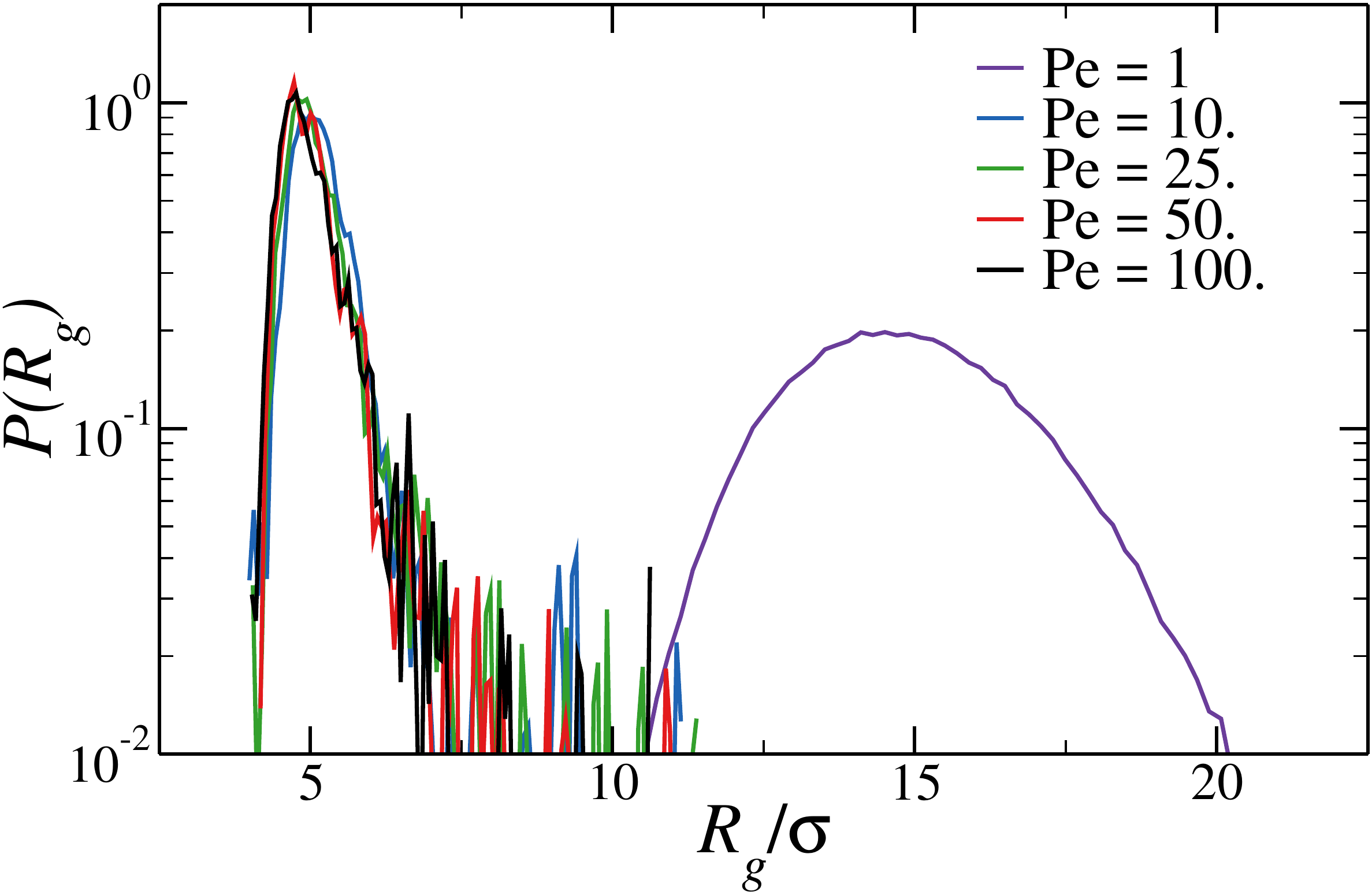}
\caption{Distribution of the gyration radius of self-propelled rings for $N =$ 70 (left panel), B) $N =$ 500, (right panel), and several values of ${\rm Pe}$.}
\label{fig:Rgdistr}
\end{figure}

We further inspect the distributions of the gyration radius, for different values of $N$ and ${\rm Pe}$ in Fig.~\ref{fig:Rgdistr}. Fixing $N =$70, we notice that the mode of the distribution increases, upon increasing ${\rm Pe}$; the distributions become progressively narrower, i.e. fluctuations are suppressed. For $N =$ 500, distributions overlap for ${\rm Pe} > $ 1.  

\subsection{Bond correlation function}

\begin{figure*}[t]
\includegraphics[width=0.45\textwidth]{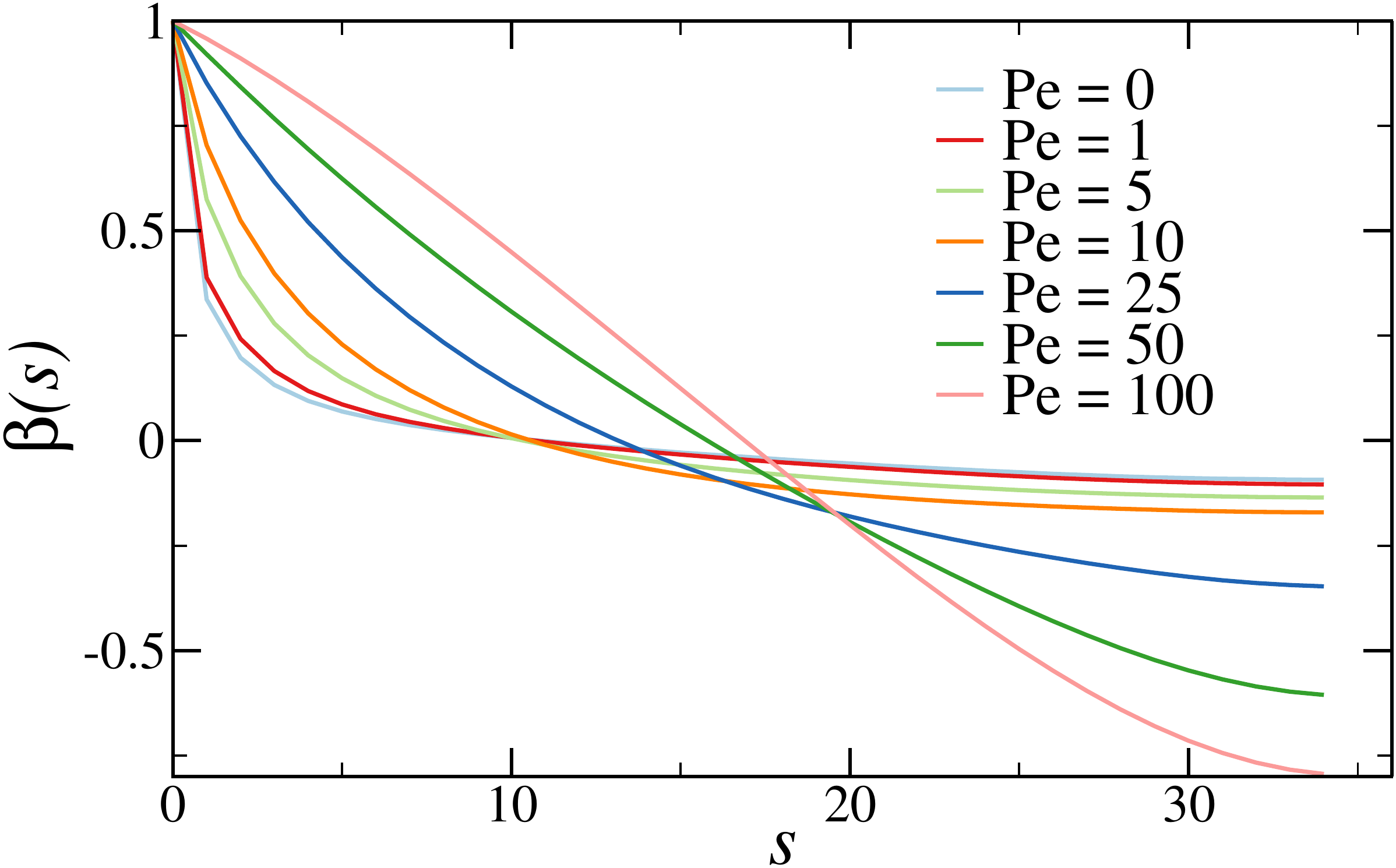}
\hspace{0.1cm}
\includegraphics[width=0.45\textwidth]{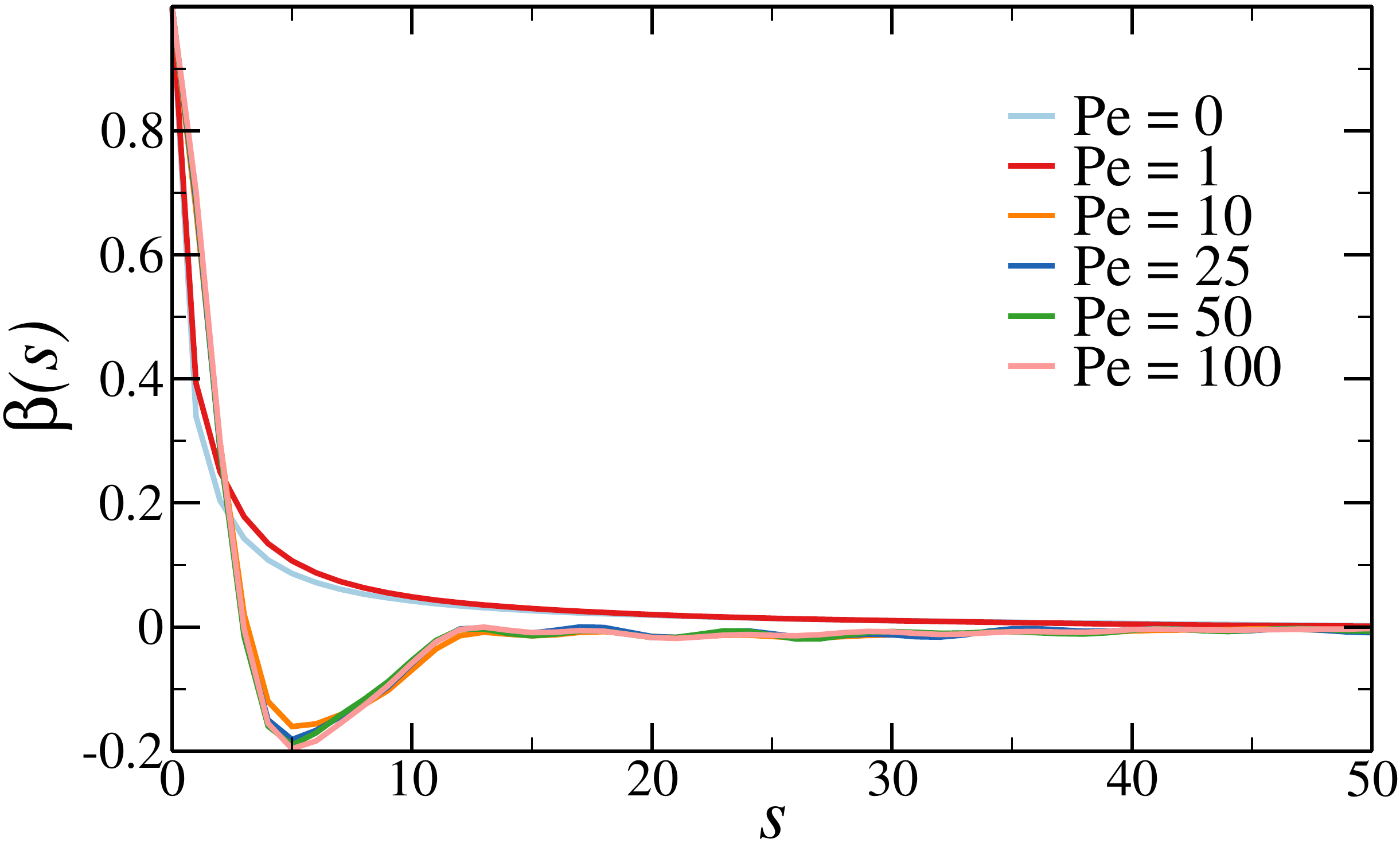}
\caption{Bond-bond correlation as function of the contour distance $s$; fixed $N =$ 70 (left panel), fixed $N =$ 500 (right panel) and several values of ${\rm Pe}$.}
\label{fig:boncorr}
\end{figure*}

We report here further examples of bond-bond correlations along the contour of the ring, (see Fig.~\ref{fig:boncorr}).

We focus here on the effect of increasing ${\rm Pe}$ for rings of different sizes $N =$ 70 (short) and $N =$ 500 (long). Notice that the function, for short rings, shows a progressively marked minimum at $s = N/2$ upon increasing ${\rm Pe}$, akin of progressively more rigid rings. On the contrary, for long rings the functions at ${\rm Pe} >$ 1 completely overlap and show a common minimum at $s \simeq$ 5. This hints at the fact that, aside of having typical "spires" or "wraps" of $~$5 beads, the collapsed section is randomly organized; notice, though, that this fact also supports the universality of the route to collapse we described in the main text.

\subsection{Tangleness}

\begin{figure}[!h]
\centering
\includegraphics[width=0.45\textwidth]{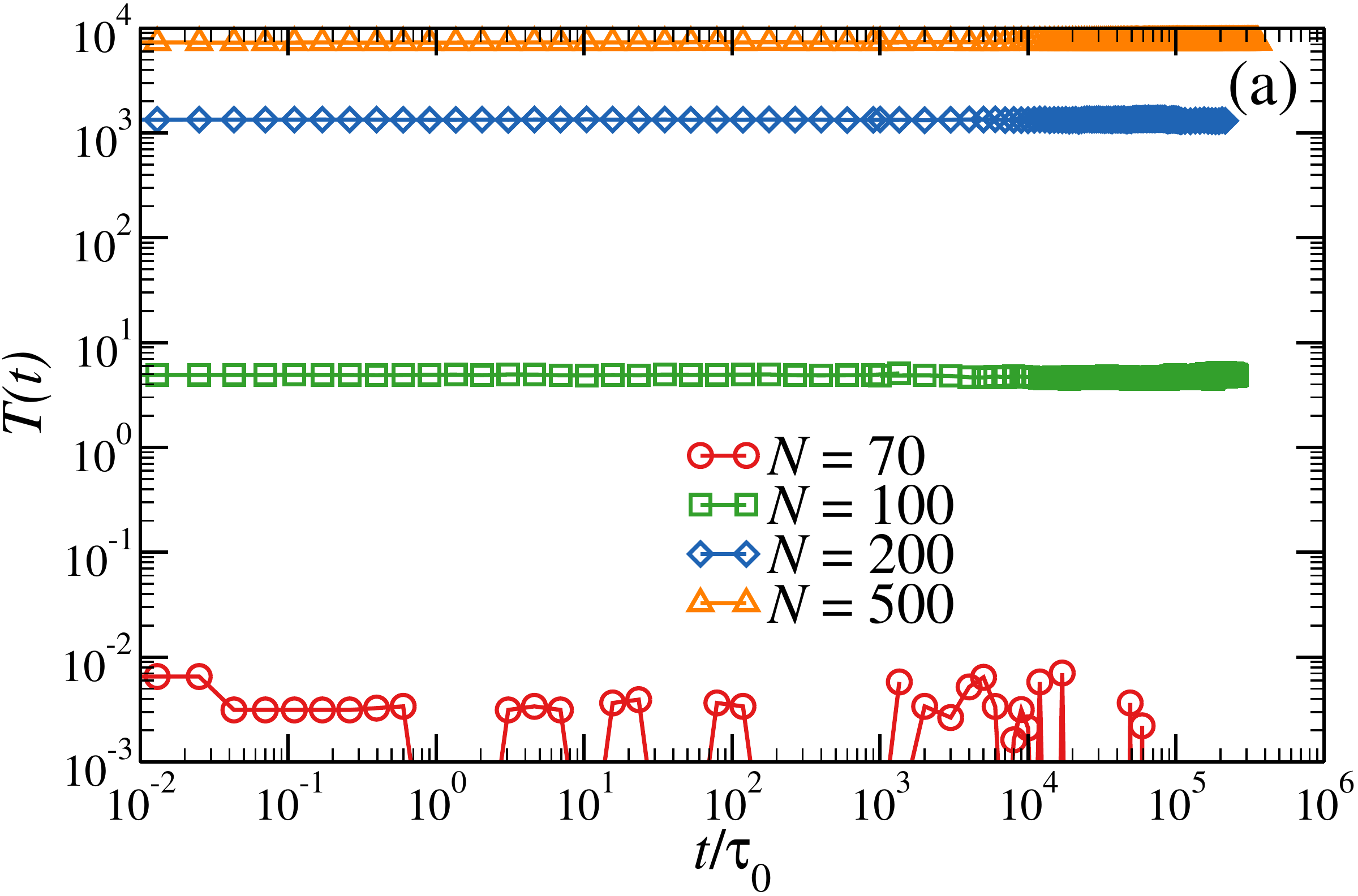}
\includegraphics[width=0.45\textwidth]{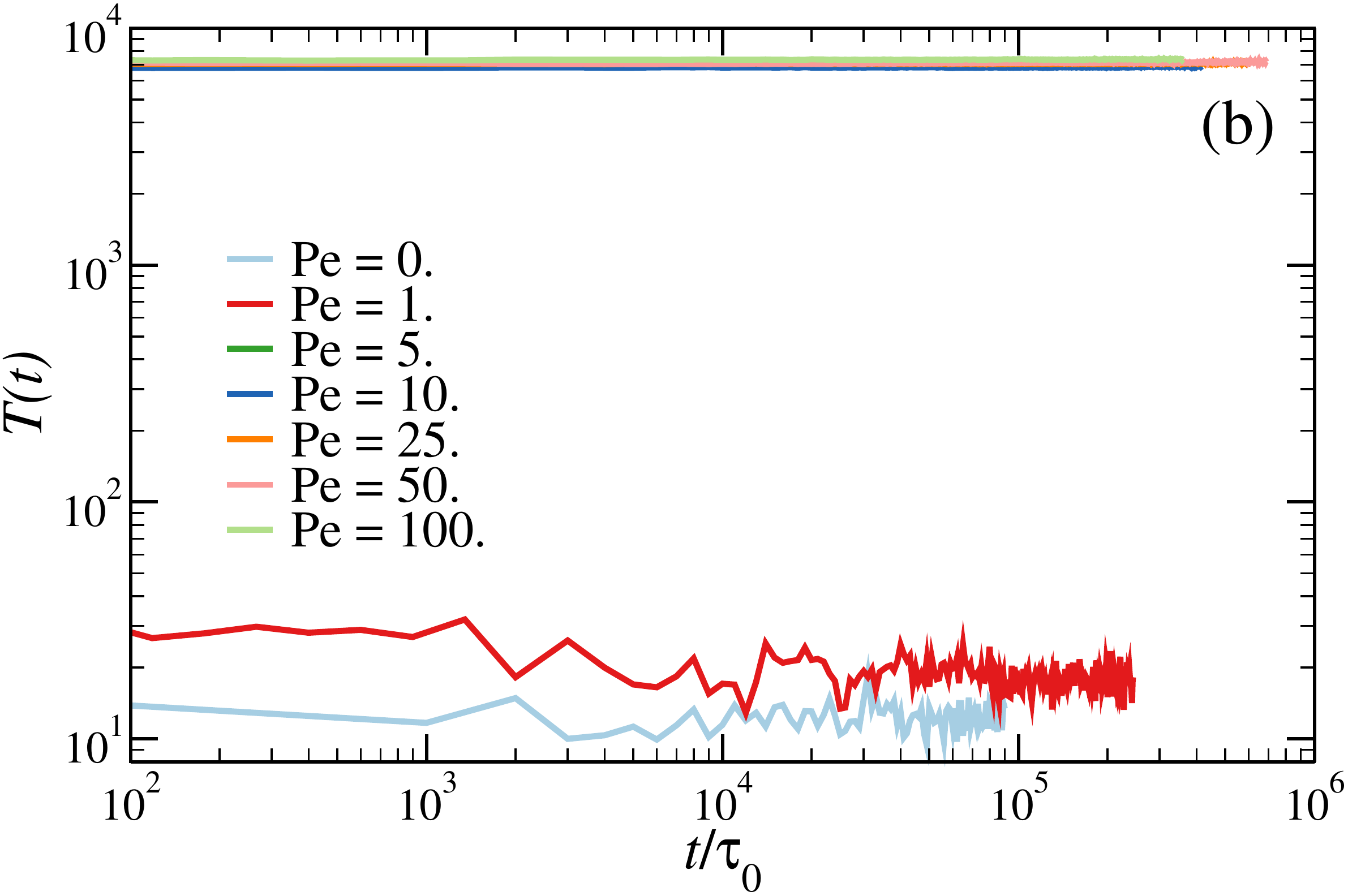}
\caption{Tangleness $T(t)$, defined in Eq.~(\ref{eq:tangl}), as a function of time in the steady state for (a) rings of different length $N$ at fixed ${\rm Pe} =$ 100, (b) rings of fixed $N =$ 500 and different values of ${\rm Pe}$.}
\label{fig:tangleness2}
\end{figure}

In Fig.~\ref{fig:tangleness2}, we show $T(t)$ as function of time during the steady state. As for ${\rm argmin}(\beta(s))$, also $T(t)$ remains constant, the particular value attained is characteristic of the ring length and increases upon increasing $N$ (see panel a). Such increase, given the definition of $T(t)$, is reasonable in a collapsed state as, increasing $N$, larger contour distances between monomers may be achieved. Further, the tangleness greatly increases upon increasing ${\rm Pe}$ at fixed $N$ (panel b): clearly, this massive difference reflect the fact that the system is in a collapsed state for sufficiently high values of ${\rm Pe}$. 

\section{Collapsed rings internal dynamics at the steady state}

\subsection{Neighbour survival}

We report here further data regarding the neighbour survival $S(t)$, i.e. the the average fraction of original neighbours as function of time.

\begin{figure}[!h]
\centering
\includegraphics[width=0.45\textwidth]{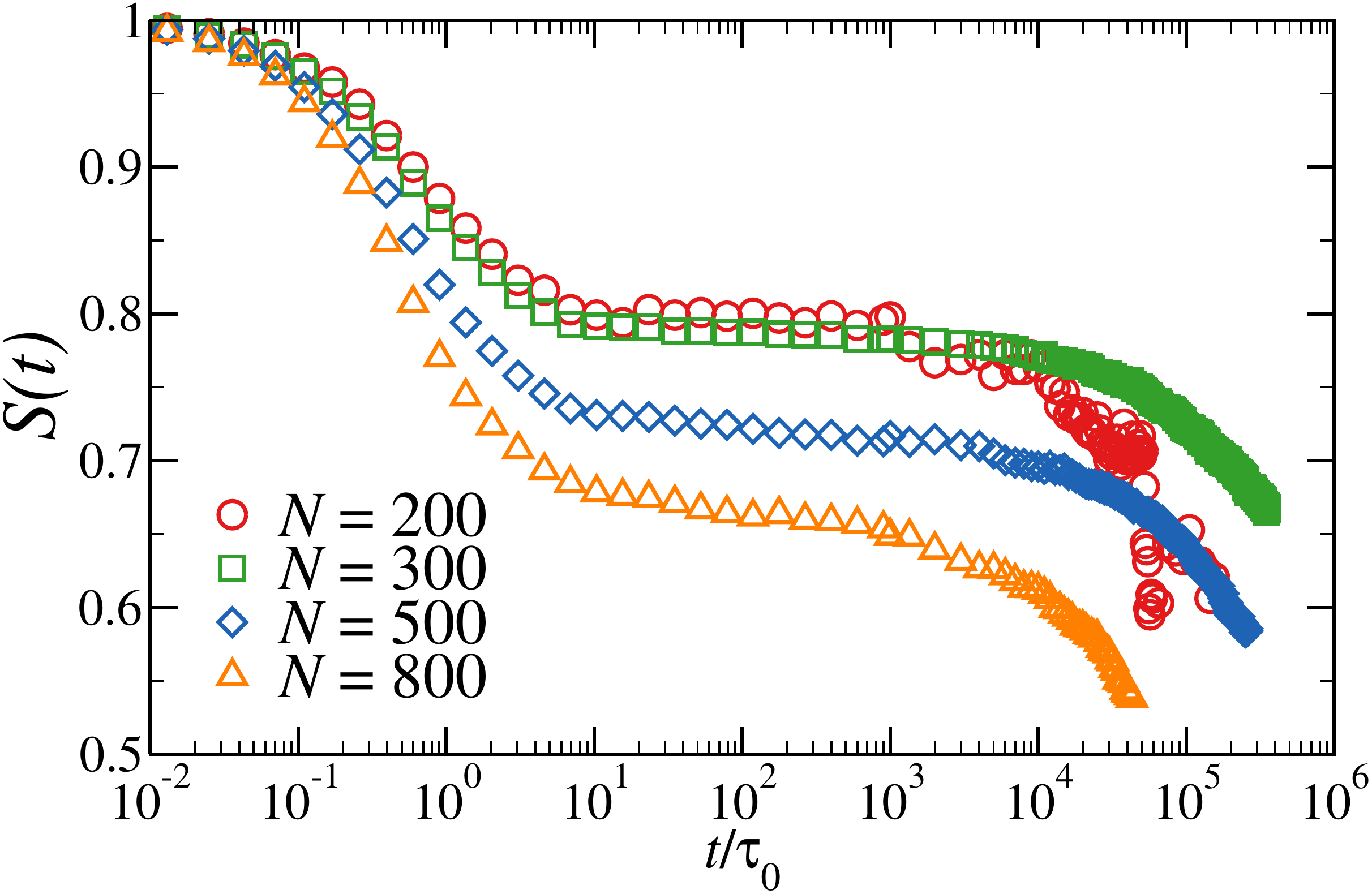}
\caption{Fraction of survived neighbours as function of time for rings of length $N =$ 500 and ${\rm Pe} =$ 0 (left panel), ${\rm Pe} =$ 25 (right panel).}
\label{fig:neigh_N}
\end{figure}

In Fig.~\ref{fig:neigh_N}, we report $S(t)$ for rings of different length. We can appreciate how the length of the plateau in time does not change for 200 $< N <$ 500 and only slightly diminishes for $N =$ 800. 

In the main text, the cut-off value chosen for this quantity is $r_c =$ 1.2 $\sigma$, which is a bit larger than the potential cut-off value 1.032$\sigma$.
\begin{figure}[!h]
\centering
\includegraphics[width=0.45\textwidth]{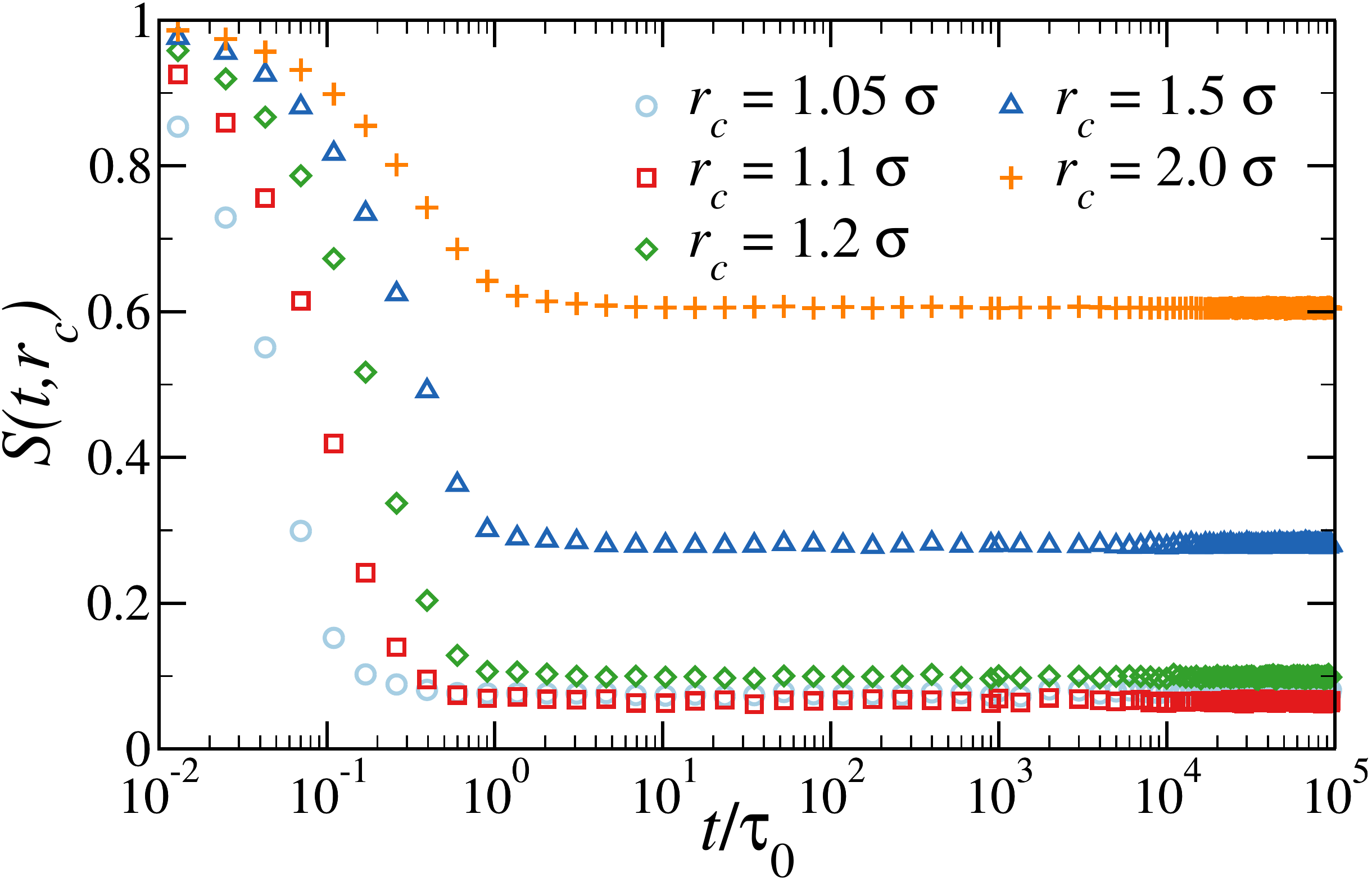}
\includegraphics[width=0.45\textwidth]{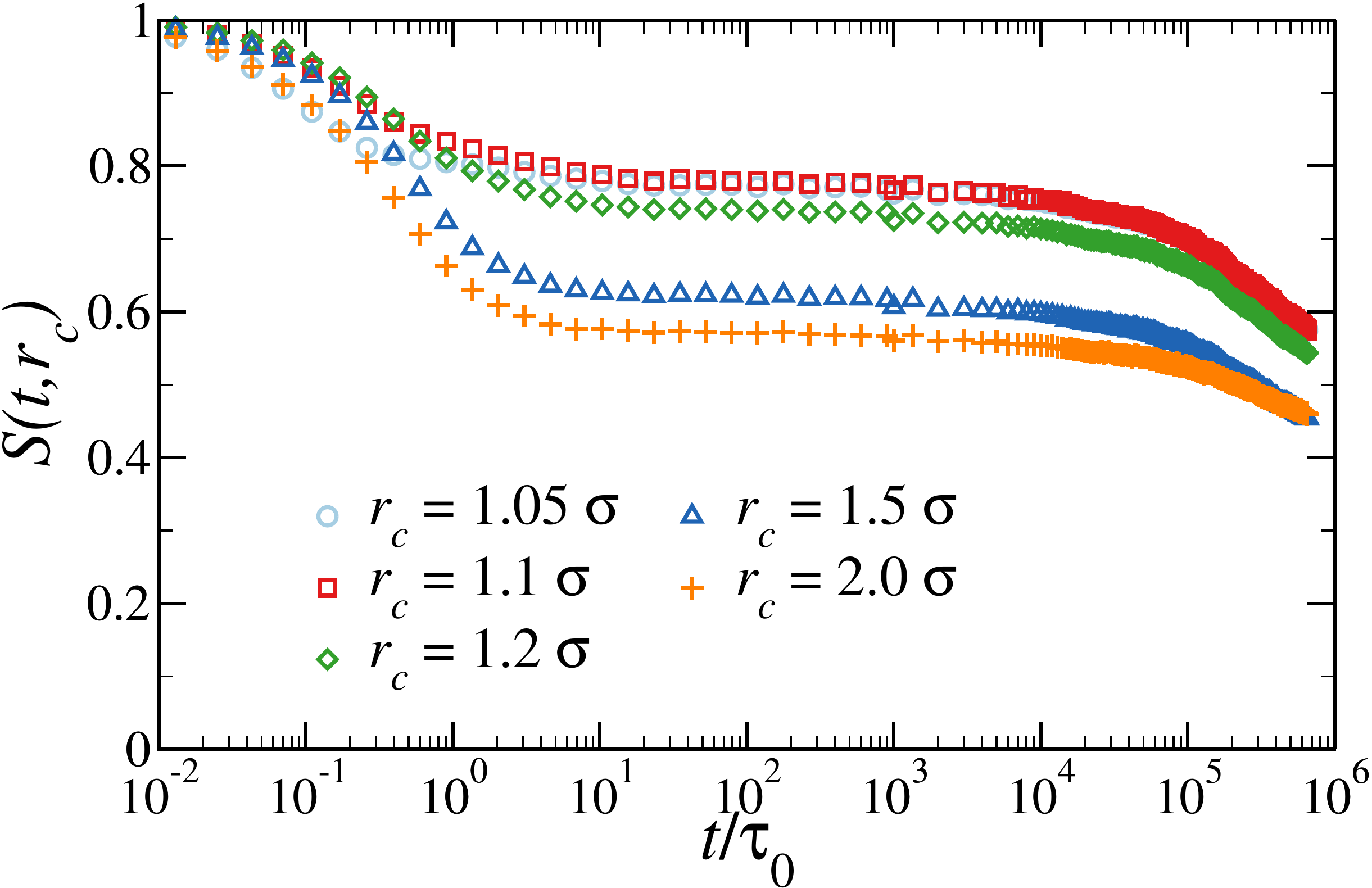}
\caption{Fraction of survived neighbours as function of time for rings of length $N =$ 500 and ${\rm Pe} =$ 0 (left panel), ${\rm Pe} =$ 25 (right panel) and different values fo the cut-off radius $r_c$, used in the definition of the neighbour search.}
\label{fig:neigh_rc}
\end{figure}
We check the effect of the cut-off radius $r_c$ on $S(t)$ in Fig.~\ref{fig:neigh_rc}. As we employ this observable to characterize the collapsed state, we focus to a specific length $N =$ 500 and on a specific activity ${\rm Pe} =$ 25; we also report the passive case ${\rm Pe} =$ 0. In both cases, we find that the shape of the curve remains the same for all $r_c$ investigated. We notice that, for the passive rings (left panel), the value of the final plateau grows upon growing $r_c$ as beads progressively further along the backbone are identified as neighbours and remain permanently close by. On the contrary, in the collapsed case the value of intermediate plateau diminishes. Choosing a large cut-off radius in the collapsed state implies to select more neighbours, for any given monomer. This means that increasingly more monomers have neighbours belonging to the rather mobile dangling sections; small rearrangements influence more monomers. Thus quantitatively the overall permanence diminishes but, qualitatively, the plateau remains.   

\subsection{Self-intermediate scattering function}

We report here the self-intermediate scattering functions $F_s(k)$ for active rings; $F_s(k)$ is usually employed to characterize the dynamics of arrested states, as it reveals the characteristic time scales of the system. 

\begin{figure}[!h]
\centering
\includegraphics[width=0.45\textwidth]{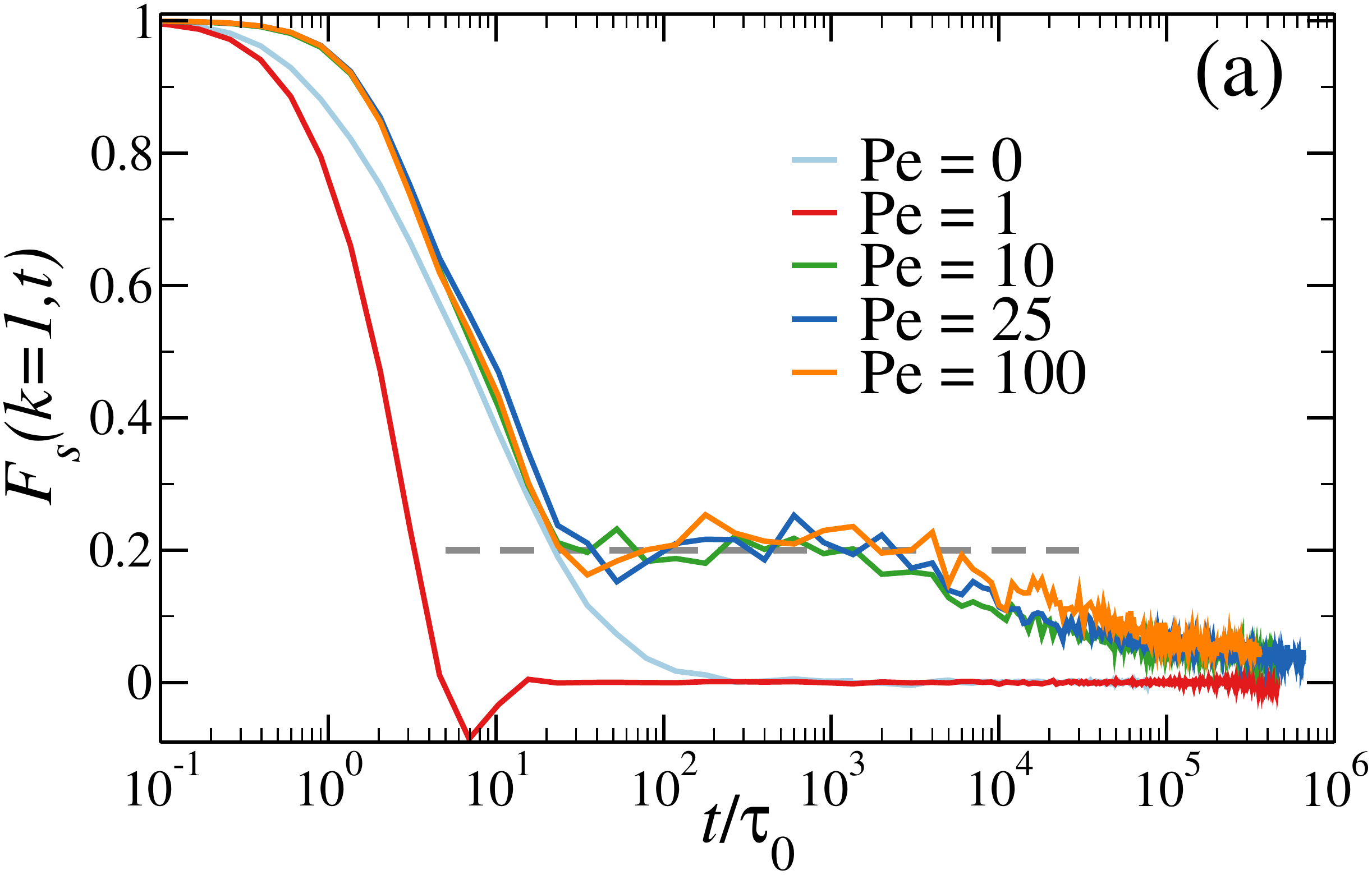}
\includegraphics[width=0.45\textwidth]{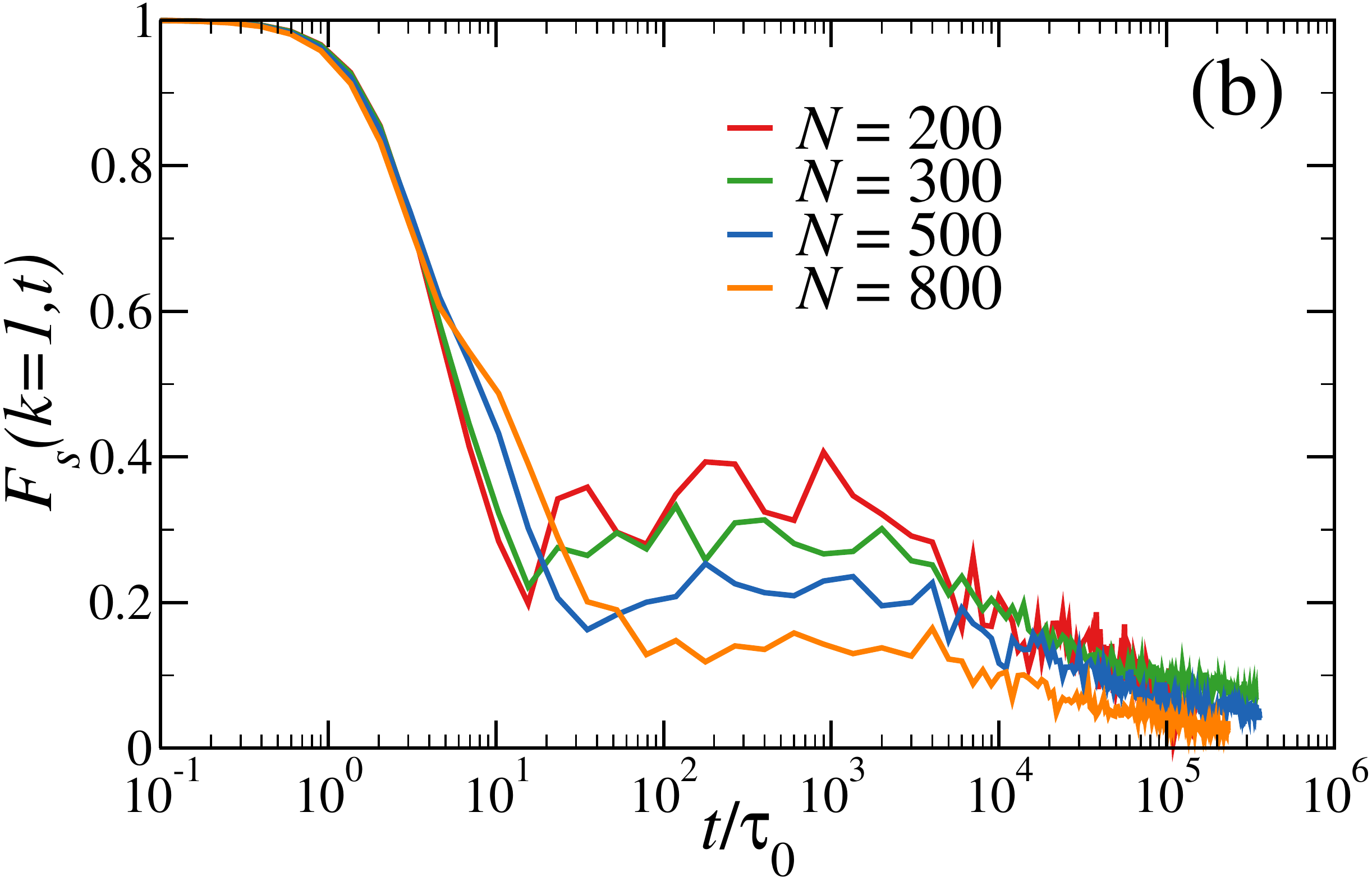}
\caption{Self intermediate scattering function for (a) fixed $N =$ 500 and different values of ${\rm Pe}$ and (b) fixed ${\rm Pe} =$ 100 and several values of $N$.}
\label{fig:Fs_k}
\end{figure}

\begin{figure}[!h]
\centering
\includegraphics[width=0.45\textwidth]{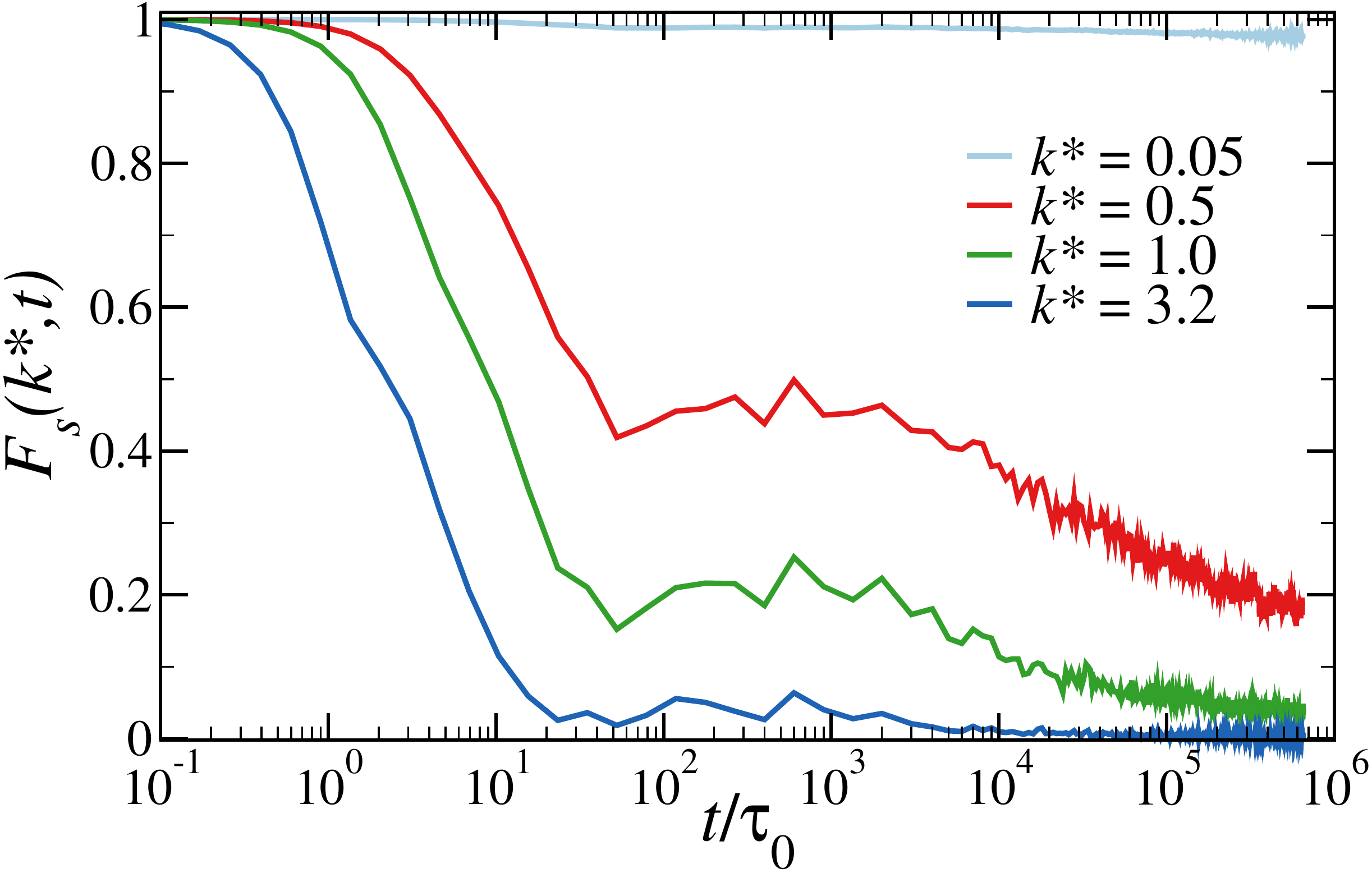}
\caption{Self intermediate scattering function for rings of fixed $N =$ 500 and ${\rm Pe} =$ 25 and different values of $k$.}
\label{fig:self_k}
\end{figure}

We report, in Fig.~\ref{fig:Fs_k} the intermediate scattering function for either rings of fixed length $N =$ 500 (panel a) or fixed activity ${\rm Pe} =$ 100(panel b). In panel (a), $F_s(k)$ for the non-collapsed cases ${\rm Pe} =$ 0, 1 shows a single decay; interestingly, for ${\rm Pe} =$ 1 it also shows a negative dip, observed in self-propelled systems\cite{kurzthaler2016intermediate}. When rings collapse (${\rm Pe} >$ 1 for $N =$ 500), $F_s(k)$ shows a double decay, a well known hallmark of arrested systems.\\In Fig.~\ref{fig:Fs_k}b, we compare the self-intermediate scattering function for rings of different length at fixed ${\rm Pe}$ (all cases considered feature collapsed configurations in the steady state). Observe that the double decay is present for all $N$ considered and, in this case, the length of the plateau is independent on $N$.\\
When calculating the self-intermediate scattering function, one usually chooses to fix the magnitude of the wave vector $k$ as the value at which the maximum of the static structure factor occurs. As the form factor of a macromolecule does not show, usually, any well defined peak, we show here the effect of using $k$-vectors of different magnitude (see Fig.~\ref{fig:self_k}). 

We focus on collapsed rings ($N =$ 500, ${\rm Pe} =$ 25 in Fig.~\ref{fig:self_k}). We observe that, if $k$ is very small, the relaxation time probed is the one of the overall macromolecule; it is thus extremely long and not fully developed during the time of the simulation. On the other hand, for $k >$ 1 the dynamics probed concerns length scales smaller than the monomer size $\sigma$ and we thus observe only a fast relaxation. Otherwise, for $k \approx$ 1 we observe the presence of a two step relaxation process, as reported in the main text.    

\subsection{Ring time correlation functions}

The last two quantities we investigate in order to characterise the arrest of the collapsed rings are two time correlation functions: the correlation of the "half-ring" vector, i.e. the vector connecting any monomer with its furthest peer $N/2$ far away along the contour, and the correlation of the vector normal to the surface of the ring, defined in Section "Observables".

\begin{figure}[!h]
\centering
\includegraphics[width=0.45\textwidth]{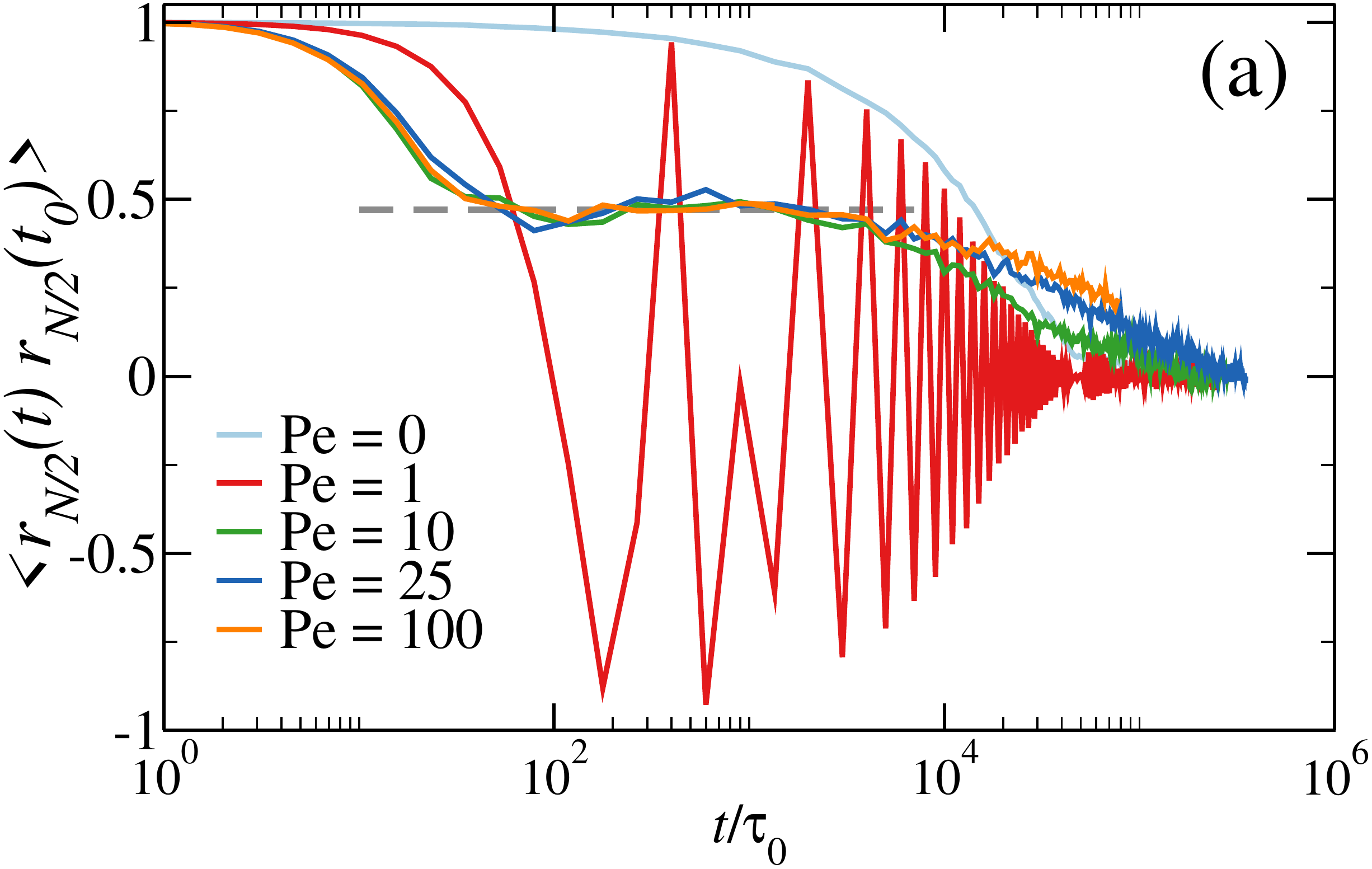}
\includegraphics[width=0.45\textwidth]{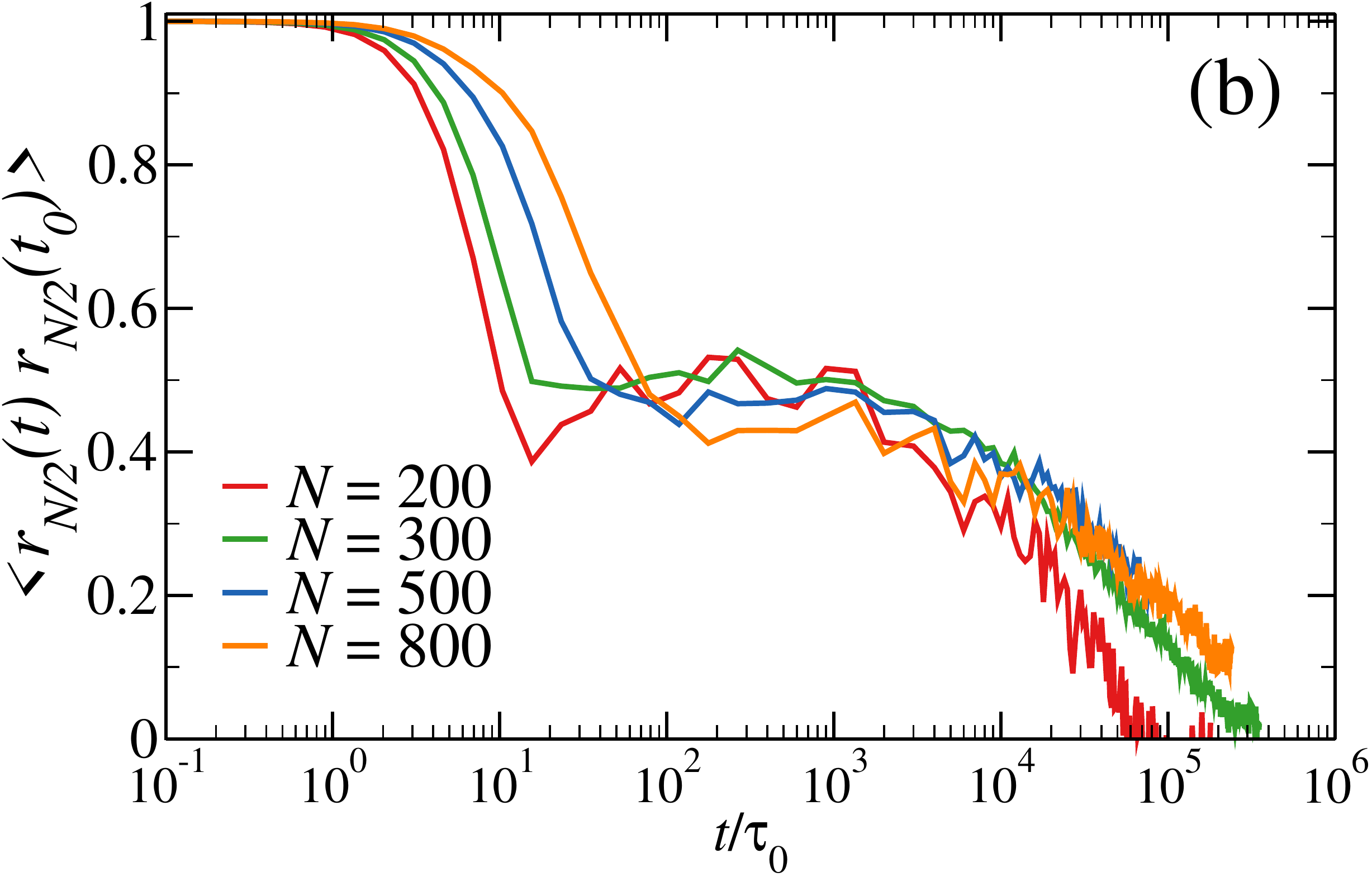}
\caption{Time correlation of the half-ring vector for rings of fixed length $N =$ 500 and different ${\rm Pe} =$ 0 (panel a), fixed ${\rm Pe} =$ 100 and different $N$ (panel b).}
\label{fig:ree_corr}
\end{figure}

\begin{figure}[!h]
\centering
\includegraphics[width=0.45\textwidth]{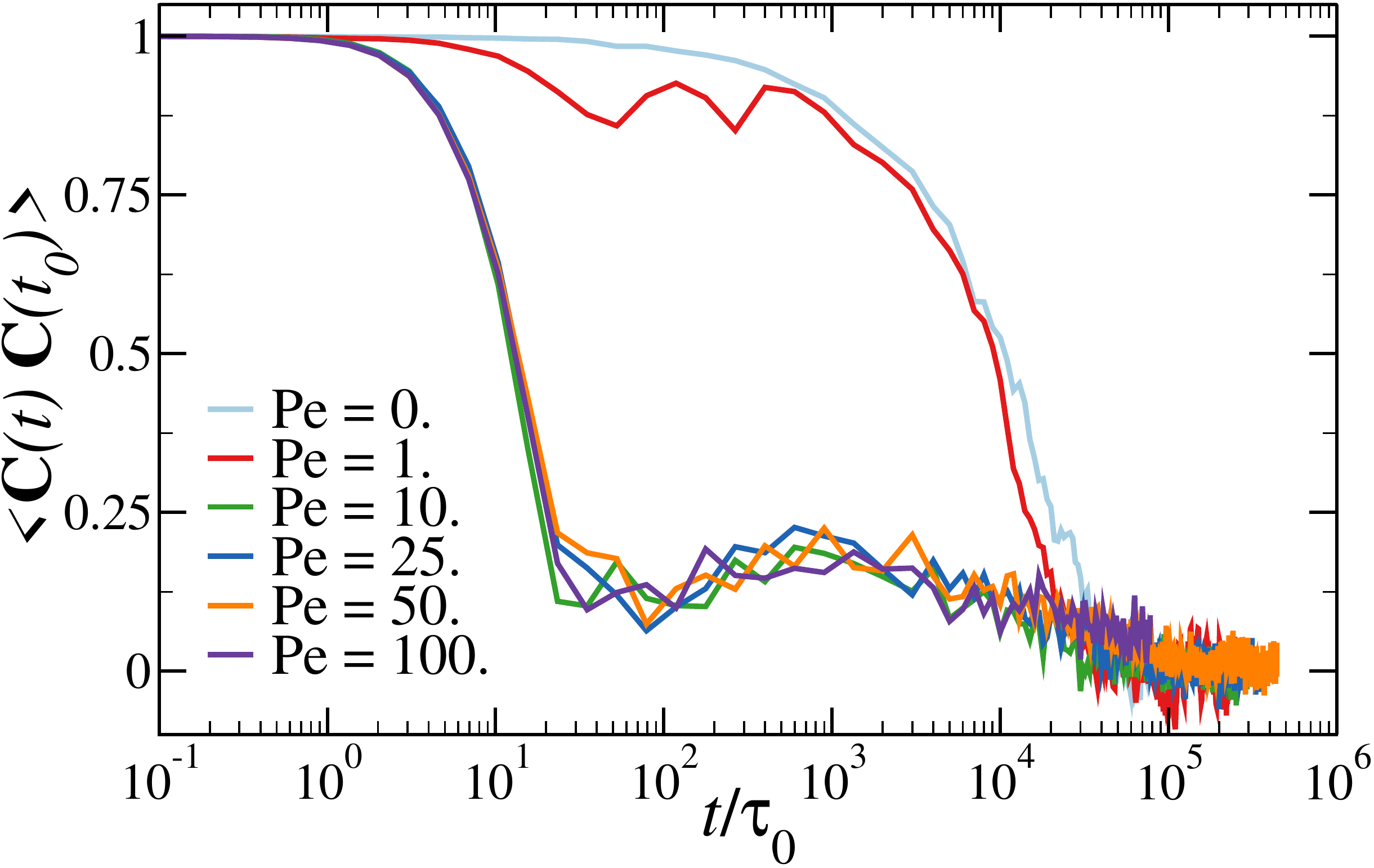}
\includegraphics[width=0.45\textwidth]{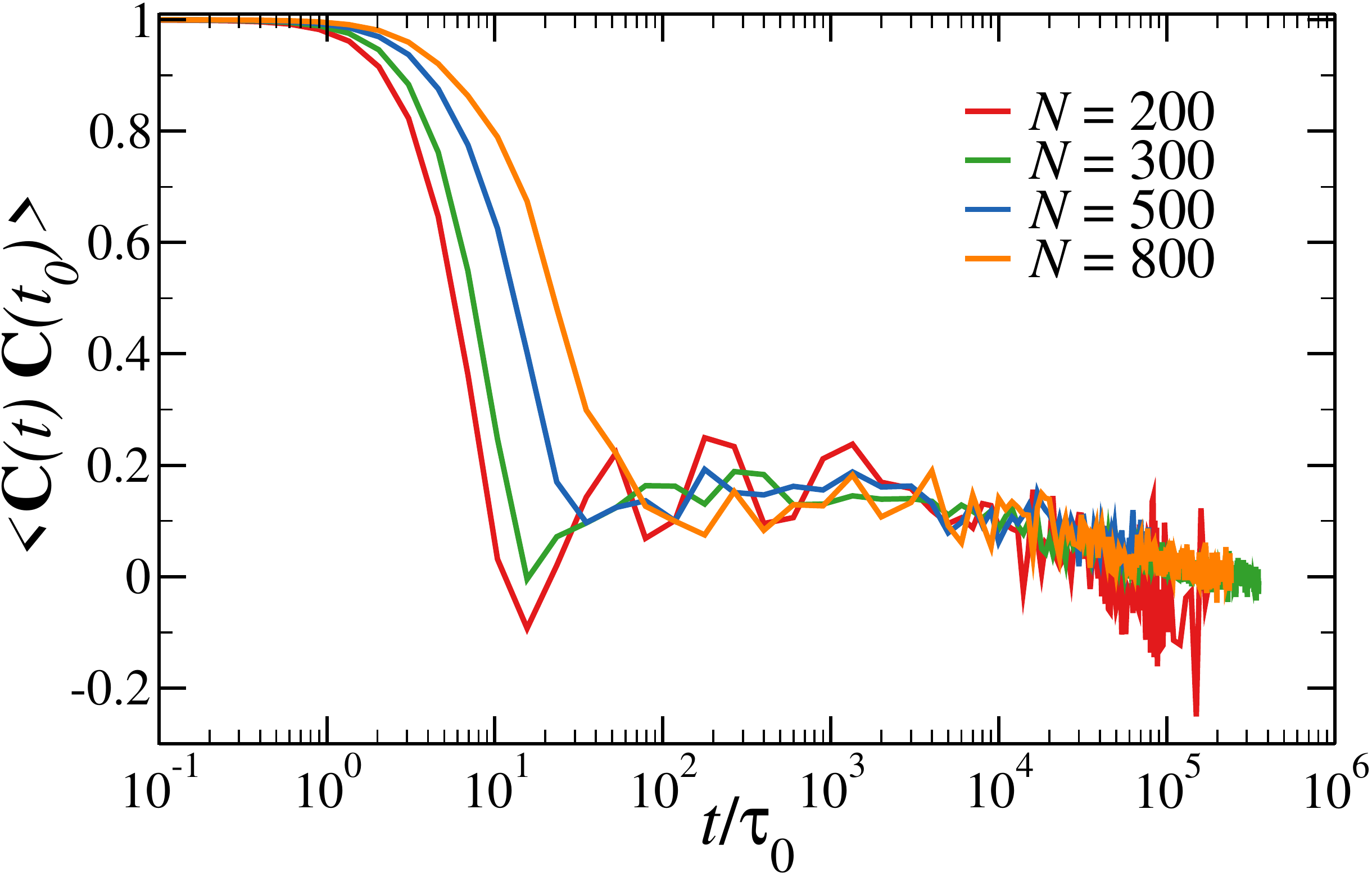}
\caption{Time correlation of the characteristic vector (see Eq.~(\ref{eq:v_ring})) for ring of fixed $N =$ 500 and several values of ${\rm Pe}$ (panel a) and fixed ${\rm Pe} =$ 100 and different $N$ (panel b).}
\label{fig:vring_corr}
\end{figure}

We report, in Fig.\ref{fig:ree_corr} the time correlation of the "half-ring" vector $r_{N/2}$. For the passive case, the function exhibits a single decay. For ${\rm Pe} =$ 1, we observe an oscillatory behaviour, fairly regular, whose intensity decays over time, following the decay of the passive ring. This behaviour reveals an overall rotation of the ring, fairly regular, powered by the activity. A single rotation happens, roughly, over $10^2 \tau_0$. In the case of the collapsed rings, interestingly we notice again the presence of a double decay; in Fig.~\ref{fig:ree_corr}b, we notice it is present for all $N$ investigated.\\
Further, we report in Fig.~\ref{fig:vring_corr} the correlation time of the vector normal to the surface of the ring, defined in Eq.~(\ref{eq:v_ring}).  

In Figure~\ref{fig:vring_corr}a, we again notice that ${\rm Pe} =$ 0,1, cases that do not feature any collapse, are quite similar among each other. Indeed the effect of the activity, for ${\rm Pe} =$ 1, is simply to rotate the ring and, for this particular observable, does not induce any drastic change. For all the other values of ${\rm Pe} $ investigated, we again observe a double decay, as well as, in Fig.~ \ref{fig:vring_corr}b, for all the different values of $N$ reported (which, again, all collapse).\\
For these time correlations, the double decay indicates the presence of fast-evolving sections, which we ascribe to the sections outside of the collapsed part of the ring (dangling sections in the main text) and of very slow regions (or possibly a single region) that freeze the evolution of the dynamics for a period of time and further impede the relaxation of the whole ring at very long times.

\subsection{Velocity distributions}

\begin{figure}[!h]
\centering
\includegraphics[width=0.45\textwidth]{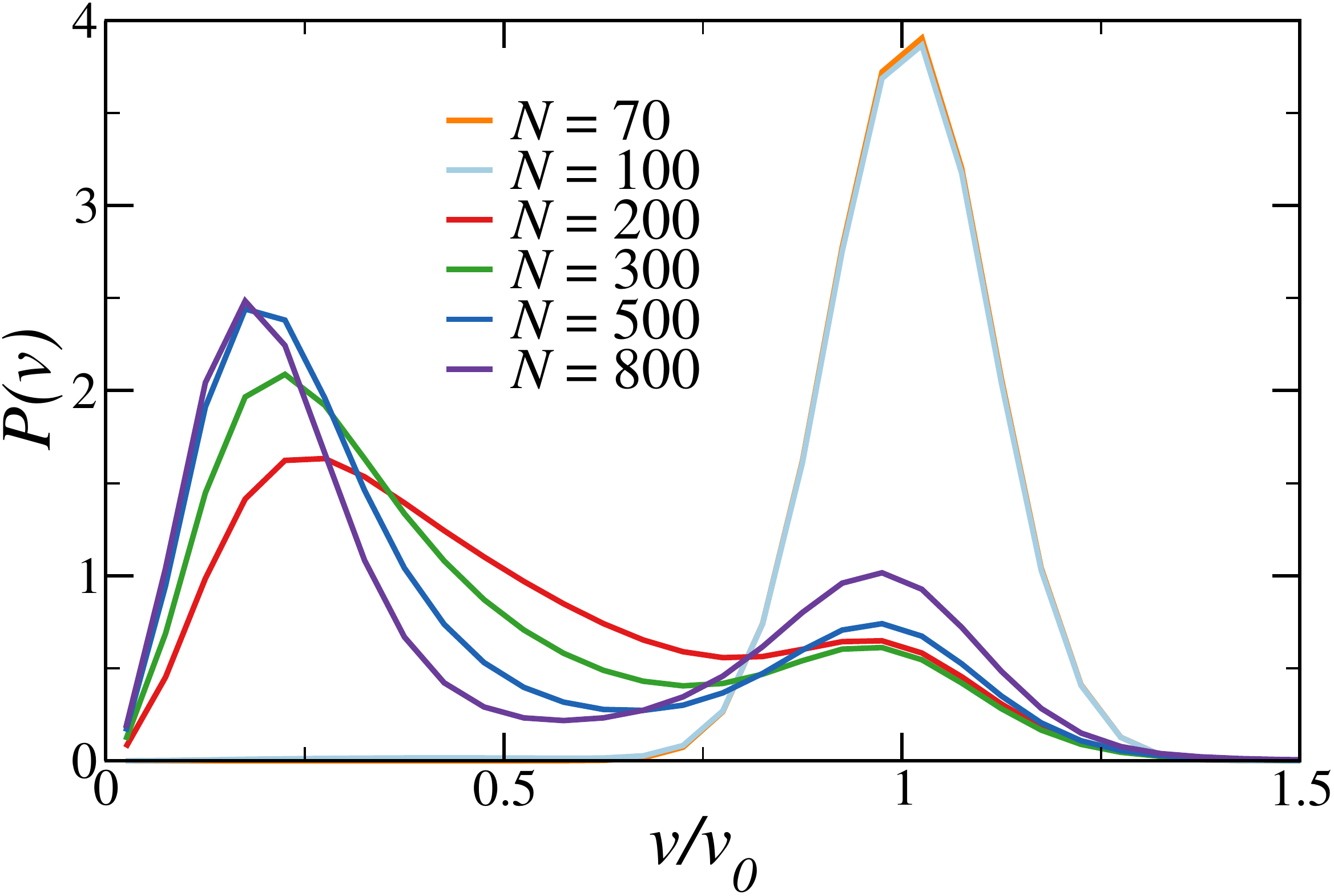}
\includegraphics[width=0.45\textwidth]{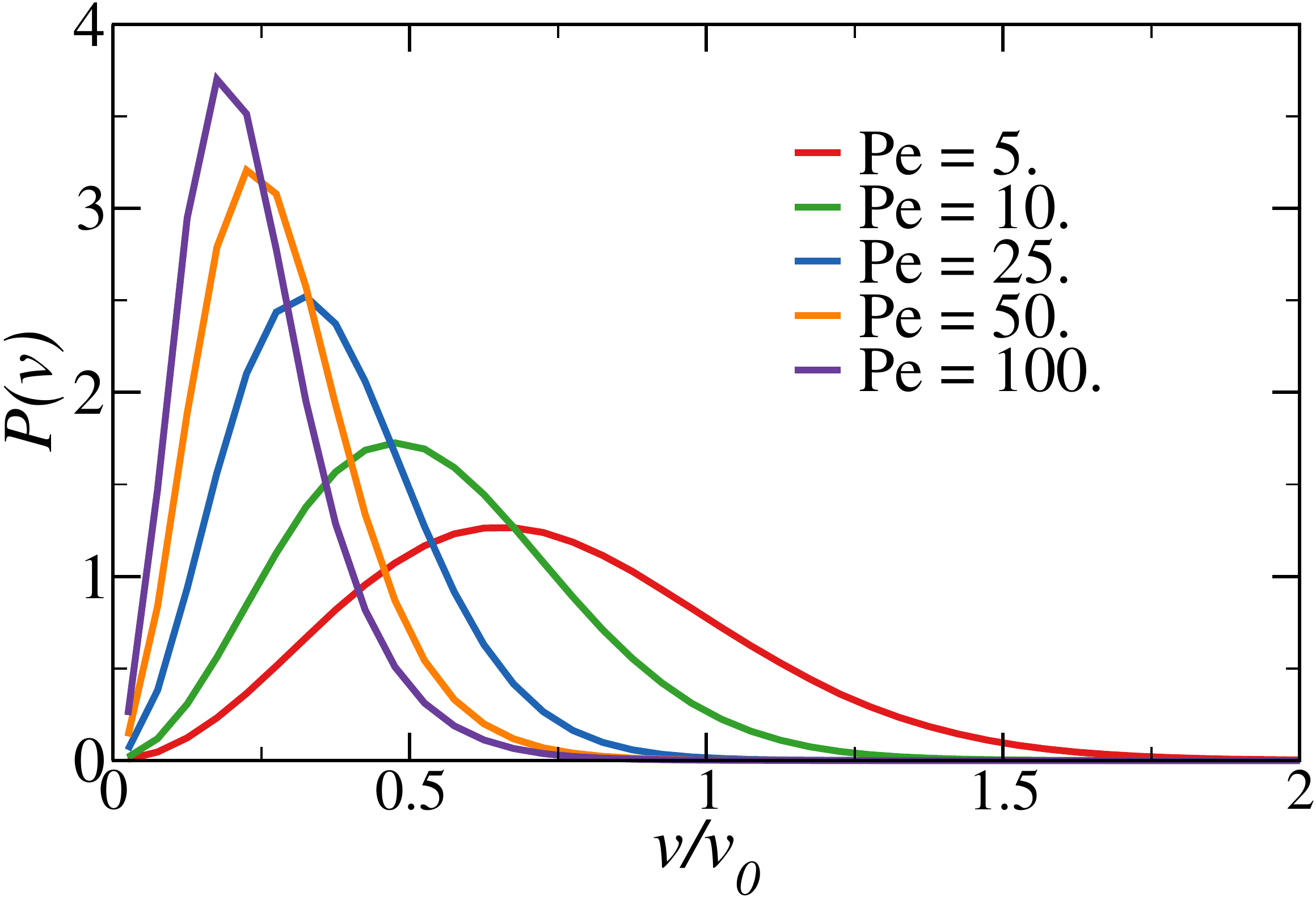}
\caption{(right panel) Monomer velocity distributions for rings of fixed ${\rm Pe} =$ 100 and several values of $N$. (Left panel): Monomer velocity distributions at the steady state for fixed $N =$ 500, several ${\rm Pe}$; distributions are limited to monomers belonging to the collapsed section. }
\label{fig:veldistro}
\end{figure}

We show here the distribution of the monomer velocities for rings of different lengths (see Fig.~\ref{fig:veldistro}, left panel).

For rings that are not collapsed, the distribution is peaked around $v = v_0$, i.e. the mean velocity of the active monomers; for collapsed rings, we observe for all $N$ the presence of two peaks, as described in the main text. Interestingly, both the first and the second peak grow increasing $N$; this suggests that, for longer rings, dangling sections are more common than for shorter rings and, in general, monomer segregate into very slow or very mobile groups. \el{Finally, in right panel of Fig.~\ref{fig:veldistro} we observe the distribution of the velocity of individual monomers for the same systems reported in the main text (Fig.~4a); here we consider, using the cluster algorithm described in the Section "Clustering Algorithm and Neighbour Search", only the monomers belonging to the collapsed sections. Indeed, we confirm that the slow monomers are the ones belonging to the collapsed sections; by exclusion, fast monomers belong to the dangling sections.}

\subsection{Angular velocity at the steady state for large rings}

\begin{figure}[!h]
\centering
\includegraphics[width=0.45\textwidth]{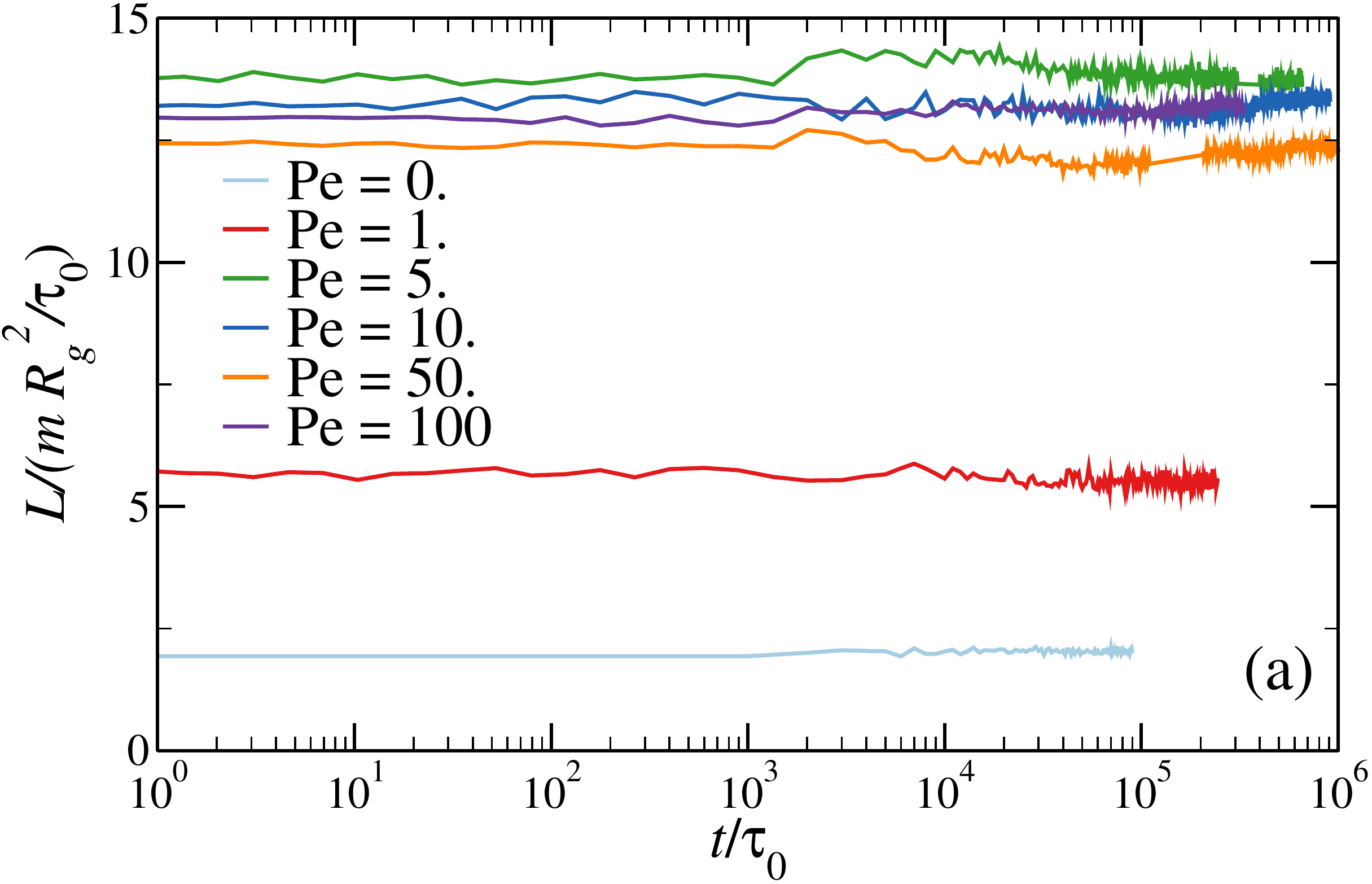}
\includegraphics[width=0.45\textwidth]{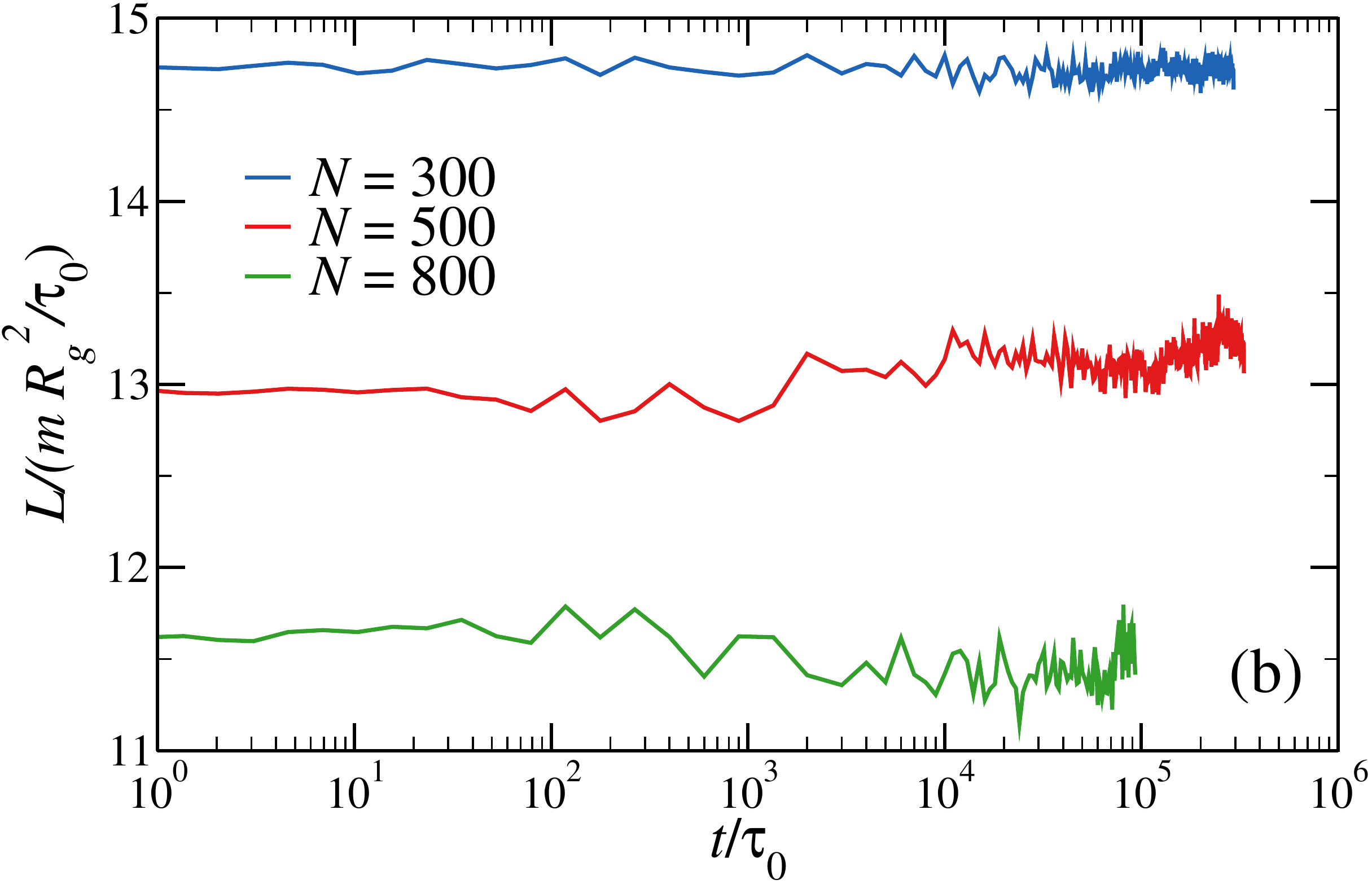}\\
\caption{Average of the angular momentum magnitude over the average size of the ring $L/(m \, R_g^2/ \tau_0)$  as function of time for a) fixed $N =$ 500, several values of ${\rm Pe}$; b) fixed ${\rm Pe} =$ 100, several $N$.}
\label{fig:angvel}
\end{figure}

\el{We compute the angular momentum in the \pa{frame of reference of the} centre of mass  
\begin{align}
    \mathbf{L} = \sum_{i=1}^{N} \mathbf{r}_i \times \mathbf{p}_i\,,
\end{align} where $\mathbf{r}_i$ and $\mathbf{p}_i$ are the 
\pa{distance from the center of mass} and the linear momentum of the monomer $i$, respectively. We report, in Fig.~\ref{fig:angvel}, the average magnitude of the angular momentum $L = \langle |\mathbf{L}| \rangle$, 
normalized by by $m \, R_g^2 / \tau_0$. Such quantity 
\pa{provides} an estimation of the state of angular \pa{velocity} of the object, as the angular momentum is 
normalized by the average size of the ring. In Fig.~\ref{fig:angvel}a, we compare $L/(m \, R_g^2/ \tau_0)$ for rings of fixed length $N =$ 500 and different values of ${\rm Pe}$. We observe that the average angular momentum magnitude is constant in time and its value is larger for ${\rm Pe} >$ 5, in comparison with ${\rm Pe =}$ 0,1 cases: notice that all rings at ${\rm Pe} >$ 5 are collapsed. In Fig.~\ref{fig:angvel}b) we compare $L/(m \, R_g^2/ \tau_0)$ for rings with a fixed value of ${\rm Pe} =$ 100 and three values of $N =$ 300, 500, 800; here all three cases display collapsed conformations at the steady state for the chosen value of ${\rm Pe}$. We observe that smaller rings have a larger rotational motion; possibly, they are powered by a proportionally larger active torque as the collapsed section is less disordered than for larger rings. The results suggest that, indeed, collapsed rings possess a higher rotational motion with respect to non-collapsed rings of the same size.}

\end{document}